\documentclass[a4paper,fleqn,usenatbib]{mnras}
\usepackage{newtxmath}
\usepackage[T1]{fontenc}
\usepackage{ae,aecompl}
\usepackage{amsmath}	% Advanced maths commands
\usepackage{amssymb}	% Extra maths symbols
\usepackage{epsfig}

{\newif\ifnotend
\notendtrue
\def\veclist{ABCDEFGHIJKLMNOPQRSTUVWXYZabcdefghijklmnopqrstuvwxyz.}
\def\top#1#2.{#1}
\def\tail#1#2.{#2.}
\loop\expandafter\xdef\csname v\expandafter\top\veclist\endcsname%
{{\noexpand\bf\expandafter\top\veclist}}
\edef\veclist{\expandafter\tail\veclist}
\if\veclist.\notendfalse\fi\ifnotend\repeat}

\newcommand {\vlos}{v_{\rm los}}
\newcommand {\kms}{\,{\rm km}\,{\rm s}^{-1}}
\newcommand {\magn}{\,{\rm mag}}

\newcommand {\yr}{\,{\rm yr}}
\newcommand {\Gyr}{\,{\rm Gyr}}
\newcommand {\K}{\,{\rm K}}
\newcommand {\pc}{\,{\rm pc}}
\newcommand {\kpc}{\,{\rm kpc}}
\newcommand {\RA}{{\rm RA}}
\newcommand {\DEC}{{\rm DEC}}
\newcommand {\marcs}{{\rm mas}}
\newcommand {\marcsyr}{{\rm \, mas \, yr^{-1}}}
\newcommand {\gl}{{l}}
\newcommand {\gb}{{b}}
\newcommand {\mul}{\mu_l}
\newcommand {\mub}{\mu_b}
\newcommand {\Var}{{\rm Var}}
\newcommand {\Cov}{{\rm Cov}}

\newcommand {\sigW}{{\sigma_W}}
\newcommand {\SNR}{{\rm SNR}}
\newcommand {\Ug}{U_g}
\newcommand {\Vg}{V_g}

\newcommand {\Vc}{V_c}

\newcommand {\Vh}{V}

\newcommand {\epar}{e_{\parallel}}
\newcommand {\siglos}{\sigma_{\rm los}}
\newcommand {\sigb}{\sigma_{\rm bin}}
\newcommand {\epari}{e_{\parallel,i}}

\newcommand {\sigloslamost}{\sigma_{{\rm los, LAM}}}
\newcommand {\siglosrave}{\sigma_{{\rm los, RAVE}}}
\newcommand {\degs}{{\rm deg}}
\newcommand {\B}[1]{{\boldsymbol{#1}}}
\newcommand {\Rsun}{{R_{0}}}

\newcommand {\vsun}{{\B{\upsilon_\odot}}}
\newcommand {\Usun}{{U_{\!\odot}}}

\newcommand {\Wsun}{{W_{\!\odot}}}
\newcommand {\Vsun}{{V_{\!\odot}}}

\newcommand {\zsun}{{z_{\!\odot}}}

\renewcommand{\deg}{{^{\circ}}}

\newcommand {\vpe}{v_{\perp}}
\newcommand {\feh}{\hbox{[Fe/H]}}
\newcommand {\dex}{\,{\rm dex}}
\newcommand {\Teff}{T_{\rm eff}}
\newcommand {\Vmsechs}{{{\rm V}\alpha 9 {\rm s} 8 \lambda \zeta }}
\newcommand {\Msun}{\,{\rm M}_\odot}

\title[Distances and Kinematics in Gaia/TGAS]{Assessing distances and consistency of kinematics in Gaia/TGAS}
\author[R. Sch\"onrich \& M.Aumer]{Ralph Sch\"onrich\thanks{E-mail: ralph.schoenrich@physics.ox.ac.uk}
        and Michael Aumer\\
        Rudolf Peierls Centre for Theoretical Physics, 1 Keble Road, Oxford, OX1 3NP, UK}

\date{Draft, \today}
\pubyear{2017}

\begin{document}
\label{firstpage}
\pagerange{\pageref{firstpage}--\pageref{lastpage}}
\maketitle

\begin{abstract}
We apply the statistical methods by Sch\"onrich, Binney \& Asplund to assess the quality of distances and kinematics in the RAVE-TGAS and LAMOST-TGAS samples of Solar neighbourhood stars. These methods yield a nominal distance accuracy of $1-2\%$. Other than common tests on parallax accuracy, they directly test distance estimations including the effects of distance priors. We show how to construct these priors including the survey selection functions (SSFs) directly from the data. We demonstrate that neglecting the SSFs causes severe distance biases. Due to the decline of the SSFs in distance, the simple 1/parallax estimate only mildly underestimates distances.
We test the accuracy of measured line-of-sight velocities ($\vlos$) by binning the samples in the nominal $\vlos$ uncertainties. We find: a) the LAMOST $\vlos$ have a $\sim -5 \kms$ offset; b) the average LAMOST measurement error for $\vlos$ is $\sim7\kms$, significantly smaller than, and nearly uncorrelated with the nominal LAMOST estimates. The RAVE sample shows either a moderate distance underestimate, or an unaccounted source of $\vlos$ dispersion ($\epar$) from measurement errors and binary stars. For a subsample of suspected binary stars in RAVE, our methods indicate significant distance underestimates. Separating a sample in metallicity or kinematics to select thick-disc/halo stars, discriminates between distance bias and $\epar$. For LAMOST, this separation yields consistency with pure $\vlos$ measurement errors. We find an anomaly near longitude $\gl\sim(300\pm60)\deg$ and distance $s\sim(0.32\pm0.03)\kpc$ on both sides of the galactic plane, which could be explained by either a localised distance error or a breathing mode.
\end{abstract}


\section{Introduction}
The Gaia satellite mission \citep[][]{Perryman01, GaiaDR1, GaiaDR1b} is the most important survey of this decade for the fields of Galactic and stellar astronomy. It is retrieving precision astrometry for of order one billion stars throughout the Galaxy, coupled to spectroscopic observations that allow determination of stellar parameters and line-of-sight velocities for a large subset of these stars. In addition, this effort is coupled to a plethora of ground-based spectroscopic surveys, including RAVE, Gaia-ESO, APOGEE, SEGUE, GALAH, LAMOST, and WEAVE.

Since the first release of Tycho-Gaia Astrometric Solution (TGAS) data \citep[][]{GaiaDR1, Lindegren16} there has been a variety of papers examining the quality of the data, ranging from the original validation paper \citep[][]{Lindegren16} to studies with standard candles, like red giants and red giant clump stars \citep[][]{deRidder16, Davies17}. A thorough compilation of standard validations of TGAS astrometry is found in \cite{Arenou17}. They discuss significant deviations of the median parallax from $0$ for remote objects in different regions of the sky, e.g. for objects in the Magellanic Clouds. Most such quality tests for parallaxes $p$ rely on distant objects, which should have $p = 0''$ (to within at most a few $\mu {\rm as}$), i.e. quasars, galaxies, or objects in neighbouring galaxies. For such tests the main caveats are potential contamination with Galactic sources, and to a minor extent their different geometry, and different spectral distribution. 

For Milky Way studies, it is vital to have a direct validation also of larger parallax measurements. However, well-constrained standard candles like Cepheids are not numerous enough to allow for resolved, high-precision testing. Most other types of standard candles, like RR Lyrae stars, Red Clump stars, or Blue Horizontal Branch stars are more distant than the local dwarf star samples. More importantly, their luminosities depend on additional parameters, like helium abundance, age, or metallicity. While relatively well-controlled via the period-dependence in pulsating stars, this is a major issue for the more numerous Red Clump stars. \cite{Girardi01} and \cite{Salaris02} have found dependencies in excess of $0.1 \magn$, severely limiting their value to test Gaia parallaxes at high precision. 

A key point for these validations is that even a perfectly unbiased parallax measurement does not yet imply unbiased distances. Each parallax measurement will still be uncertain, and so to estimate stellar distances, one has to translate the parallax error distribution into a distance distribution and combine the full information on this likelihood distribution \citep[][]{Stromberg27, Astraatmadja16} with models for the density of stars in each region, and (if stellar parameters are not directly used) with the selection function. Each of these factors can bias the derived stellar kinematics. They are frequently neglected in attempts to test and validate Gaia data on standard candles \citep[see e.g.][]{deRidder16, Davies17}. For everyone studying stellar kinematics, the main question is not if the parallaxes are unbiased, but if the derived stellar distances are unbiased. In this paper we will apply a statistical distance estimator (Sch\"onrich, Binney \& Asplund 2012, hereafter SBA) that does exactly this.

\begin{figure}
\epsfig{file=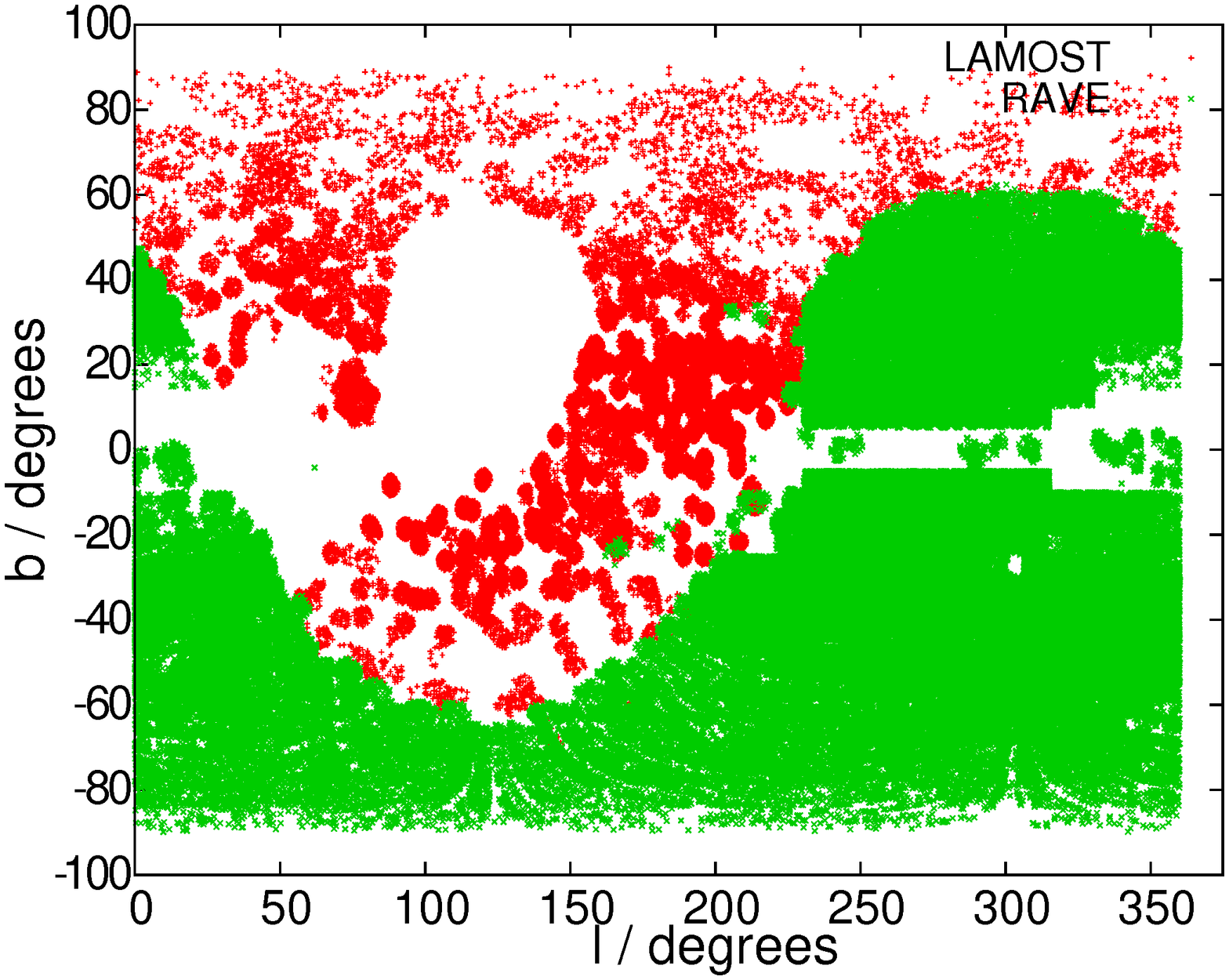,angle=-90,width=\hsize}
\caption{Positions of stars in the RAVE-TGAS (green) and LAMOST-TGAS (red) samples in galactic latitude $\gb$ vs. longitude $\gl$.}\label{fig:ravelamost}
\end{figure}

The RAVE-TGAS sample, released with RAVE DR5 \citep[][]{Kunder16}, is the first major combined sample of ground-based spectroscopic data from the Radial Velocity Experiment \citep[RAVE, ][]{RAVE} with combined astrometry from Gaia \citep[][]{GaiaDR1, Lindegren16}, Tycho-2 \citep[][]{Tycho2}, and Hipparcos \citep[][]{Perryman97, vanLeeuwen}. The RAVE-TGAS sample is also particularly interesting, since RAVE spectra have similar resolution and wavelength coverage to those of the Gaia spectrograph. The LAMOST survey  \citep[Large Sky Area Multiobject Fiber Spectroscopic Telescope, ][]{LAMOSTTEL, LAMOSTDR1} also overlaps with TGAS. As seen in Fig. \ref{fig:ravelamost}, it complements RAVE on the northern sky and is of great use to study the outer disc regions of the Milky Way.

The aim of this paper is to derive stellar distances from the Gaia parallaxes, and to validate the resulting kinematics. We make these data publicly available.\footnote{Please find the datasets with distances and kinematics and our source code at \small{http://www-thphys.physics.ox.ac.uk/people/RalphSchoenrich/data/ tgasdist/data.tar.gz}
or request them directly from the authors.} For the validation we use the method of SBA, which employs correlations between either radial ($U$) or azimuthal ($V$) with vertical ($W$) velocities and combinations of galactic coordinates. The correlation between $U$ and $W$ velocity is affected by the vertical turn of the velocity ellipsoid above the plane, while the correlation between $V$ and $W$ velocities in disc samples can be affected by spiral breathing modes \citep[][]{Debattista14, Faure14}, or the Galactic warp \citep[][]{Dehnen98, Poggio17}. That the Milky Way has some warp and possible vertical waves is known both from observations in gas \citep[][]{Burke57, Kerr57} and stars \citep[][]{Djorgovski89, Xu15}, but a possible imprint on local stellar kinematics demands scrutiny, since observational errors are known to cause similar apparent correlations between heliocentric velocities (SBA). We will discuss our finding of a warp signal in a second paper.

The structure of this paper is as follows: In Section \ref{sec:Data} we describe the data sets we use, the merging of data from different sources, and our quality cuts.  Section \ref{sec:Distances} provides a description how we calculate distances from parallaxes. In Section \ref{sec:Theory} we describe how our distance estimator, can be used to measure to high accuracy the mean distance bias in a sample and the $\vlos$ source dispersion $\epar$. In Section \ref{sec:Application} we apply this method to the RAVE-TGAS and LAMOST-TGAS samples. We show how a separation of the samples into subsets can be used to validate and measure the line-of-sight velocity uncertainties given in a pipeline. We also show how we can derive the effective spatial selection function for each sample and how our distance statistics can be used to validate the results. Our conclusions are found in Section \ref{sec:Conclusions}.

\begin{figure}
\epsfig{file=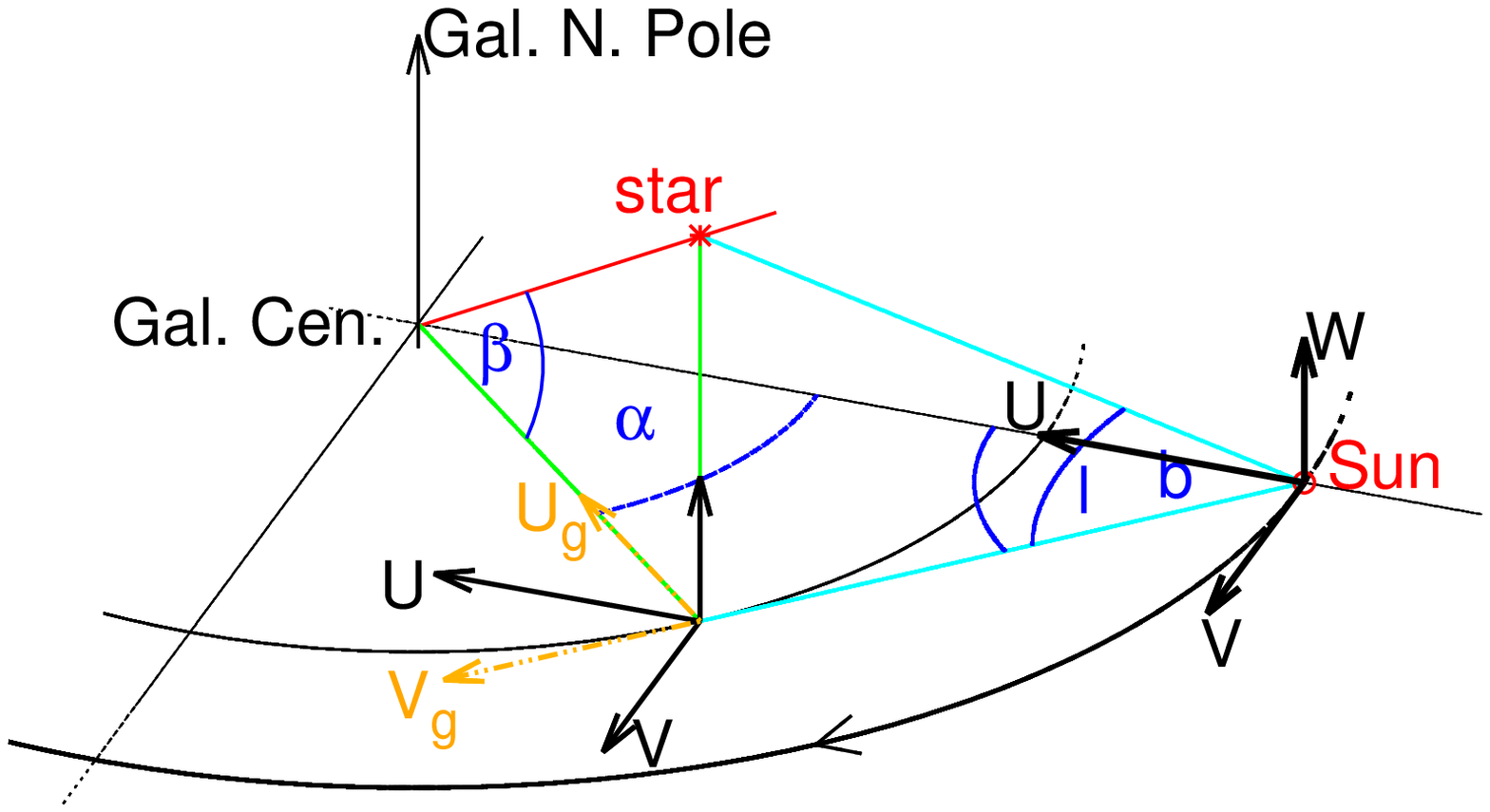,angle=-90,width=\hsize}
\caption{Schematic visualisation of coordinates, velocity components in a local heliocentric Cartesian frame ($U$,$V$,$W$) vs. local galactocentric cartesian frame ($U_g$,$V_g$,$W$) and angle definitions for the observation of a star from the position of the Sun. In this plot the Galactic disc rotates clockwise. The angle between $\Ug$ and $U$ is $\alpha$.}\label{fig:definitions}
\end{figure}

\section{Data and definitions}\label{sec:Data}

\subsection{Coordinate frame and definitions}\label{sec:definitions}

For all subsequent discussions, please refer to Fig. \ref{fig:definitions} for definitions of our coordinate frame. We use either the heliocentric cartesian frame with velocities ($U$,$V$,$W$) radially towards the Galactic centre, azimuthally in the direction of Galactic rotation, and vertically out of the plane, or the galactocentric cylindrical coordinate frame, marked by indices $g$, naming the velocity vector ($\Ug$,$\Vg$,$W$). The galactocentric radius is termed $R$, the distance to the Galactic Centre $r$, Galactic longitude and latitude are $\gl$ and $\gb$. The angle $\beta = \tan^{-1}(z/R)$ is the angle between the Galactic plane and the connection line star-Galactic centre. The angle between the Sun-centre and the star-centre connection lines is called $\alpha$. We use the Sun's galactocentric distance $\Rsun = 8.27 \kpc$, the Sun's total azimuthal velocity $\vsun = \Vc + \Vsun = 248 \kms$ in the Galactic rest frame, where $\Vc$ is the local circular speed of the disc, and the Solar motion with respect to the LSR is $(\Usun, \Vsun, \Wsun) = (13,12.24,7.24)\kms$. These values are chosen in concordance with \cite{S12}, \cite{McMillan11}, \cite{McMillan17}, and \cite{S10}. Following \cite{Joshi07}, we place the Sun at $z = 20 \pc$ above the Galactic midplane.

\subsection{TGAS data}\label{sec:TGAS}

For this work, we crossmatch TGAS data from Gaia DR1 \citep[][]{GaiaDR1, Lindegren16}  both with RAVE DR5 \citep[][]{Kunder16} and with DR2 of LAMOST \citep[][]{LAMOSTDR1}. As can be seen in Fig. \ref{fig:ravelamost}, the samples would together provide a full sky coverage. While LAMOST DR2 has far more stars, the catalogue covers significantly fainter stars giving a smaller common footprint with TGAS. More importantly, LAMOST's line-of-sight velocity determinations are about one order of magnitude less precise than in the RAVE survey, so on LAMOST our statistics will rather test their $\vlos$ measurements than the distances. To ensure sufficient quality of the data, we adopt a few general cuts:
\begin{itemize}
\item A derived distance $s > 30 \pc$. \cite{Lindegren16} report that TGAS is biased against stars with a proper motion larger than $\approx 3.5 \arcsec\yr^{-1}$. To prevent kinematic bias (a cut in proper motions might feign a distance underestimate) we drop the few stars with $s < 30 \pc$, where a proper motion of $1 \arcsec\yr^{-1}$ would imply a transverse velocity $|\vpe| \approx 140\kms$. 
\item Limiting $-100 < \Vg / \kms < 310$: again this cut in heliocentric azimuthal velocity affects only a handful of stars, and we have tested that our results are robust against the precise choice. However, damaged data tend to assemble in the extreme wings, so we drop them.
\item Limiting the line-of-sight motion $|\vlos| < 500 \kms$: excising another few objects is justified by indications of erroneous velocity determinations at extreme $|\vlos|$ in RAVE.
\end{itemize}

\subsection{RAVE}\label{sec:RAVE}

The RAVE DR5 sample contains $520\,701$ entries. However, about $15\%$ of the entries are multiple measurements of the same object. Of the $68\,822$ duplicate entries, we choose the one that has a smaller radial velocity (RV) error given by the RAVE pipeline, leaving us with $451\,879$ unique stars. To test the crossmatch between the RAVE DR5 sample with TGAS, we have compared two different strategies: a) using the already crossmatched positions in \cite{Kunder16} and demanding a precise match in position to the Gaia catalogues, which gives us $210\,415$ crossmatched stars (by identifier we find $210\,368$ unique stars), and b) using the positions in RAVE vs. TGAS positions and proper motions, demanding that the position and proper motion in TGAS predicts within $3$ arcseconds the original position in the RAVE catalogue. The latter exercise gives us $215\,640$ stars, of which a few dozens are apparent binaries in the Gaia catalogue. The Gaia data are provided at epoch $2015.0$, and use the International Celestial Reference System \citep[ICRS, ][]{Arias95, Feissel98}. To transform the celestial coordinates to Galactic coordinates $\gl$ and $\gb$, we use the values $(\alpha_{\rm G}, \delta_{\rm G}, l_{\Omega}) = (192.85948, +27.12825, 32.93192)\deg$ provided in \cite{ESAHip}.\footnote{See also \href{https://gaia.esac.esa.int/documentation//GDR1/Data_processing/chap_cu3ast/sec_cu3ast_intro.html}{gaia.esac.esa.int} Section 3.1.7 for reference.} We have compared our results from the different selections and crossmatches, and since these choices do not significantly affect our results, we report here only results for the original RAVE DR5 crossmatch. We further exclude all stars that are flagged as cluster members in the RAVE dataset, or have a line-of-sight velocity error given as $\siglosrave = 0 \kms$. If not explicitly stated otherwise, we use the quality cut $\siglosrave < 5 \kms$, which will be justified in Section \ref{sec:vloserr}. A large number of stars in this cross-matched sample will fail our variable cut for the precision of the measured TGAS parallax. E.g. the RAVE subsample under our quality restrictions with parallaxes ($p_0$) better than $20 \%$, i.e. $p_0 / \sigma_p > 5$, has only $88\,464$ members. 

An important classification of RAVE spectra is given by the flags from \cite{Matijevic12}. These flags list the $20$ closest matches to each RAVE spectrum for different classes of stars, e.g. normal stars (marked with 'n'), chromospherically active stars, or suspected binaries. Throughout this paper we use these flags to define three subsets of RAVE: a) all stars in RAVE irrespective of their classification, b) unflagged stars, i.e. stars that have all $20$ flags set to 'n', and c) suspected binaries, which have at least one flag set to 'b'.

\subsection{LAMOST}\label{sec:LAMOST}

To crossmatch the LAMOST sample with TGAS, we use the TGAS proper motions and positions to calculate back to the suspected position in LAMOST at epoch $2000$. For a proper match we demand $\sqrt{\Delta_{\RA}^2 + \Delta_{\DEC}^2} < 1.2 \times 10^{-3} \degs$ (or equivalently $< 4.3 \arcsec$), between the predicted position and the position given in the LAMOST catalogue. Repeat observations in LAMOST are purged by taking for each object in TGAS the best match in position, or at equal match, the latest entry in the LAMOST catalogue. This leaves us with $107\,663$ stars in the crossmatched sample. To avoid kinematic biases in the selection function, which could affect our statistics, we remove the region of $0 < \RA \, \degs^{-1} < 67$ and $42 < \DEC \, \degs^{-1} < 59$ from the LAMOST sample, since \cite{LAMOSTDR1} report that this region contains a subsample of plates that have a proper motion selection, limiting the stellar proper motions to smaller than $7 \marcsyr$. There might be minor contamination of in-plane fields with kinematic selections, but since the distance statistics we choose for this analysis are not sensitive to stars in the Galactic Plane, we judge the above exclusion region to be safe. Demanding a parallax uncertainty better than $20 \%$ and a signal to noise ratio $\SNR > 30$, the usual line-of-sight velocity cut $|\vlos| < 500 \kms$ and a galactocentric azimuthal velocity of $-100 < \Vg / \kms < 310$, we have a remaining sample of $34\,384$ stars. The signal to noise cut does not strongly affect the sample size, since the cross-match with Gaia includes mostly bright objects; with a cut at $\SNR > 5$ the sample size would only increase to $38\,834$ stars. 

We will show in Section \ref{sec:vloserr} that line-of-sight velocities in LAMOST have to be corrected by adding $\delta \vlos = 5 \kms$. Further, we will limit the sample to a nominal $\vlos$ measurement error $\sigloslamost < 27 \kms$, and replace their estimate by using $\siglos = 7 \kms$, motivated by Section \ref{sec:lamostlos}.

\begin{figure}
\epsfig{file=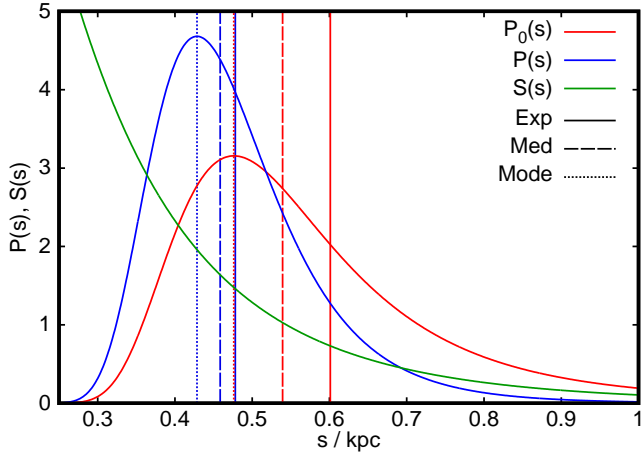,angle=-90,width=\hsize}
\caption{A typical posterior probability distribution $P(s)$ (solid blue line) for a star with $\sigma_p / p \sim 0.2$. In the same plot we show the selection function $S(s)$ for the full RAVE sample (green) in arbitrary units, and the posterior probability distribution $P_0(s)$ (red), when we neglect $S(s)$. The vertical lines show expectation value (solid), median (dashed), and mode (dotted lines) for each distribution.}\label{fig:distdist}
\end{figure}

\section{Distance determinations}\label{sec:Distances}

For the sake of well-defined error distributions in our distance determination, we will rely solely on parallaxes. Since the reported parallax errors are all larger than $0.2 \, \marcs$, a $10\%$ (or $20 \%$) quality cut on the parallax will limit the sample to within $\sim 0.5 \kpc$ ($1 \kpc$), with some weak dependence on the prior used. With parallax errors still a significant fraction of the measured parallax, it is not advisable to use just the inverse parallax as a distance. For the derivation of distances from parallaxes, see \cite{Stromberg27}, \cite{SB14}, or \cite{Astraatmadja16}. Since we are using exclusively the parallaxes to assess stellar distances, the magnitude based selection function of the sample has to be written into the prior, so that the posterior distribution in distance is:
\begin{equation}\label{eq:Ps}
P(s) = N^{-1} s^2 G(p, p_0, \sigma_{p}) \rho(s(p),l,b) S(s(p)) $,$
\end{equation}
where
\begin{equation}
N =  \int ds \, s^2 G(p, p_0, \sigma_{p}) \rho(s(p),l,b) S(s(p))
\end{equation}
is the normalisation. $G(p, p_0, \sigma_{p})$ is the observational likelihood of the parallax given the measurement, $(p_0, \sigma_{p})$, $s^2$ is the geometric factor, i.e. surface of the observational cone at fixed distance $s$, $\rho(s,l,b)$ is the supposed density of stars in the direction of galactic longitude and latitude $(l,b)$, and $S$ is the selection function, which we approximate as a function of distance $s$ only.
The expectation value for the distance of each star is then just:
\begin{equation}\label{eq:distex}
\left<s\right> = N^{-1} \int ds \, s^3 G(p, p_0, \sigma_{p}) \rho(s(p),l,b) S(s(p)) $.$
\end{equation}
For the density prior $\rho$, we use the simple thin/thick disc + halo decomposition from equation (32) in \cite{SB14}, i.e. our density model simply reads as:
\begin{equation}\label{eq:rho}
\rho(R,z) = e^{-\frac{(R-\Rsun)}{R_d}} e^{-\frac{|z|}{z_0}} + a_t e^{-\frac{(R-\Rsun)}{R_d}} e^{-\frac{|z|}{z_{0,t}}} + a_h \left(\frac{r}{\Rsun}\right)^{-2.5} $,$
\end{equation}
where $R$ is the galactocentric radius in cylindrical coordinates, $\Rsun$ is the (cylindrical) galactocentric radius of the Sun, $R_d$ is the scale-length of the disc, chosen at $R_d = 2.5 \kpc$, $z$ the altitude above the plane, $z_0 = 0.3 \kpc$ and $z_{0,t} = 0.9 \kpc$ are the scale-heights of the thin and thick disc, $r$ is the galactic centre distance used for a simple spherical halo model, $a_t = 0.12$ is a local thick disc normalisation set in concordance with \cite{Ivezic08}, $a_h = 0.001$ normalises the halo component. We note that the placement of the Sun above the plane (we assume it at $\zsun = 20 \pc$ has virtually no influence on the distance estimates. When assuming a vertical position of the Sun either at $\zsun = 0$ or $\zsun = 50 \pc$, every single star in the entire RAVE sample experiences a relative distance change smaller than $0.01$, the rms dispersion of the fractional distance differences is $3 \cdot 10^{-4}$ and the mean offset is $8 \cdot 10^{-5}$. This is to some part due to the dominance of the selection function, to another part due to the inclination of most sightlines against the vertical direction.

Of great importance are, however, the selection function $S(s)$ and the choice, which statistical quantity we use. Fig. \ref{fig:distdist} shows the posterior distance distribution $P(s)$ according to eq. \ref{eq:Ps} with a solid blue curve for a typical RAVE star with $\sigma_p / p \sim 0.2$. As a comparison, we show with a dashed red line the posterior distance distribution $P_0(s)$ when we would neglect the selection function $S(s)$ (depicted in arbitrary units with a dashed green line). The selection function favours at this distance more nearby stars, and so the posterior distribution loses most of its tail towards long distances. This strongly reduces the distance expectation value from $\langle s \rangle_0 \sim 0.6 \kpc$ down to $\langle s \rangle \sim 0.47 \kpc$. The median and mode (which is not a sensible statistical quantity due to its dependence on the axis scaling) of the distribution are shifted in the same direction, but react slightly less. Since the parallax probability distribution is symmetric, the resulting posterior probability in distance is strongly skew, and so the expectation value is a lot larger than the median and in particular the mode/maximum. 

In short we have to take two things from this: i) The mode is not a sensible quantity to use and both median and mode underestimate the expectation value. ii) The selection function strongly affects all distance estimates.

The selection function $S$ would typically require the use of a full population synthesis model: It is apparent that the magnitude limits of the RAVE survey will introduce a strong bias towards nearby and more luminous stars, which ultimately favours younger stellar populations. Stellar metallicities affect brightness and colour, thus influencing the selection function as well. While $S$ will hence in general not be isotropic, we approximate $S(s)$ as a function of distance $s$ only. We discuss its derivation in Section \ref{sec:prior}.

\begin{figure*}
\epsfig{file=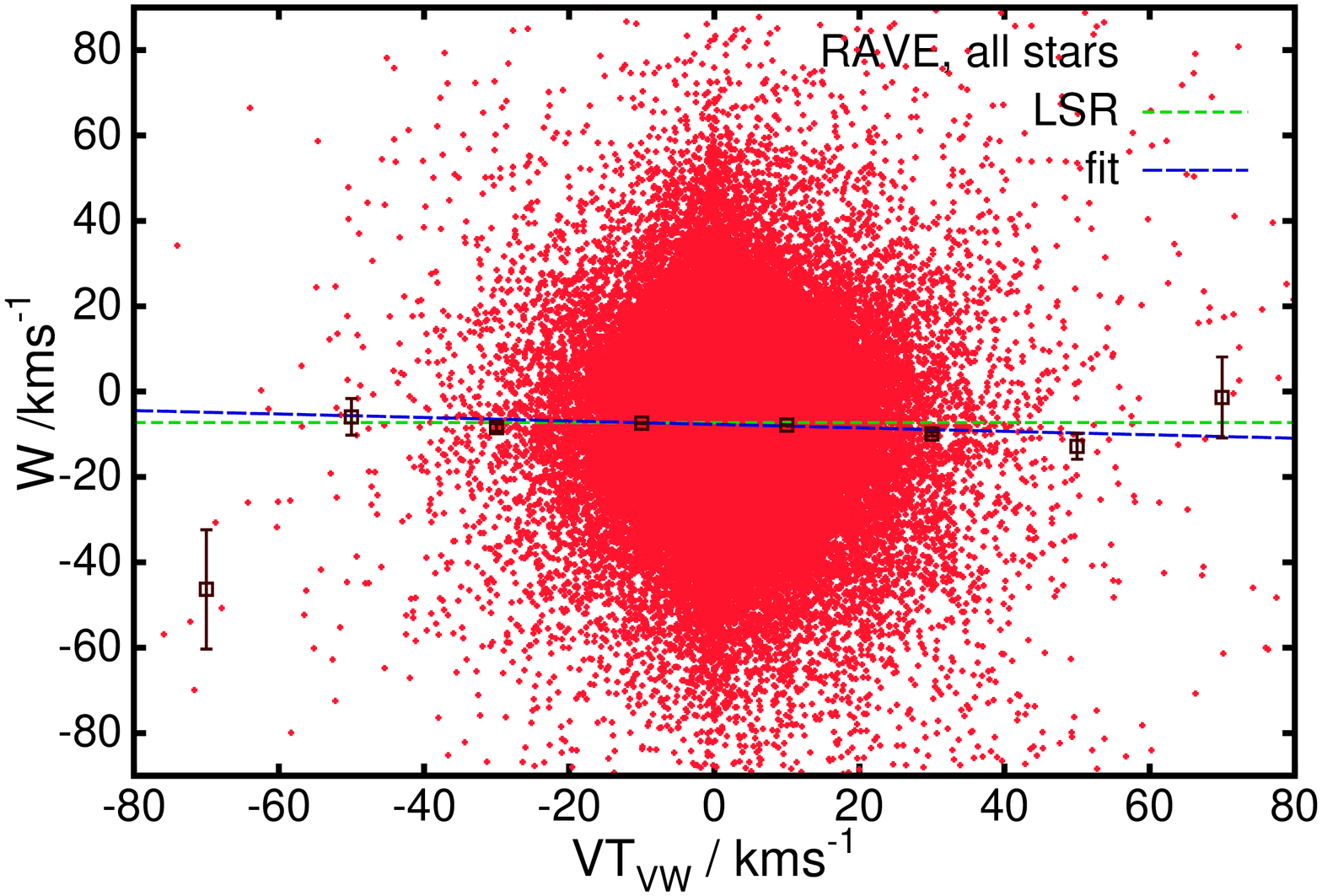,angle=-90,width=0.49\hsize}
\epsfig{file=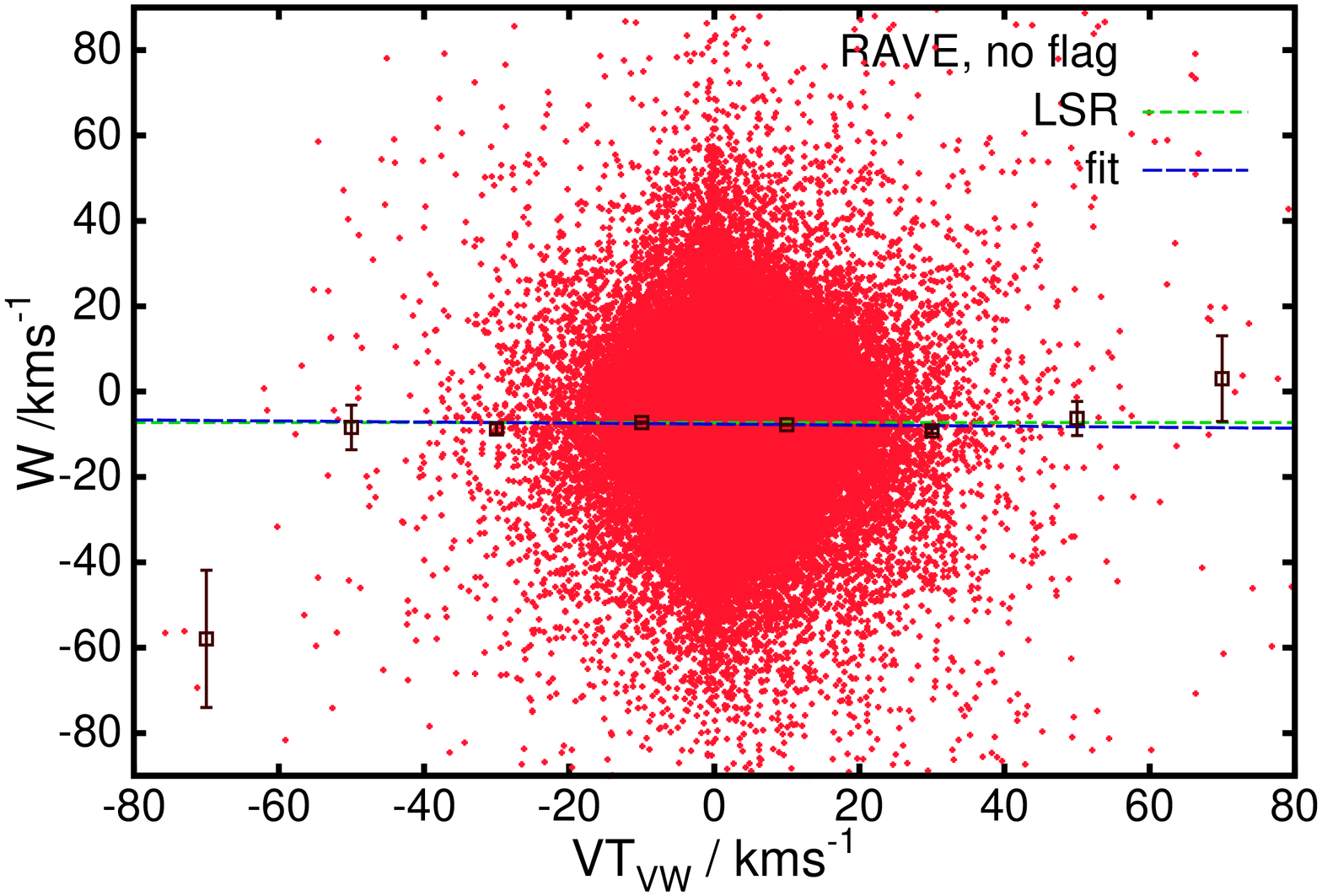,angle=-90,width=0.49\hsize}
\epsfig{file=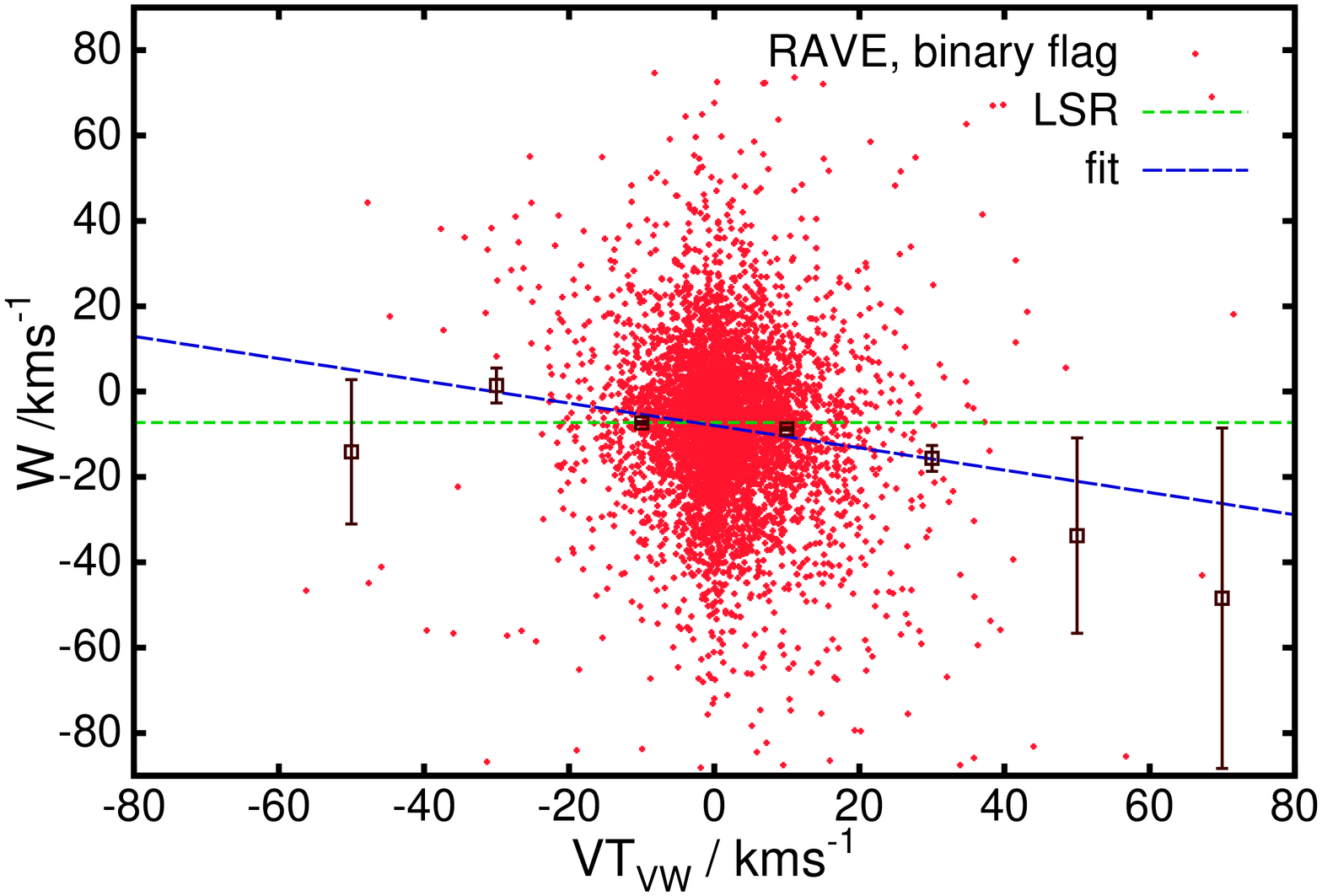,angle=-90,width=0.49\hsize}
\epsfig{file=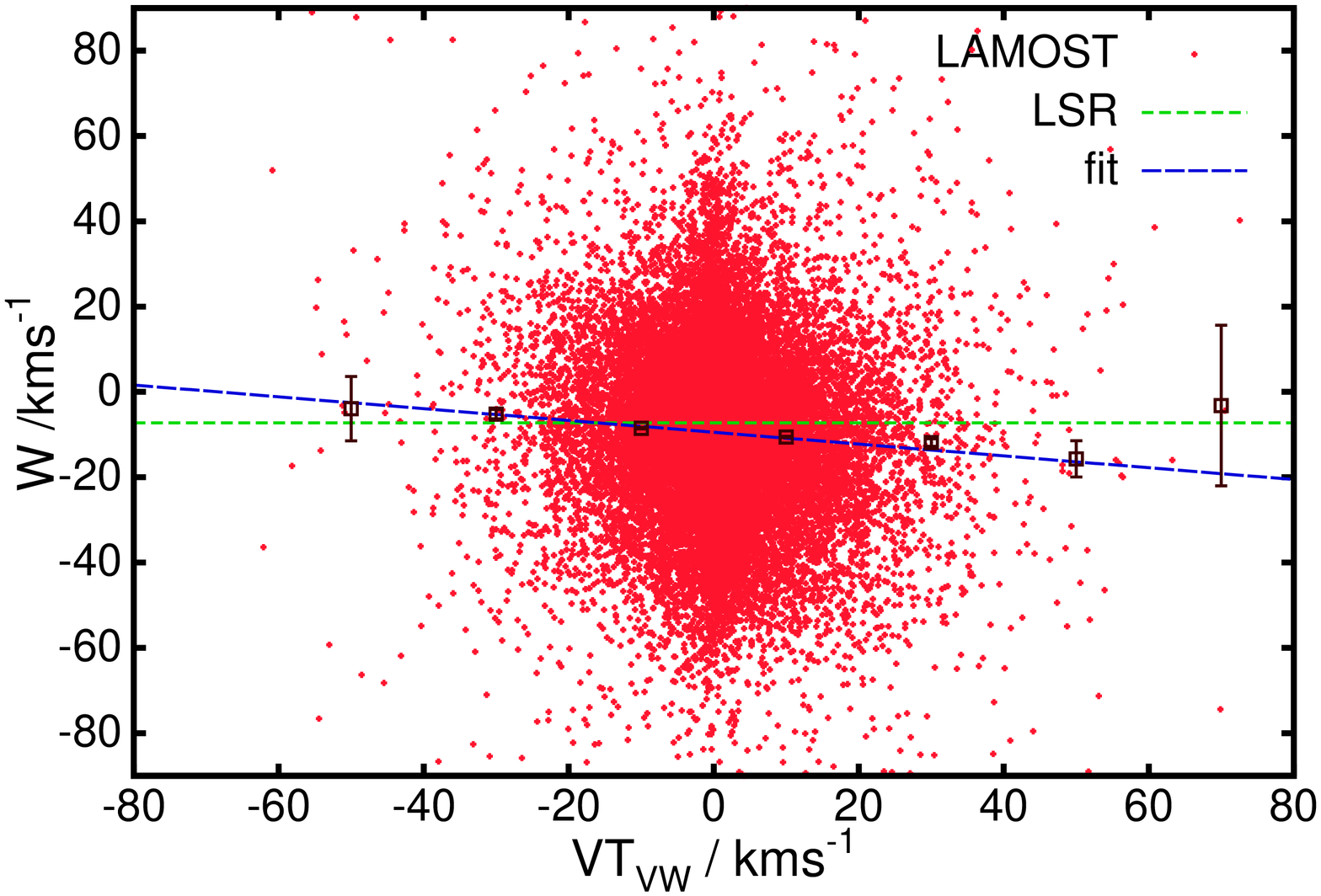,angle=-90,width=0.49\hsize}
\caption{Trends in $W$ vs. $\Vh T_{VW}$ in all RAVE stars (top left), all unflagged RAVE stars (top right), all RAVE stars with at least one binary flag (bottom left), and all LAMOST stars with $\sigloslamost < 25 \kms$ (bottom right). The dark red errorbars show the mean values of $W$ when binning the sample along the x-axis. The green short dashed line marks the reflex motion of the Sun, the blue long-dashed line the linear fit to the data. Despite the curious outliers at extreme azimuthal velocities in the RAVE sample, the slopes in all four plots are significant and given in Table \ref{tab:distf}.}\label{fig:scatter}
\end{figure*}

\section{Assessing distances and line-of-sight velocity distribution errors}\label{sec:Theory}

To assess distances we make use of the statistical distance method of SBA, where a full derivation of all equations can be found. In addition to the theoretical justification, we have performed numerous validations on the original method. Since we are using a restricted approach in this paper, we present in the Appendix a short validation on N-body galaxy simulations, which are structurally similar to the Milky Way. 

This method exploits position dependent correlations between different heliocentric velocity components. A stellar motion relative to the Sun $(U_0,V_0,W_0)$ results in observable line-of-sight velocity $\vlos$ and proper motions $\mul$ and $\mub$, depending on the distance $s_0$ to the star, calculated with the translation matrix ${\vM}$. Since this translation matrix is orthogonal, we have 
\begin{equation}
\begin{pmatrix} U_0 \\ V_0 \\ W_0 \end{pmatrix} = \vM \begin{pmatrix} s \mul \\ s \mub \\ \vlos \end{pmatrix} \,\quad $and$ \,\quad 
\begin{pmatrix} s \mul \\ s \mub \\ \vlos \end{pmatrix} = \vM^T \begin{pmatrix} U_0 \\ V_0 \\ W_0  \end{pmatrix}
\end{equation}
If our distance estimate $s$ for the star is wrong by a fraction $f$, the terms containing the proper motions will be stretched by $(1 + f) = s/s_0$, and our estimated velocity vector $(U,V,W)$ for the star will be altered to
\begin{equation}
\begin{pmatrix} U \\ V \\ W \end{pmatrix} = (\vI+f\vT)\begin{pmatrix} U_0 \\ V_0 \\ W_0 \end{pmatrix} $,$
\end{equation}
where $\vT = \vM\vP\vM^{T}$, $\vI$ is the identity matrix, and $\vP$ is the identity for the proper motion terms, but zero in the $\vlos$ term. For the following discussion, we only need the terms of $\vT$ connecting the vertical motion to all velocity components, i.e.
\begin{equation}\label{eq:angleterms}
\begin{array}{l}
T_{UW} = \cos(\gl)\sin(\gb)\cos(\gb) $,$ \\
T_{VW} = \sin(\gl)\sin(\gb)\cos(\gb) $, and$ \\
T_{WW} = \cos^2(\gb) $.$ \\
\end{array}
\end{equation}

Take for example the motion of the Sun against a sample of halo stars. The Sun has an azimuthal velocity of roughly $\Vsun \sim 250 \kms$, while halo stars have no net azimuthal motion. $T_{VW}$ is maximal when we look up out of the galactic plane at an angle $\gb \sim 45\deg$ in or against the direction of solar motion. To get a picture on how this term links vertical heliocentric velocities $W$ with azimuthal velocities $V$, imagine approaching horizontally the wall of your room, while looking at the corners of the floor and the ceiling. Both corners will be blue-shifted in their line-of sight velocity, but you also observe an angular motion. If your distance estimate to the corners is correct, the conclusion will be a zero net motion of the room. However, if the distance is over-estimated, the angular motion of the upper corner will be mistakenly translated into a net upwards motion, and similarly, the lower corner will appear to move downwards. Similarly, if we overestimate stellar distances in our Galactic halo example, we will detect a net upwards motion (positive $W$) maximal near $(\gl,\gb) \sim (90 \deg, 45 \deg)$, and $(\gl, \gb) \sim (270 \deg, -45 \deg)$ and a downwards motion (negative $W$) around $(\gl,\gb) \sim (90 \deg, -45 \deg)$, and $(\gl, \gb) \sim (270 \deg, 45 \deg)$, and opposite bias for a distance underestimate. This motion will be proportional to the azimuthal velocity difference between the Sun and the stars. 

As discussed in SBA, there are similar connections between each velocity pair. However, since we are dealing with a disc sample with a complicated selection function and strive for maximum accuracy, we will drop all terms that are to first order affected by galactic rotation and streaming ($U$ vs. $V$ motion). Since proper motion errors are minimal, we can use the more precise non-linear estimator from SBA.

As a quick test for distance errors in a sample of stars, we can assume that $(U,V,W)$ is sufficiently close to the real velocity vector $(U_0,V_0,W_0)$ and fit a straight line to $W$ vs. $T_{VW}V$, via the regression model
\begin{equation}\label{eq:key}
W_i = \gamma T_{VW,i} V_i - \Wsun + \epsilon_i $,$
\end{equation}
where $W_i$ and $V_i$ are the measured vertical and azimuthal velocity components for the $i$th star, and $T_{VW,i}$ is given in equation (\ref{eq:angleterms}). The fit parameters $\Wsun$ and $\gamma$ represent the vertical motion of the Sun, and the average fractional distance error, while $\epsilon_i$ represents the random term from stellar velocity dispersion and measurement errors, which should have zero expectation value. These fits are shown in Fig. \ref{fig:scatter} for different subsamples of RAVE-TGAS, and for LAMOST-TGAS. The dark red errorbars show means of $W$ when binning the sample in $VT_{VW}$, while the blue dashed lines show the fitted regression line from equation (\ref{eq:key}). It is evident that in particular LAMOST and the suspected binary stars in RAVE show a large slope $\gamma$, indicating either a major distance bias, or a problem with line-of-sight velocity determination.

We note that a robust linear estimator can be derived from equation (\ref{eq:key}) by replacing the individual azimuthal velocities with their expectation value, i.e. $V_i \to \left<V\right>$. This estimator is particularly useful for remote samples of halo stars with a large mean heliocentric $\Vh$. However, for disc stars this estimator is about one order of magnitude less sensitive than our full estimator, and so for this work of little use compared to the full distance estimator, which we will now describe.

The regression in equation (\ref{eq:key}) can be formalised to an estimate of the fractional distance error $f_{UV}$. We adopt equation (19) from SBA:
\begin{equation}\label{eq:fest}
f = \frac{\Cov(W, y)}{\Var(y) + \langle\left( T_{VW}^2 + T_{UW}^2\right) \sigW^2\rangle} $.$
\end{equation}
where $y = T_{UW}U + T_{VW}V - T_{WW}\Wsun$ is the ``baseline'' on which the rise in $W$ is measured, $\Cov$ is the covariance, and $\Var(y)$ is the variance of $y$. To measure the distance bias in practice, we multiply all distances with the same factor, until the estimated $f$ is zero. Error bars shown throughout the paper are derived by varying the distance correction, until $f$ is at the $1 \sigma$ confidence limit. The denominator of this term matters only for the error determination. The essential task is to find the distance correction factor, for which $\Cov(W,y) = 0$. 

To construct an estimator that is more stable against outliers in $W$, we cap the vertical velocities at $|W| = 200 \kms$, i.e. assign all stars beyond that value the limit of $\pm 200 \kms$. In statistical terms, we are essentially using an M-estimator for our linear regression. We have checked by varying this ceiling between $150$ and $400 \kms$ that our results are not affected by outliers.

In some situations, sample kinematics/measurements may be erroneous, or the data may be affected by streams. In this case it is useful to look at the separate statistics from the $U \to W$ and $V \to W$ terms. These separate terms have analogously the estimators:
\begin{equation}\label{eq:restricted}
f_{V} = \frac{\Cov(W, T_{VW} V)}{\Var(V T_{VW}) + <T_{VW}^2 \sigW^2>} \quad $, and$
\end{equation}
\begin{equation}
f_{U} = \frac{\Cov(W, T_{UW} U)}{\Var(U T_{UW}) + <T_{UW}^2 \sigW^2>} \quad $.$
\end{equation}

\subsection{Bias corrections}

There are three further contributions to these covariances: the tipping of the velocity ellipsoid outside the Galactic plane, observational uncertainties in proper motions, and both $\vlos$ measurement errors and intrinsic $\vlos$ dispersion from binaries. We summarize the last two items as $\vlos$ source variance $\epar^2$. To get an unbiased distance estimator we have to subtract these contribution from the measured covariance:
\begin{equation}
\Cov(W, T_{VW} V) = \Cov_{\rm measured}(W, T_{VW} V) - \Cov_{\rm bias} $.$
\end{equation}
We now separate the bias term $\Cov_{\rm bias}$ into its three sources: the bias from the tipping velocity ellipsoid $\Cov_{\rm ve}$, the measurement error on proper motions $\Cov_{\perp}$ and the $\epar$ term $\Cov_{\parallel}$:
\begin{equation}\label{eq:bias}
\Cov_{\rm bias}(W, T_{VW} V) = \Cov_{\rm ve} + \Cov_{\perp} + \Cov_{\parallel} $.$
\end{equation}

Let us first discuss the minor contribution from the tipping of the velocity ellipsoid. As we can see in Fig. \ref{fig:definitions}, some part of the galactocentric radial motion contributes to the heliocentric azimuthal velocity, when a star is observed away from the connecting line between Sun and Galactic centre. This way, the inclination of the non-isotropic velocity ellipsoid in galactocentric coordinates translates into a minor correlation between the velocities in the heliocentric frame. The velocity ellipsoid corrections from equations (40f.) of \cite{SBA} for the $U \to W$ and $V \to W$ terms read:
\begin{equation}
\begin{array}{l}
\Cov_{\rm ve}(W, T_{UW} U) = \frac{1}{4} \left<\sin(2\gb)\cos(\gl) \cos(\alpha) \sin(2 \beta) \left(\sigma_U^2 - \sigma_W^2 \right) \right> $,$ \\
\Cov_{\rm ve}(W, T_{VW} V) = -\frac{1}{4} \left<\sin(2\gb)\sin(\gl) \sin(\alpha) \sin(2 \beta) \left(\sigma_U^2 - \sigma_W^2 \right) \right> $,$ 
\end{array}
\end{equation}
where $\sigma_U^2$ and $\sigma_W^2$, are the radial and vertical velocity dispersions. For definitions of the angles, see Fig. \ref{fig:definitions} and Section \ref{sec:definitions}. Here we assume that the velocity ellipsoid is pointing roughly towards the Galactic centre \citep[][]{Siebert08, Binney14}. $\sigma_U$ and $\sigma_W$ are measured directly from the data. Since our samples are local and close to the plane (small $\alpha$ and $\beta$), the resulting bias on the distance estimators is significantly less than $0.01$ for $f_{U}$, and less than $0.002$ for $f_{V}$. 

Adapting equations (23ff.) from SBA, we can calculate the bias in the covariance from proper motion errors:
\begin{equation}
\begin{array}{l}
\Cov_{\perp}(W, T_{UW} U) = - \sum_i T_{UW,i}^2 e_{\perp, i}^2 \\
\Cov_{\perp}(W, T_{VW} V) = - \sum_i T_{VW,i}^2 e_{\perp, i}^2 \\
\end{array}
\end{equation}
where again $i$ is the index running over all stars in the sample. $e_{\perp, i}$ is the uncertainty in the proper motion of star $i$ in the latitudinal  ($\gb$) direction. If uncorrected, this term will feign a slight distance overestimate, but due to the excellent proper motions in TGAS, the resulting bias on the fractional distance error is below $0.3 \%$.

\subsection{Line-of-sight velocity variance}

The $\vlos$ errors require significantly more caution. Consider the effect on $f_V$, an analogous discussion will apply to $f_U$. Just where $T_{VW}$ is maximal, i.e. at $\gl \sim 90 \deg, 270 \deg$ and at an elevation of $\gb \sim 45 \deg$ above the Galactic plane, an error in line-of-sight velocity determination will be evenly split into the derived $V$ and $W$ velocities, leading to a maximal correlation. This term will, if uncorrected, push the statistics towards a distance underestimate:
\begin{equation}\label{eq:vloscorr}
\begin{array}{l}
\Cov_{\parallel}(W, T_{UW}U) = \sum_i T_{UW,i}^2 \epari^2 $ and $ \\
\Cov_{\parallel}(W, T_{VW}V) = \sum_i T_{VW,i}^2 \epari^2 $,$ \\
\end{array}
\end{equation}
where $\epari^2$ is the source $\vlos$ variance (we analogously call $\epar$ the source $\vlos$ dispersion), i.e. all $\vlos$ variance that does not derive from the stellar velocity ellipsoid. For the full estimator $f$ both $\Cov$ terms have to be added. We conceptually split $\epar$ into two contributions:  
\begin{equation}\label{eq:epar}
\epar^2 = \siglos^2 + \sigb^2 $,$
\end{equation}
where $\siglos$ is the intrinsic measurement error in each survey, and $\sigb$ is the line-of-sight velocity dispersion from the orbital motion of multiple stellar systems (which we will address simply as ``binary'' contamination). The source $\vlos$ variance is the most important bias in LAMOST, with a distance bias near $10\%$. The impact of $\siglosrave$ on the distance estimators is nearly negligible with less than $0.3 \%$, but $\sigb$ might be important. 
While the general problem of binary velocity dispersions has been known for a long-time \citep[][]{Olszewski96, Hargreaves96}, we could not locate a good prediction for the Milky Way. A strong expected age-dependence \citep[][]{Gieles10}, and probably some metallicity dependence, make it difficult to translate results from dwarf spheroidal galaxies to the MW. Depending on the binary fraction and distribution, the dispersion contributed to the line-of-sight velocities could be of order $3 \kms$ with a large uncertainty \citep[][]{Vogt95}, though some open clusters appear to yield smaller values. We also note that our statistics will not pick up the full line-of-sight velocity dispersion, since the proper motions will also be affected to some extent. The difference is mostly in the long baseline in time for Gaia - short-period binary systems (with periods significantly shorter than the time elapsed between Hipparcos and Gaia) should fully affect $f$, whereas we expect that very long period binary systems will have a vanishing effect. However, while long-period systems are more numerous, most of the contribution to the velocity distributions by binary systems is expected to stem from short-period systems \citep[see e.g.][]{deRijcke02}.

This problem will be further examined in Sections \ref{sec:vloserr} and \ref{sec:lamostlos}.

Can we discriminate the contribution of $\epar$ from a general distance error? To some extent, yes, with a sufficiently large number of stars in our sample. The bias from $\epar$, $\Cov_{\parallel}/N$, for any sample of $N$ stars should be the same between subsamples with the same geometry (i.e. distribution in values of $T_{VW,i}$), and same $\vlos$ measurement quality and binary contamination (i.e. similar $e_{\parallel, i}$). In contrast with this, the covariance term created by a distance error depends on the length of the baseline, which can be seen by isolating $\Cov(W,T_{VW}V)$ or $\Cov(W,y)$ in equation (\ref{eq:fest}). 

To summarize, these equations allow for a degenerate estimate between the mean distance error and an unknown bias in the line-of-sight velocities from binaries, but this degeneracy can be broken by selecting different subsamples with different baseline lengths $(\Var(T_{VW} W))$, either by cuts in metallicity or (more dangerous) by enforcing cuts on the galactocentric azimuthal velocity $V_g$.\footnote{We iterate the warning that while it is feasible to cut the sample in $V_g$, cuts in heliocentric $V$ are illicit, since they produce a biased cut in the actual velocity ellipsoid, biasing the sample in $U_g$ and hence, via the inclination of the vellocity ellipsoid in $W$.}

To account for uncertainties in $\Cov_{\rm bias}$, we add $30\%$ of the distance bias from the tipping velocity ellipsoid and the proper motion errors as independent terms to the error budget in $f$. The geometric terms of the source $\vlos$ dispersion $\epar$ are not uncertain, and so we add $10 \%$ of this bias to the error budget.

\subsection{Further biases and caveats}\label{sec:caveats}

In a realistic galaxy, we expect further biases on our distance estimator from Galactic structure. The main features will be 
\begin{itemize}
\item streams from accretion events,
\item the Galactic warp correlating $V$ and $W$ motions near the line of nodes,
\item (vertical) breathing modes from bar and spiral pattern, and
\item disc streaming motions near resonances.
\end{itemize}
We do not expect a major bias from disc streaming, because those motions are predominantly in the galactic plane \citep[][]{Dehnen00, Peres17}, and our distance estimator is only biased by vertical motions.

If a stellar stream through the sample, it may have a large vertical motion $W$. This does not yet imply even a local bias on our distance estimator, because even a correlation between $W$ and e.g. $V$ velocities is not enough. The stream has to correlate $W$ significantly with $T_{VW}V$ to affect the $f_{V}$ part. Even if the condition is fulfilled, good spatial coverage will cancel the bias to first order: If we have a bias at $b > 0$, an equivalent patch in the southern galactic hemisphere $b < 0$ will cancel the impact of the stream, because the stream's $W$ and $V$ velocity will be similar, but the geometric term $T_{VW}V$ reverses its sign. In addition, the $f_{U}$ estimator will not be affected or react differently, and tend to cancel the effect. In short, streams in a sample with good spatial coverage will not be a problem, and in worst case be detected by internal discrepancy between $f_{U}$ and $f_{V}$ \cite{Helmi17} have already suggested that the TGAS sample has no dominant single stream or small number of identifiable streams.

Similarly, the Galactic warp may be a problem for the $f_{V}$ term, but since the correlation between $W$ and $V$ will be similar in all directions  \citep[][]{Dehnen98}, $f_{V}$ is only biased, if our sample looks predominantly into one quadrant of the sky. This is a major caveat, however, for dissecting samples in $\gl$ and $\gb$.

Vertical breathing modes \citep[][]{Faure14} are more dangerous, because their $W$ motion reverses in parallel with a sign reversal of $T_{VW}$ between $\gb > 0$ and $\gb < 0$. This can have some impact, if the vertical motion feature is not symmetric in the field of view. However, the effect can be detected by comparing $f_{V}$ with $f_{U}$. 

A summary of performance tests of our method on N-body simulations of Milky-Way like galaxies, which reasonably match both Galactic structure and local kinematics, is discussed in the Appendix. On full sky samples measured at a solar-like position, the distance estimators perform at the level of their statistical error, i.e. the systematic effects discussed in this section must be below $1 \%$ on the distance estimator. The worst case scenario that we could find was a simulation with an unrealistically long bar. In that case the substructure introduced no concerning bias on the full sample, but a scatter of $\sim 4 \%$ in $f_V$ and $\sim 2 \%$ in $f$ when we restricted the sample to separate hemispheres in $\gl$.

We further note that (cf. the discussion in SBA) $f$ is an intrinsically quadratic estimator. Stars with distance overestimates have a longer measured baseline in $y$, and so achieve a slightly larger weight in the distance estimator than distance underestimates. So, for a set of different fractional distance errors $f_i$ for a set of stars, the fractional distance estimate is in fact $f = 1 - \sqrt{\sum_i <{(1+f_i)}^2>}$. While of some importance with spectro-photometric distances, the effect is not important here due to the good parallax quality and hence moderate random scatter of distance estimates in TGAS.

\begin{table*}
\caption{Distance correction factors assuming the given errors for different subsamples of RAVE and LAMOST stars. Column 2 shows the number of stars, $\gamma$ is the slope in the simple linear regression from equation (\ref{eq:key}), $f_V$ is the distance estimator from equation (\ref{eq:restricted}), $\Delta f$ is the bias correction from eq. \ref{eq:bias} to $f$, and $f$ is the full distance estimator from equation (\ref{eq:fest}). The shorthands for the used samples are: RAVE: using RAVE stars with parallax errors better than $20 \%$; ${\rm RAVE}_{10}$: RAVE stars with parallax errors better than $10\%$; all: all stars; nf: no flags; nf,s: no flags and excluding the distance region $0.3 < s/\kpc < 0.35$; b: at least one binary flag; LAMe1: LAMOST with given line-of-sight uncertainties $\sigloslamost < 25 \kms$, LAMe2: LAMOST with $\sigloslamost < 15 \kms$. A positive $f$ indicates an average distance overestimate by a $1+f$.}
\setlength{\tabcolsep}{20pt}
\begin{tabular}{l||r|r|r|r|r|}
sample & $\#$ stars & $\gamma \quad \quad$ & $f_{V}\quad\quad$ & $\Delta f$ & $f$ \quad \quad \\ \hline
${\rm RAVE_{all}}$ & $88464$ & $-0.041 \pm 0.007$ & $-0.027 \pm 0.007$ & $-0.009$ & $-0.013 \pm 0.005$ \\
${\rm RAVE_{10,all}}$ & $40570$ & $-0.063 \pm 0.010$ & $-0.041 \pm 0.010$ & $-0.004$ & $-0.024 \pm 0.007$  \\
${\rm RAVE_{nf}}$ & $59923$ & $-0.012 \pm 0.008$ & $-0.009 \pm 0.009$ & $-0.011$ & $0.001 \pm 0.006$ \\
${\rm RAVE_{nf,s}}$ & $52865$ & $-0.026 \pm 0.009$ & $-0.022 \pm 0.009$ & $-0.011$ & $-0.006 \pm 0.007$ \\
${\rm RAVE_{10,nf}}$ & $22574$ & $-0.030 \pm 0.014$ & $-0.014 \pm 0.015$ & $-0.005$ & $-0.010 \pm 0.009$ \\
${\rm RAVE_{10,nf,s}}$ & $18710$ & $-0.045 \pm 0.015$ & $-0.032 \pm 0.016$ & $-0.005$ & $-0.017 \pm 0.010$ \\
${\rm RAVE_{b}}$ & $5929$ & $-0.261 \pm 0.027$ & $-0.170 \pm 0.024$ & $-0.006$ & $-0.123 \pm 0.016$ \\
${\rm RAVE_{10,b}}$& $2803$ & $-0.332 \pm 0.041$ & $-0.207 \pm 0.037$ & $-0.001$ & $-0.114 \pm 0.024$ \\
${\rm LAM_{e1}}$ & $28475$ & $-0.134 \pm 0.012$ & $0.131 \pm 0.035$ & $0.214$ & $0.147 \pm 0.031$ \\
${\rm LAM_{e2}}$ & $6673$ & $-0.141 \pm 0.024$ & $-0.013 \pm 0.026$ & $0.084$ & $0.011 \pm 0.019$ \\
\end{tabular}\label{tab:distf}
\end{table*}

\section{Application to the data}\label{sec:Application}

\subsection{Mean distance error in the different subsamples}

We start our discussion with an inventory of the mean distance biases in different subsamples, and will justify some of our assumptions in the later Sections. For deriving distances in these samples, we use the full distance prior from equation (\ref{eq:distex}) including the survey selection functions, which we will derive in Section \ref{sec:prior}, and full bias corrections with the given errors according to equation (\ref{eq:bias}). 

Table \ref{tab:distf} summarizes the distance statistics for different subsamples in RAVE and for LAMOST. The third column ($\gamma$) provides the slope in the simple linear fit of $W$ motion against $VT_{VW}$ given in equation (\ref{eq:key}). These simple fits are shown with blue dashed lines for four samples in Fig. \ref{fig:scatter}: all RAVE stars (top left), unflagged RAVE stars (top right), suspected binaries in RAVE (bottom left), and LAMOST (bottom right). These slopes indicate distance underestimates or significant $\vlos$ source dispersions $\epar$ in all samples, but in particular for the suspected binaries and LAMOST. The fourth and fifth columns of Table \ref{tab:distf} show the two bias-corrected distance estimators $f_V$ and $f$ from equations (\ref{eq:fest}) and (\ref{eq:restricted}). For the full RAVE sample, these terms both show a moderate, but significant distance underestimate of order $2\%$. 

Our first, naive, idea was to ascribe this distance underestimate to parallax errors in TGAS, as systematic effects are e.g. found in \cite{Arenou17}. However, when we select stars that have no flags according to \cite{Matijevic12}, i.e. are considered likely normal stars, the distance bias diminishes. In turn, when we test suspected binary systems, i.e. stars that have at least one 'b' flag, there is a very strong bias equivalent to more than $10\%$ distance underestimate. We can conclude from this that Gaia astrometry is generally of high quality, as long as we restrict the sample to apparently normal, single stars.

For the LAMOST sample, the situation is more complex. The simple linear regression test delivers a slope $|\gamma|$ in excess of $0.1$, i.e. a very strong kinematic bias. However, LAMOST $\vlos$ determinations have nominal measurement errors $\sigloslamost > 10 \kms$. In the bias corrected statistics $f$ and $f_V$, the expected impact by $\siglos$ fully accounts for this correlation in the LAMOST subsample with $\sigloslamost < 15 \kms$, resulting in a near-zero $f$. When we ease this limit to $\sigloslamost < 25 \kms$, the bias correction apparently overshoots, giving strongly positive $f$. The only viable conclusion is that LAMOST greatly overestimates its $\vlos$ measurement error. We will test this hypothesis and estimate the real $\vlos$ measurement error of LAMOST in the next section.

\begin{figure}
\epsfig{file=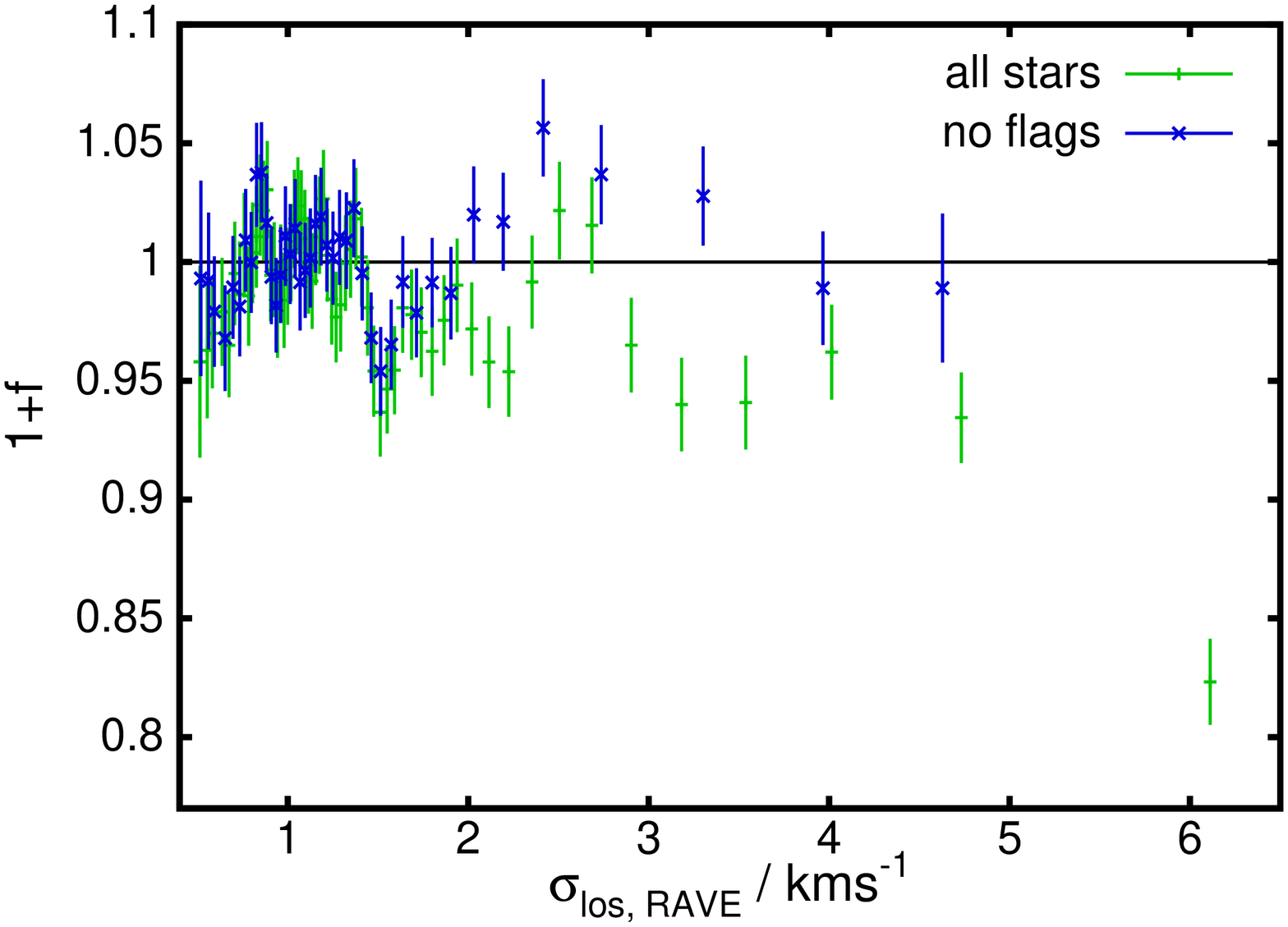,angle=-90,width=\hsize}
\epsfig{file=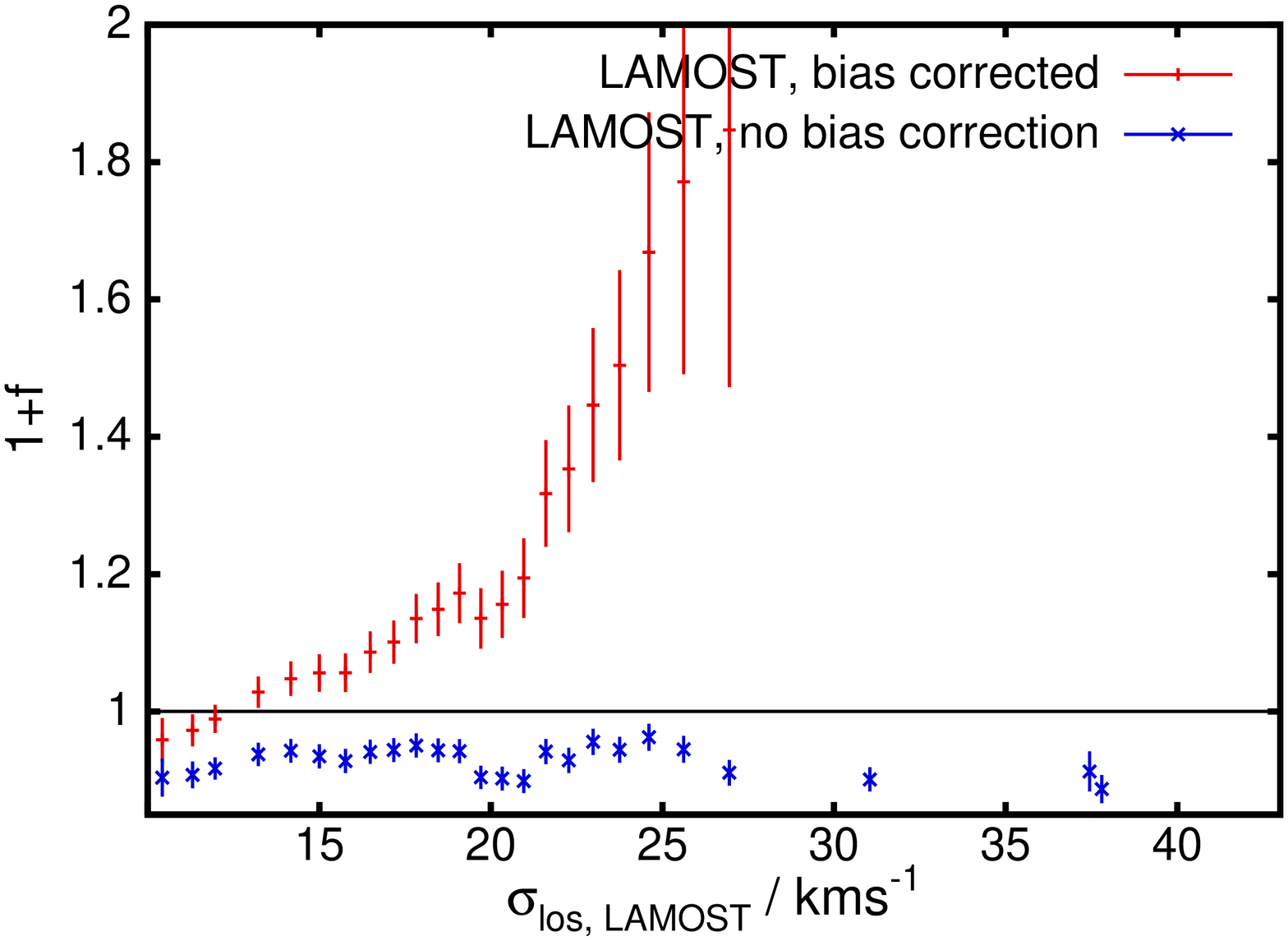,angle=-90,width=\hsize}
\caption{Testing line-of-sight velocity uncertainties $\siglos$ as provided by the RAVE (top panel) and LAMOST surveys (bottom panel). Both samples are binned in $\siglos$ using a mask of $4500$ stars sliding by steps of $1500$ stars, so every third data point is independent. The small $\vlos$ errors of RAVE result in small bias corrections, so here we show only corrected estimates for $f$. The entire RAVE sample is shown with green error bars, the subsample of unflagged stars is shown in blue. Note the better stability of the cleaned sample towards larger $\siglos$. Due to the larger errors in LAMOST, the bottom plot is on a different scale. Here we plot the bias corrected distance statistics (red error bars) vs. the uncorrected distance statistics (blue). The last data points for the corrected sample are above the figure margin.}\label{fig:rvatest}
\end{figure}

\begin{figure}
\epsfig{file=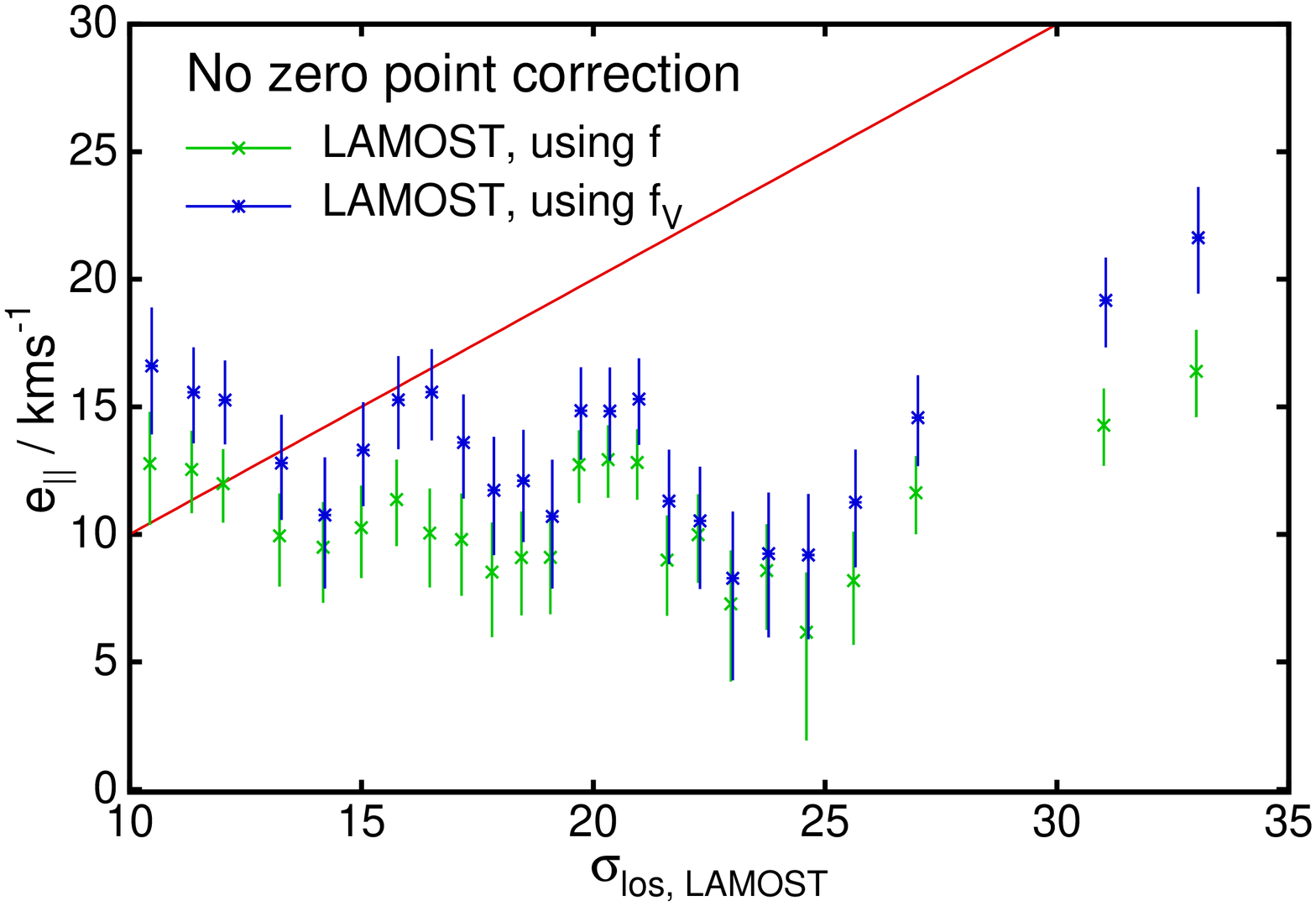,angle=-90,width=\hsize}
\epsfig{file=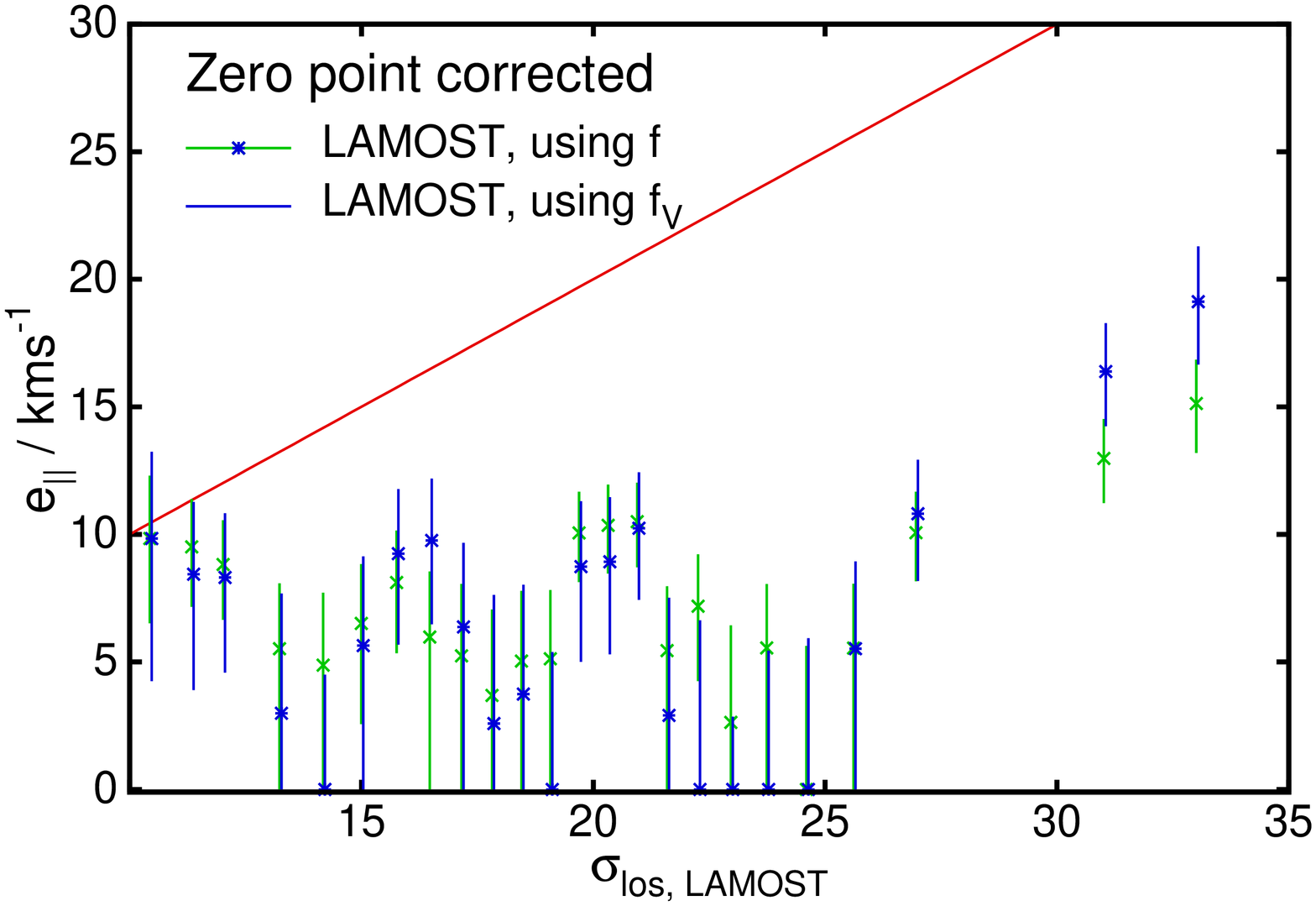,angle=-90,width=\hsize}
\epsfig{file=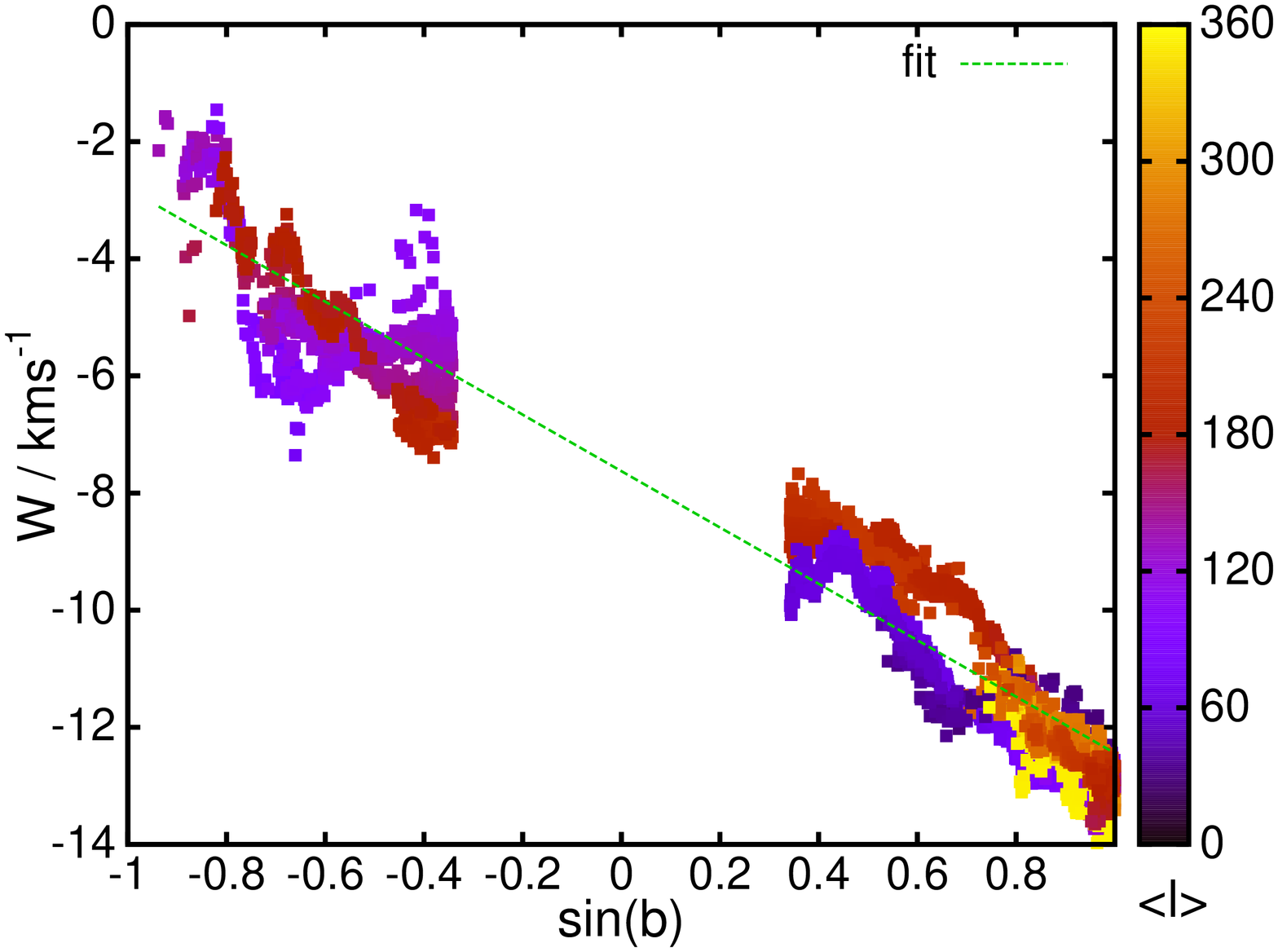,angle=-90,width=\hsize}
\caption{Top panel: Measuring the $\vlos$ source dispersion $\epar$ in subsamples of stars ordered by the measurement uncertainty as provided by LAMOST in each subsample. Since for these values $\epar^2 \gg \sigb$, the values on the y-axis can be directly compared to $\sigloslamost$. We mark the identity line in red. The plot uses again a sliding mask, i.e. every third datapoint is independent. In the middle panel, we show the same statistics, but using a general correction of LAMOST $\vlos$ by the offset $\delta \vlos = 5 \kms$. The bottom panel shows the average vertical velocity $W$ vs. average $\sin \gb$ when we take around each star all objects with $\sin\gb$ values within $0.1$ and longitude $\gl$ within $\pi/4$. The trend-line (green dashed) was obtained by fitting equation (\ref{eq:xi}) on the LAMOST sample; its slope indicates a $5 \kms$ global bias in LAMOST $\vlos$ measurements, which we corrected to create the middle panel.}\label{fig:rvatestlamost}
\end{figure}

\begin{figure}
\epsfig{file=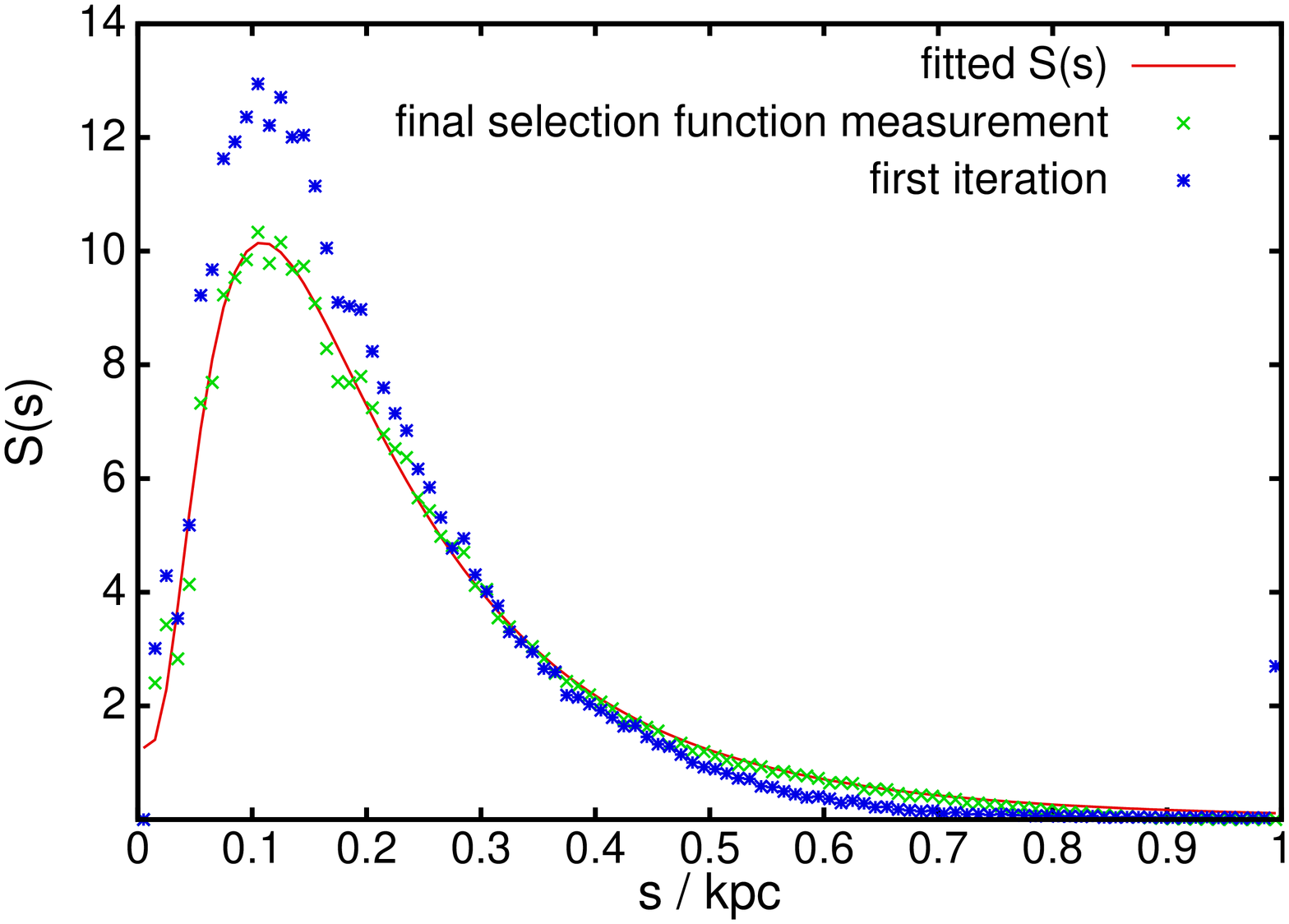,angle=-90,width=0.9\hsize}
\epsfig{file=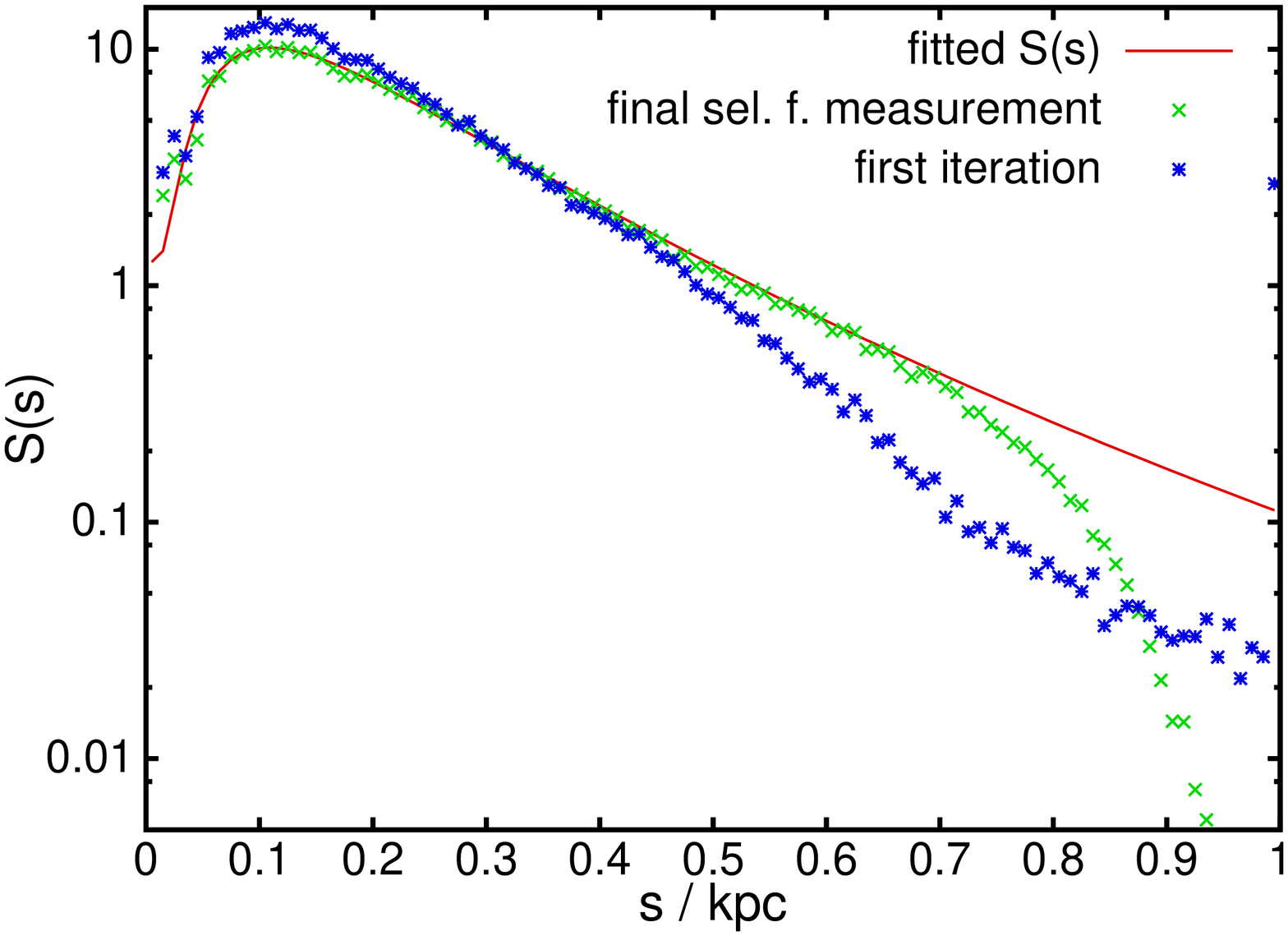,angle=-90,width=0.9\hsize}
\epsfig{file=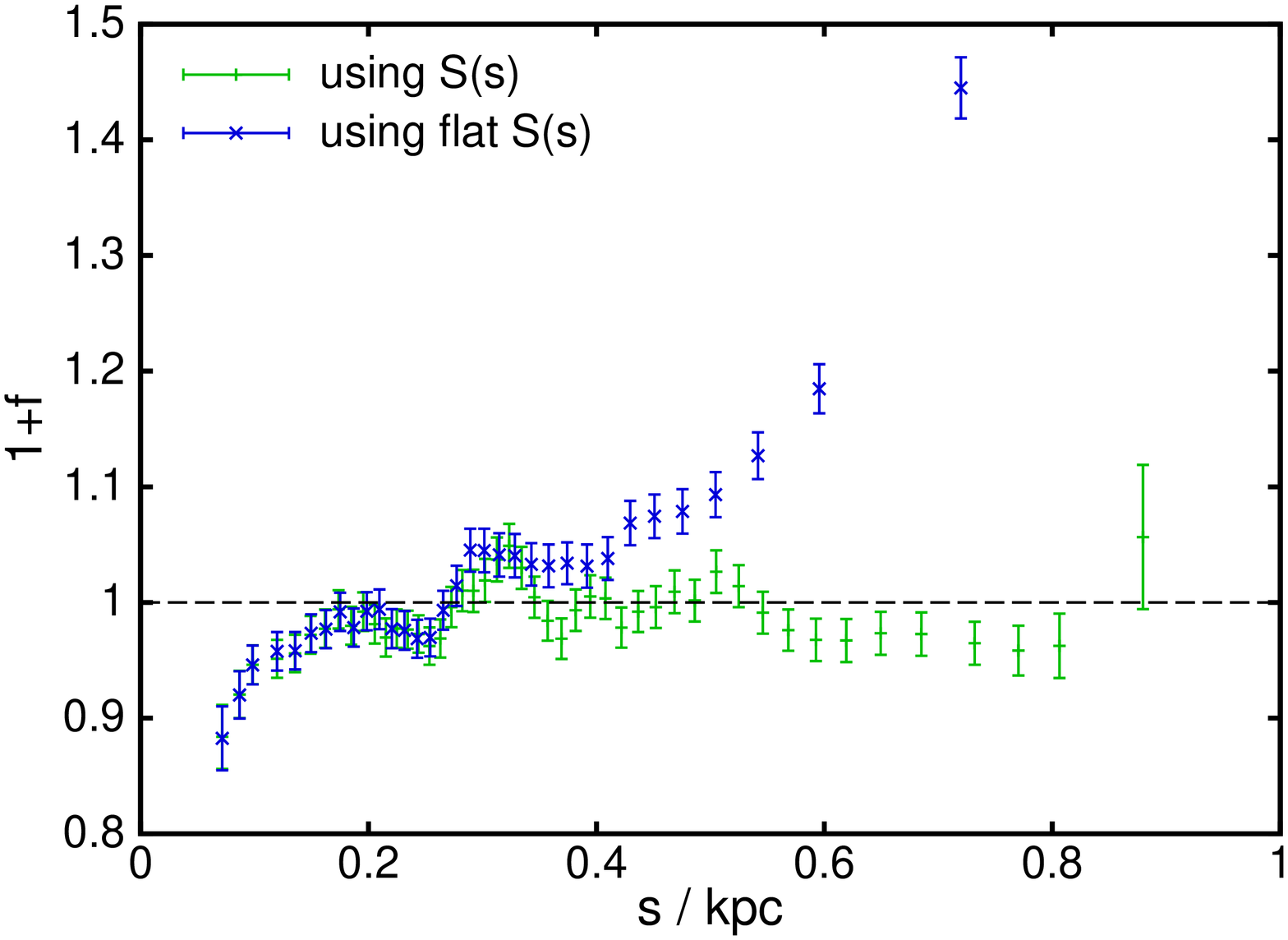,angle=-90,width=0.9\hsize}
\caption{Derivation of the RAVE selection function using the entire RAVE-TGAS sample, the top panel in linear scale, the middle panel in log-scale. The blue points show the first measurement for S(s) when starting with a trial flat selection function. The selection function $S(s)$ converges as shown within less than $5$ iterations to its final value. The red line gives the smooth fitting function applied. The break-down of the selection function on the right-hand end of the distribution is caused by the quality cuts on the parallax accuracy, not to be fitted. The bottom panel compares the distance statistics between using the final $S(s)$ (green) and a flat $S(s)$ (blue) binned in $\langle s \rangle $. The last (blue) data points in the flat $S(s)$ are not shown as they have $1+f > 2$.}\label{fig:priortestall}
\end{figure}

\subsection{Testing line-of-sight velocity errors}\label{sec:vloserr}

We can use our method to test line-of-sight velocity errors. Even if we do not fully trust the parallax distances, Gaia-TGAS astrometry and hence our distance estimates are independent of the stellar parameters and line-of-sight velocity determinations. We can thus reasonably assume that the real distance bias in the sample should be nearly the same between samples of stars grouped according to their $\siglos$ given by the spectroscopic surveys. A minor caveat is the potential correlations of line-of-sight velocity errors with stellar apparent magnitude, metallicities and ages, which introduce a small bias to the selection function.

The top panel of Fig. \ref{fig:rvatest} shows $1+f$ for subsamples of RAVE grouped by $\siglosrave$ and binned to samples of $4500$ stars, sliding the mask by $1500$ stars each, so every third data point is independent. We use the given error estimates from the RAVE pipeline, i.e. we set $\epar = \siglosrave$. The contribution to $f$ from $\siglos$ is small in this region, reaching $\delta f \sim 1.5 \%$ at $\epar = 5 \kms$. Green errorbars show the statistics for the entire RAVE sample, while the blue error bars depict the RAVE sample with no flags from \cite{Matijevic12}. There are some stars in the RAVE sample with given $\siglosrave > 10 \kms$, and the sharp decline in the $1+f$ estimate seen in the full sample continues for those. Since their distance error should not be substantially different, this implies that beyond $\siglosrave > 5 \kms$, $\vlos$ estimates are significantly more uncertain than stated and should not be used at all. It is apparent that the unflagged sample shows excellent stability; the little aberration around $\siglosrave \sim 1.7 \kms$ is likely just a statistical fluctuation (about $10$ error bars should have their $1 \sigma$ confidence intervals not crossing $1.0$). The entire sample, however, shows a clear abnormality for $\siglosrave > 2 \kms$. Apparently the flagged stars contain line-of-sight velocity uncertainties beyond the provided errors, but the contamination is comparably small for $\siglosrave < 2 \kms$. There is some tentative indication for problems with the line-of-sight velocities of stars with very small errors $\siglosrave < 0.7 \kms$, possibly from some erroneous $\siglosrave$ determinations going in hand with $\vlos$ measurement errors.

The same analysis on LAMOST (bottom panel of Fig. \ref{fig:rvatest}) demonstrates the usefulness of the SBA method. Note that the x-axis in this plot just starts beyond the range of the top panel, i.e. in the entire sample, the correction of $f$ for $\epar$ and hence $\vlos$ errors is crucial. Here, we show the bias corrected estimates for $1+f$ with red error bars, setting $\epar = \sigloslamost$. These are compared to the same estimate without correcting for the $\epar$ bias. If the pipeline estimates for $\sigloslamost$ were correct, the red errorbars should lie in a horizontal line, while the uncorrected estimates (blue) should trend sharply downwards. However, we observe the opposite. While the uncorrected $1 + f < 1$ indicates a significant $\vlos$ source dispersion $\epar$, the points lie in a horizontal line, indicating that there is almost no correlation between the real $\epar$ and the estimated $\sigloslamost$. We conclude that $\sigloslamost$ does not indicate the real $\vlos$ measurement error, and that the real $\vlos$ measurement errors must be far better than claimed by their pipeline. In accordance with this, the red errorbars trend sharply upwards, because the bias correction increasingly overestimates the real $\vlos$ measurement uncertainty, explaining our findings in Table \ref{tab:distf}.

\begin{table}
\caption{Fit parameters for testing the line-of-sight velocities via the dependence of vertical velocities on $\sin(\gb)$ as in equation (\ref{eq:xi}). The slope $\xi$ between $W$ and $\sin(\gb)$ can be interpreted as an average $\vlos$ measurement offset $\delta \vlos$.}
\begin{tabular}{l||r|r|}
sample & $\delta \vlos / \kms$ & $-\Wsun / \kms$  \\ \hline
${\rm RAVE_{all}}$ & $0.26 \pm 0.13$ & $-7.68 \pm 0.09$ \\
${\rm RAVE_{nf}}$ & $0.20 \pm 0.16$ & $-7.56 \pm 0.10$ \\
${\rm RAVE_{10,nf}}$ & $-0.09 \pm 0.28$ & $-8.12 \pm 0.19$ \\
${\rm RAVE_{b}}$ & $0.24 \pm 0.47$ & $-8.27 \pm 0.31$ \\
${\rm LAMOST}$ & $-4.81 \pm 0.21$ & $-7.63 \pm 0.15$ \\
${\rm LAMOST_{e15}}$ & $-4.72 \pm 0.40$ & $-7.73 \pm 0.29$ \\
\end{tabular}\label{tab:veltest}
\end{table}

The top panel of Fig. \ref{fig:rvatestlamost} shows the measurement of $\epar$ from the bias in $f$ when ordering the sample in terms of $\sigloslamost$. For this figure, we use our experience from RAVE that the distance bias should be negligible compared to the $\epar$ bias correction, and hence determine $\epar$ by demanding an estimate of $f=0$ after the bias correction. The error bars are derived by varying $\epar$ until the estimate for $f$ becomes marginally positive or negative. Since the bias correction on $f$ from equation (\ref{eq:vloscorr}) is quadratic in $\epar$, the error bars on $\epar$ are asymmetric. The plot clearly shows that $\epar$ is nearly uncorrelated with $\sigloslamost$, in fact it slightly declines. The sudden rise of $\epar$ near $\sigloslamost = 27 \kms$ indicates problems with $\vlos$ estimates beyond this point, so that we from now own will adopt the quality cut $\sigloslamost < 27 \kms$. However, there is a problem: The estimates of $\epar$ using the full set of terms in $f$ are significantly smaller than the estimates of $\epar$ when using only the azimuthal vs. vertical motions in $f_{V}$. Something must be wrong beyond a random uncertainty.

The bottom panel of Fig. \ref{fig:rvatestlamost} solves this riddle. Around each star in LAMOST we select all stars that have a similar value in galactic latitude, i.e. $\Delta \sin(\gb) < 0.1$ and lie within $\pi/4$ in $\gl$. We plot the average vertical velocity of each subsample vs. its mean value in $\sin(\gb)$; the colour indicates the mean longitude $<\gl>$ in degrees. Apart from some statistical fluctuations, the sample lies almost perfectly along a line:
\begin{equation}\label{eq:xi}
W = -\Wsun + \xi \sin(\gb) $.$
\end{equation}
However, $\sin(\gb)$ is just the projection factor of the $\vlos$ estimate into $W$, so the slope $\xi$ can be interpreted as an average global offset in the $\vlos$ determinations: $\xi = \delta \vlos$. An alternative explanation would be a vertical breathing mode in the disc, but it cannot explain the perfect trend. Table \ref{tab:veltest} compares fits of equation (\ref{eq:xi}) for different subsamples of RAVE and LAMOST. No subsample of RAVE shows an appreciable slope, ruling out a physical phenomenon. Note that the plotting technique leads to a strong autocorrelation between the plotted points (there would be about $20-30$ independent samples with an error of $\sigma_{<w>}\sim 1 \kms$), so to the eye, there might be an optical illusion of deviations from the plotted trend, which are, however, not significant. We conclude that LAMOST has an average zero-point offset $\delta \vlos \sim -5 \kms$, which we correct by adding $5 \kms$ to all LAMOST $\vlos$ estimates. 

The middle panel of Fig. \ref{fig:rvatestlamost} confirms the benefit of this correction - the two different estimators for $\epar$ are now in line. The estimated $\vlos$ source dispersion $\epar \sim 7.1 \kms$ is significantly smaller than in the top panel, and shows that the LAMOST $\vlos$ determinations are far more precise than their estimate $\sigloslamost$. We also find a comparable dichotomy of mean radial stellar velocities inwards and outwards along the connection line between Sun and Galactic Centre, which confirms the $\vlos$ offset found on vertical motion. Such an offset in $\vlos$ is not unprecedented - SEGUE/SDSS was plagued by a similar problem, and even after its correction, \cite{S12} found a residual bias, although smaller than this one. The fact that the $\vlos$ bias on SEGUE depends on colour and metallicity of stars as discussed by \cite{S12}, indicates that future revisions of the LAMOST sample should examine those dependencies as well.

To summarize our findings in this Section: We have shown that our method can be used to test the quality of $\siglos$ estimates. $\vlos$ estimates and $\siglosrave$ estimates from the RAVE survey are of high quality, with minor issues for flagged stars. A strict cut of $\siglosrave < 5 \kms$ should be applied. The LAMOST $\vlos$ determinations are far more precise than suggested by their nominal $\sigloslamost$. We adopt the quality cut $\sigloslamost < 27 \kms$. One should adopt a general error of $\siglos \sim 7 \kms$, and correct all $\vlos$ estimates by $+5 \kms$. 

\begin{figure*}
\epsfig{file=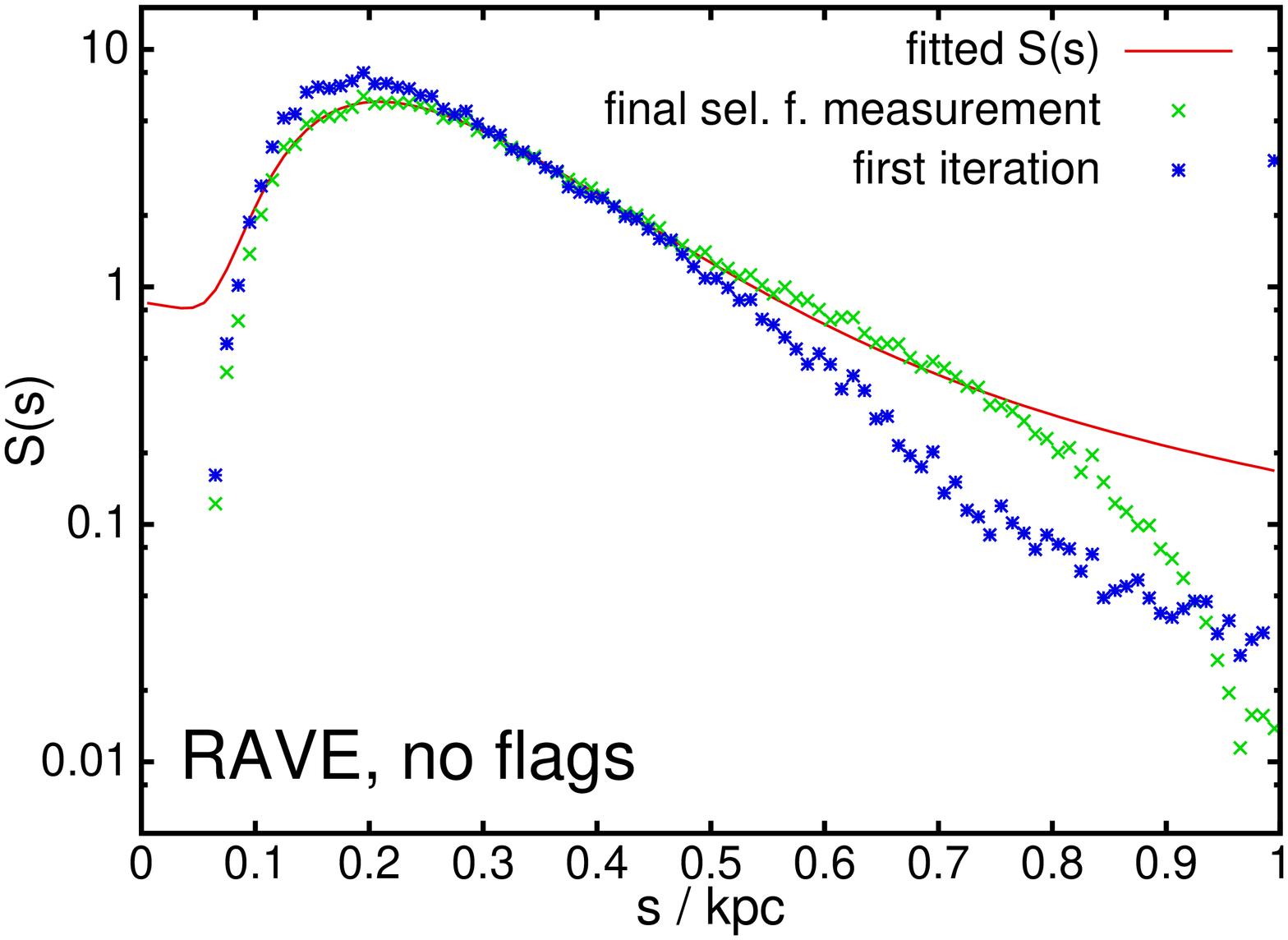,angle=-90,width=0.49\hsize}
\epsfig{file=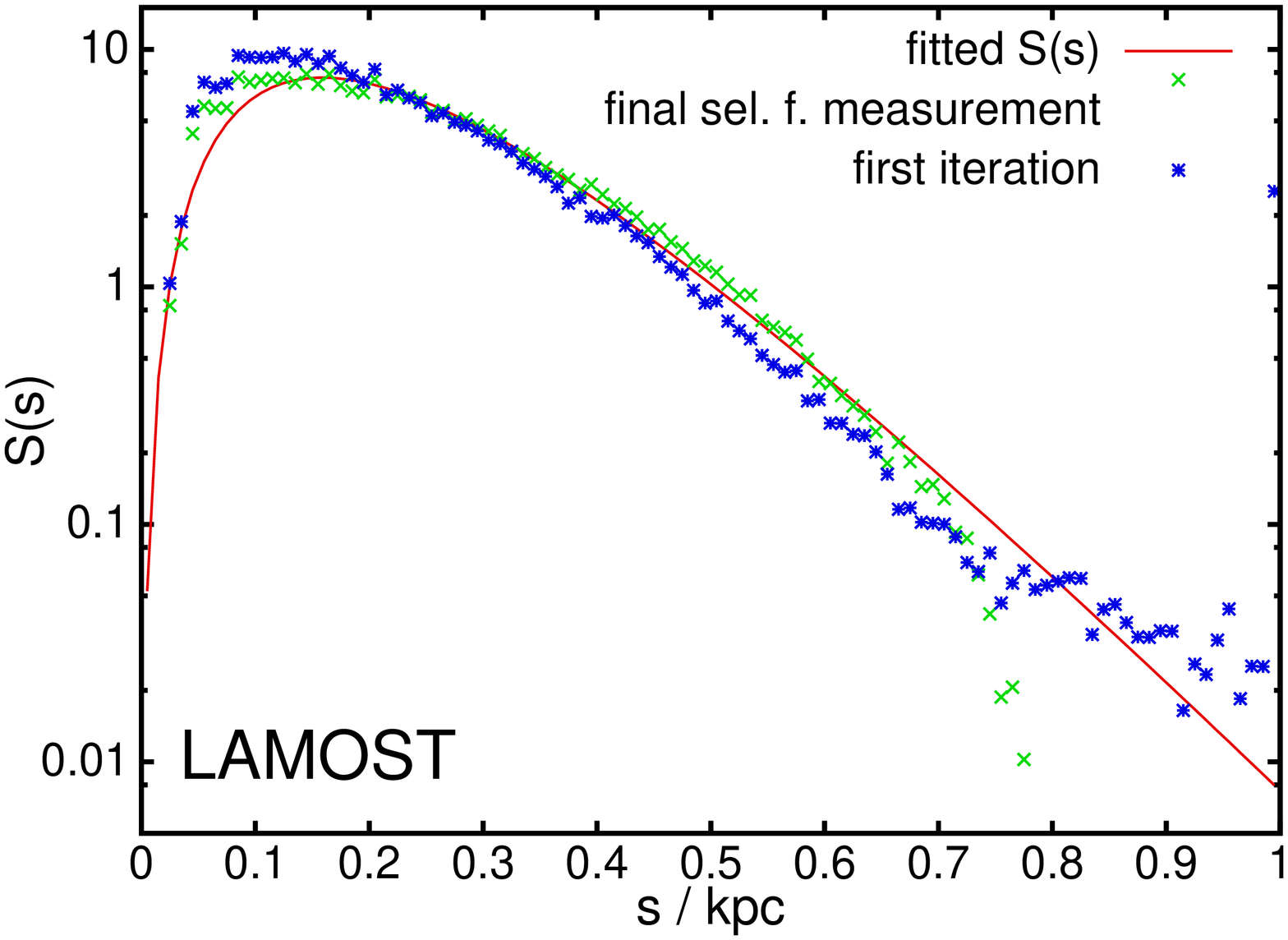,angle=-90,width=0.49\hsize}
\epsfig{file=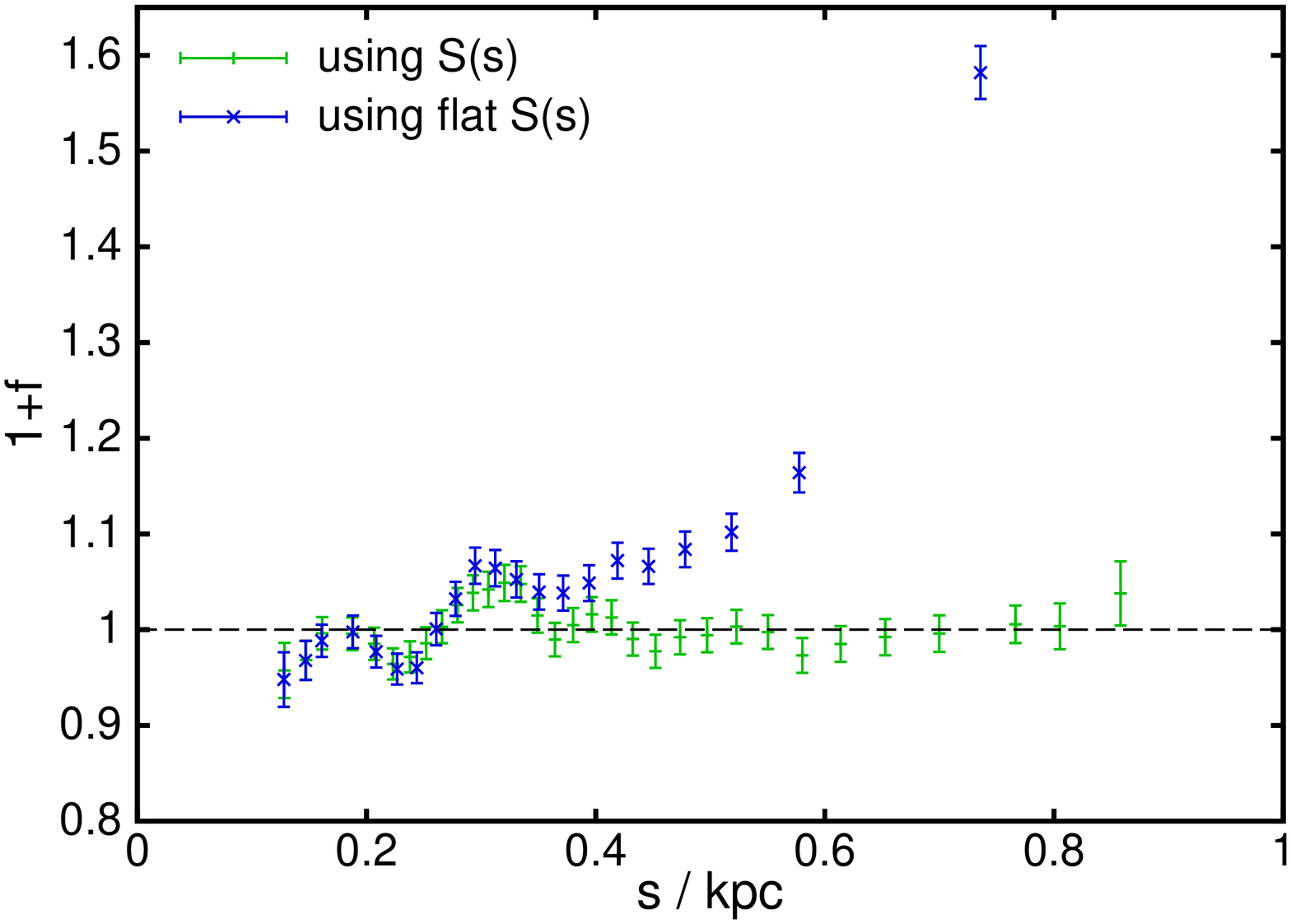,angle=-90,width=0.49\hsize}
\epsfig{file=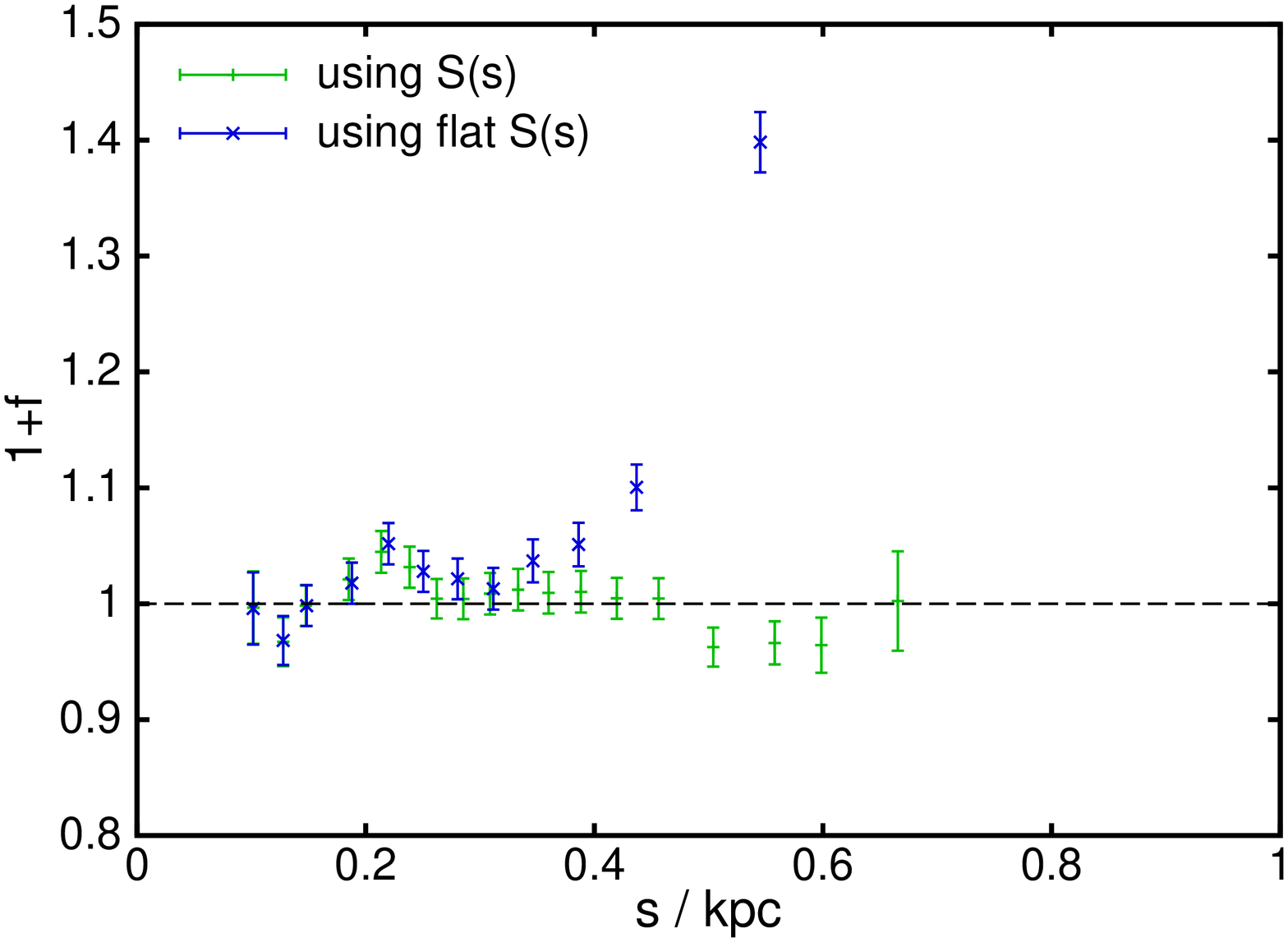,angle=-90,width=0.49\hsize}
\caption{Same as Fig. \ref{fig:priortestall}, for RAVE-TGAS stars that are unflagged (left column) and LAMOST (right column). When calculating the distance statistics (bottom row) we apply all bias corrections in RAVE, but no bias corrections in LAMOST. The last data points in the flat $S(s)$ are not shown because they are far above the range of this plot.}\label{fig:priortestrest}
\end{figure*}

\subsection{Distance dependent bias}

\subsubsection{Retrieving the prior/selection function}\label{sec:prior}

In this section we will finally derive the selection function $S(s)$ required in the distance estimate in equation (\ref{eq:distex}) and test it by measuring the distance bias $f$ in samples binned in distance $s$.

A full a priori calculation of the distance priors and selection functions $S(s)$, which enter the distance estimation in equation (\ref{eq:distex}), would demand a full chemodynamical model of the disc with all its assumptions. Population synthesis models can calculate $S(s,\tau,\feh$) and give us some indication of the shape of the selection function. However, even with this knowledge, we still cannot reliably build a selection function for our samples, since we would need to know/assume the exact distribution of stars in age and metallicity at each point in the Solar neighbourhood.

For example, the RAVE selection function described in \cite{RAVEsel} can be roughly described as a flat selection in I-band magnitudes between about $9$ and $11.3$ mag, with some wings towards lower and higher magnitudes. Running a population synthesis model as used in \cite{SB14} on this selection, demanding additionally $\Teff > 4200 \K$ for reasonable stellar parameter estimates, we obtain a relatively steep selection function $S(s,\tau)$ in distance and age: $S(s)$ at fixed metallicity and age falls off approximately exponentially with a scale-length of $0.12 \kpc$ at $s > 0.2 \kpc$, and with a flatter slope at $s > 0.5 \kpc$. At fixed metallicity and distance $S(\tau)$ behaves roughtly like $ 1/(\tau + 1.5) \Gyr$. 

Here, we will adopt a simpler, more direct approach: we can actually derive the $S(s)$ from the data. We know that with the correct prior, this should give stable values of $f$ vs. $\langle s \rangle $, just like we could demand a stable distance bias $f$ when binning in $\siglos$ in the previous section.

To obtain the full prior including the selection function, we start from the simple density model in equation (\ref{eq:rho}). For the sake of simplicity, we assume that the selection function is solely a function of distance, i.e. $S(s)$, ignoring angle-dependent deviations from the simple density model, and more importantly from the age- and metallicity structure of the disc. We can now measure $S(s)$ in a simple way: With an expected distribution of stars in distance $\rho_S(s)$, we can estimate $S(s) \sim \rho_m(s) / \rho_S(s)$, where $\rho_m(s)$ is the measured density of stars in distance. Since the distances of single stars are quite well-determined, we approximate $\rho_m(s)$ by simply binning the number of stars according to their distance expectation value $\langle s \rangle $. One caveat is that near the far end of the sample, the measurement will fall quickly below the real selection function, due to stars dropping out from the sample by the quality cut on parallax accuracy. The latter information is already used by the quality cut, and so must not be written into the prior. 

To obtain $\rho_S(s)$ we use the positions $(l_i,b_i)$ on the sky of all stars $i$ in the sample on our density model and sum up their normalised density distributions in distance $s$:
\begin{equation}
\rho_S(s) = \sum_i N'_i s^2 \rho(s, l_i, b_i) $,$
\end{equation}
where 
\begin{equation}
{N'}_i^{-1} = \int ds \, s^2 \rho(s, l_i, b_i)
\end{equation}
normalises the integral in distance $s$ over each of the individual distance distributions to $1$, and $\rho(s, l_i, b_i)$ is calculated from equation (\ref{eq:rho}). The problem is solved by an iteration. We start with a flat $S_0(s)$ in distance, calculate all distance expectation values from equation (\ref{eq:distex}), bin them in distance to obtain $\rho_m(s)$ and calculate our first estimate for $S_1(s) = \rho_m(s)/\rho_S(s)$. We fit $S_i(s)$ with a simple analytic equation. A smooth fitting function is important, because a non-smooth function might pull data towards single points yielding a false convergence. We re-insert this analytic approximation to $S$ into the distance calculation and repeat the procedure, until the selection function converges (typically within five iterations).

The top (linear) and middle (logarithmic plot) panel of Fig. \ref{fig:priortestall} show the measurement and fits to the selection function from the entire RAVE sample. The blue points show the first estimate $S_1(s)$, and green points depict the final estimate for $S(s)$, which is the selection function that reproduces itself in further iterations. The precise parallaxes on the near end result in virtually no change of the shape of $S$ at distances $s < 0.2 \kpc$. On the long distance end, the selection function pulls stars closer, steepening $S(s)$.  

As a fitting equation we employ a log-normal distribution, plus a minor exponential term to smooth the edges:
\begin{equation}\label{eq:ravesel}
S(s) = d_1(\exp(-((\ln(s/\kpc) - \ln(a))c_1)^2/2) + d_2\exp(-sc_2) \, $,$
\end{equation}
where $a,c_1,c_2,d_1,d_2$ are free fitting parameters. The fit shown in Fig. \ref{fig:priortestall} yielded $a = 0.112, c_1 =  1.37, d_1 = 9.52, d_2 = 1.3, c_2 = 6.51 \kpc^{-1}$, converging within five iterations. The fit quality is good; the discrepancy at large $s$ is caused by the quality cut on parallaxes, which must not be part of $S(s)$.

\subsubsection{Testing the selection function}

Beyond this good convergence of $S(s)$, how can we judge if this selection function is reasonable? The key lies again in the distance statistics. The bottom panel in Fig. \ref{fig:priortestall} shows $1+f$ on the y-axis, using either a flat $S$ (blue errorbars), or the derived $S(s)$ from equation (\ref{eq:ravesel}), depicted in green errorbars. At the far end of the sample, where parallaxes are relatively uncertain, a flat selection function results in a massive overestimate of stellar distances, which in turn gets signalled by $1+f \gg 1$ in the blue errorbars. The last data points for the flat $S(s)$ are not shown, because they have $1+f > 2$, outside the plotting range. The declining parallax precision hence results in a strong trend in $f$, while with the (relatively) correct selection function, we see no such trend. This is direct proof that the selection function is sound and has to be applied. Note that when binning in $1/p$, the breakdown in the distance statistics for a flat $S(s)$ would be found in almost all of the sample with $p \lesssim 5 \marcs$. 

The left-hand column of Fig. \ref{fig:priortestrest} shows the same results for the RAVE-TGAS sample when selecting only unflagged stars (left column). The fitting parameters for the RAVE selection function (equation~\ref{eq:ravesel}) are $a = 0.21, c_1 = 2.19, d_1 = 5.43, d_2 = 0.86, c_2 = 1.75 \kpc^{-1}$. The exponential scale-length is fully degenerate with the normalisation of the exponential term, and is hence undetermined by the fit. However, it has a significant impact on the long-range behaviour of $1+f$: if we would use $e \lesssim 1.5$, i.e. a scale-length of the selection function larger than about $0.7 \kpc$, $1+f$ would become significantly larger than $1$ for $s > 0.6 \kpc$. The chosen value of $e = 1.75$ is in the range predicted by population synthesis models, and a compromise between our distance statistics and the direct fit for $S(s)$, and leads to a satisfactory result in the bottom left panel of Fig. \ref{fig:priortestrest}.

On the right-hand side of Fig. \ref{fig:priortestrest} we analyse the selection function of the LAMOST survey. This selection function is not only limited by the survey's magnitude cuts, but also by our applied signal-to-noise limit of $\SNR > 30$ in the r-band. The LAMOST selection function could not be fitted by the same log-normal distribution as RAVE, so we choose
\begin{equation}
S_{\rm L}(s) = a s^2 \exp(-cs) $,$
\end{equation}
where the free-fit parameters are determined as $a = 2230 \kpc^{-2}, c = 12.6\kpc^{-1}$. The fit is not good for small $s < 0.1 \kpc$, since we made no attempt to fit the distribution in that range. As detailed above, $S$ plays a negligible role for determining distances at $s<0.1\kpc$ due to the high quality of TGAS parallaxes. On the right-hand side there is a small bump in both the RAVE unflagged sample and in the LAMOST sample compared to our fits. Since the selection function only has to capture the rough shape of the distribution, this is acceptable. Also, the distance statistics in the bottom panel indicate that we cannot allow for a steeper decline in $S(s)$ at $s > 0.5 \kpc$, since this would make the distance underestimate in this region significant.

\begin{figure}
\epsfig{file=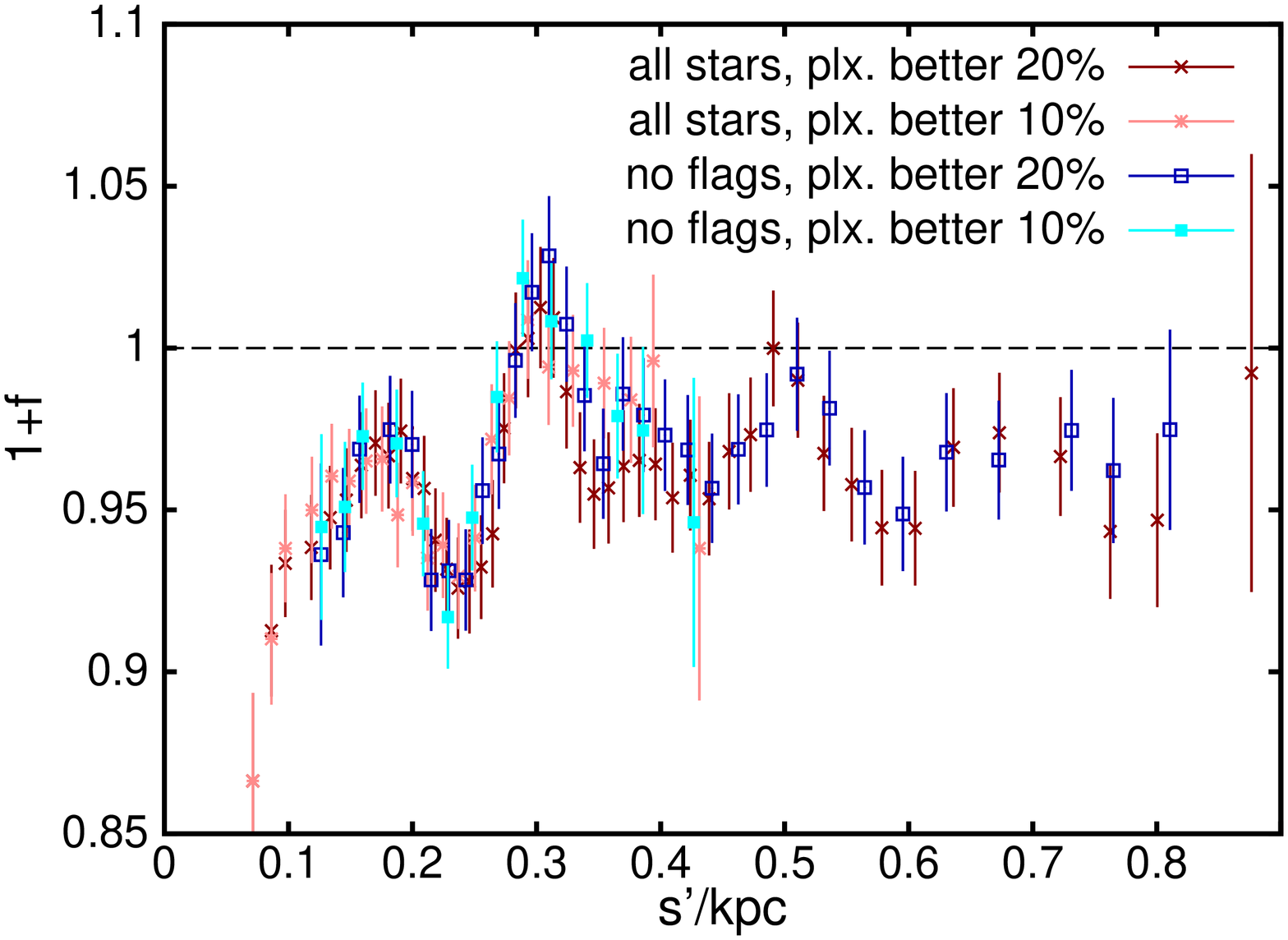,angle=-90,width=0.95\hsize}
\epsfig{file=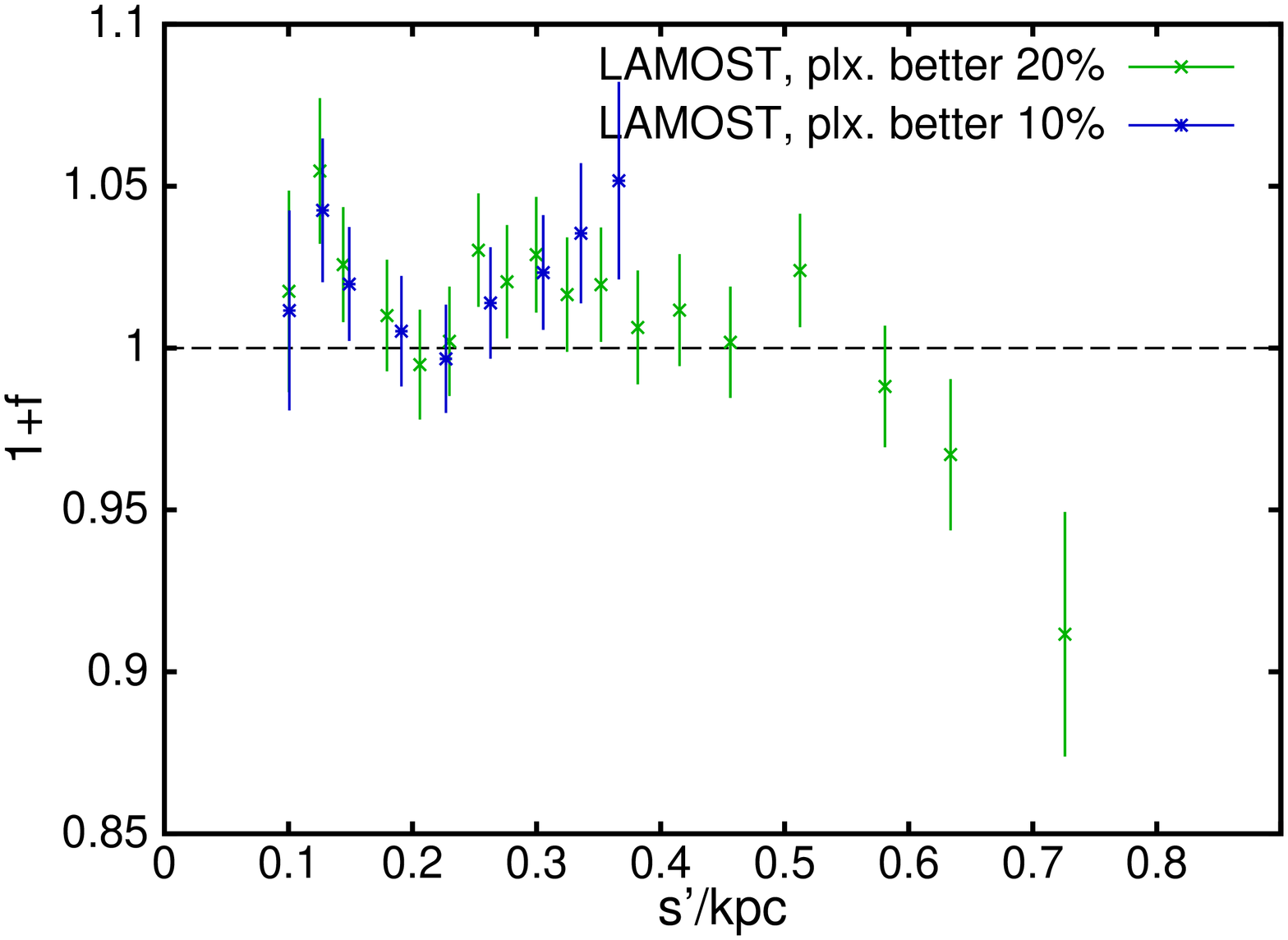,angle=-90,width=0.95\hsize}
\epsfig{file=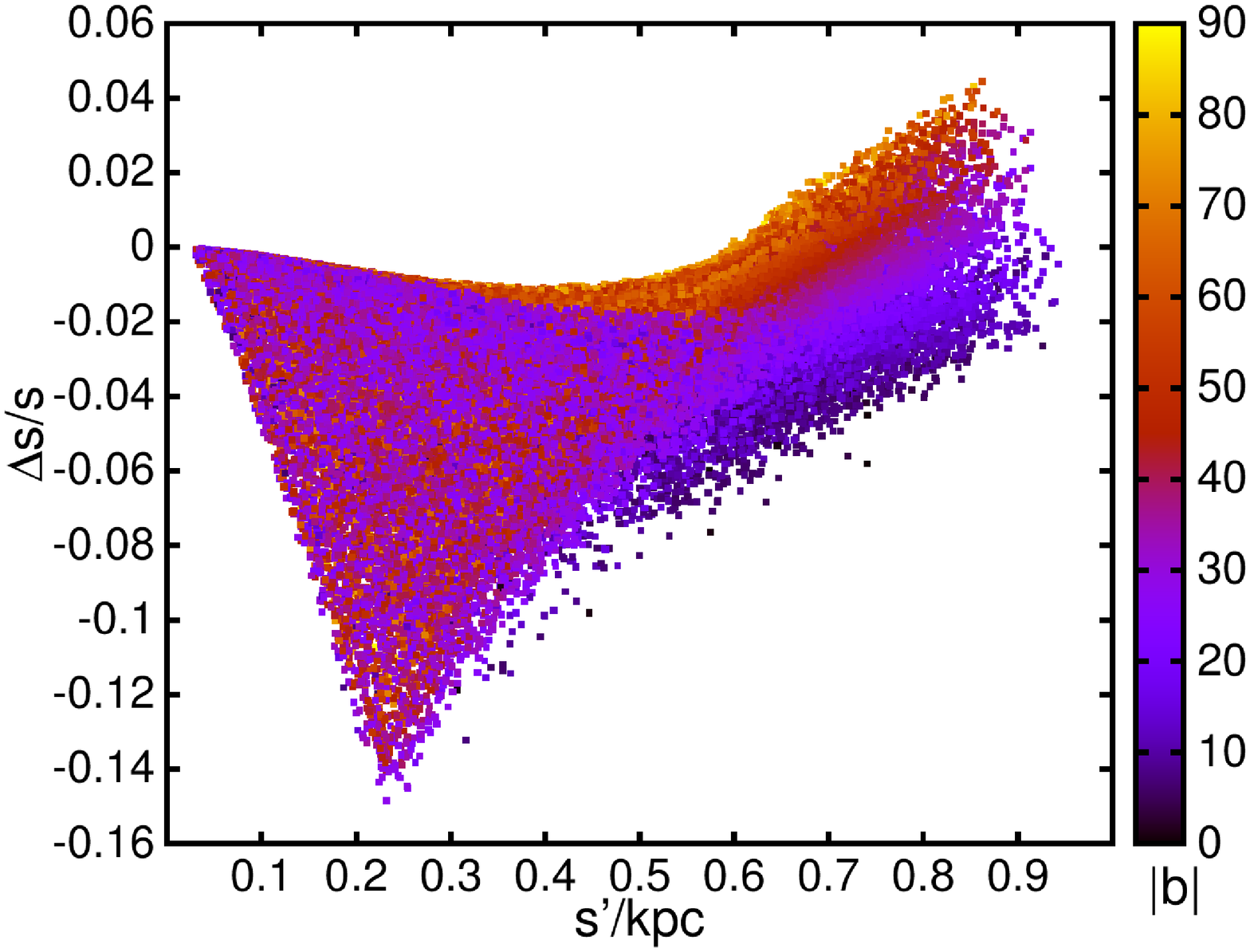,angle=-90,width=0.95\hsize}
\caption{Distance statistics when using the simple $s'=1/p$ distance estimator for RAVE (top panel) and LAMOST (middle panel). The distance statistics for LAMOST are balanced, since we use $\epar = 7 \kms$. The bottom panel shows for RAVE stars with $\sigma_p/p < 20\%$ the relative difference $\Delta s /\langle s \rangle  = (s'-\langle s \rangle )/\langle s \rangle $ between the full distance estimate $\langle s \rangle $ from equation (\ref{eq:distex}) and $s'=1/p$ coloured by the absolute $|\gb|$.}\label{fig:invpar}
\end{figure}

\subsubsection{Nature's bad joke: The simple parallax estimator}

Fig. \ref{fig:invpar} shows the estimated $1+f$ distance estimator vs. $s'$ for the RAVE-TGAS sample (top panel) and the LAMOST sample (middle panel), when we use the naive estimate $s' = 1/p$ instead of the full distance estimate from equation (\ref{eq:distex}). These statistics look surprisingly good. While the general distance underestimate in RAVE gets slightly worse and there is a trend towards small $1+f$ at large $s'$, but the naive estimate obviously avoids the catastrophic failure we registered when neglecting the selection function, as shown in Fig. \ref{fig:priortestall} and Fig. \ref{fig:priortestrest}.

Already \cite{Stromberg27} pointed out that one should be careful not to use $s' = 1/p$ blindly, but by a mood of nature, or rather the selection function, the resulting distance bias in these samples is quite small. We would expect a gradual increase of the bias towards larger distances, where the parallax quality decreases, just like it happened with $1+f$ vs. distance, when we used the flat selection function $S(s)$ in the previous section. Yet, for RAVE-TGAS $1+f$ is remarkably stable. This is achieved because the selection function in RAVE almost perfectly cancels the terms from the coordinate transformation and the density distribution (the selection function falls steeply in distance, acting in the same direction as the $s'$ estimate). The LAMOST sample shows good stability out to about $0.5 \kpc$. 

The cancellation of the distance biases is also evident from the bottom panel of Fig. \ref{fig:invpar}, where we plot for the full RAVE sample with parallaxes better than $20\%$ the relative difference in the distance estimates $\Delta s/\langle s \rangle  = (\langle s \rangle  - s')/\langle s \rangle $ for each star. $\langle s \rangle $ is the full distance estimate using equation (\ref{eq:distex}). The datapoints are coloured by their galactic latitude $|\gb|$. Since the density falls quicker towards higher latitudes, the distance underestimate by $s'$ is stronger for stars at small latitude $|\gb|$. To understand the shape of this plot, we remember the behaviour of the RAVE selection function from equation (\ref{eq:ravesel}) and the top panel of Fig. \ref{fig:priortestall}. At distances below $\sim 0.2 \kpc$ the selection function is rising and $s^2\rho_S(s)$ rises sharply as well, so that stars with even moderate parallax uncertainty show a strongly negative $\Delta s/\langle s \rangle $. However, beyond that point, the selection function and the spatial density fall steeply, bringing $s'$ closer to the full estimate.
As a result, the average $\Delta s/\langle s \rangle $ is about $-2.4 \%$ in the RAVE sample with parallaxes better than $20\%$ ($\sigma_p < 0.2 p$), and $-1.4 \%$ in the sample with parallaxes better than $10 \%$.

\begin{figure}
\epsfig{file=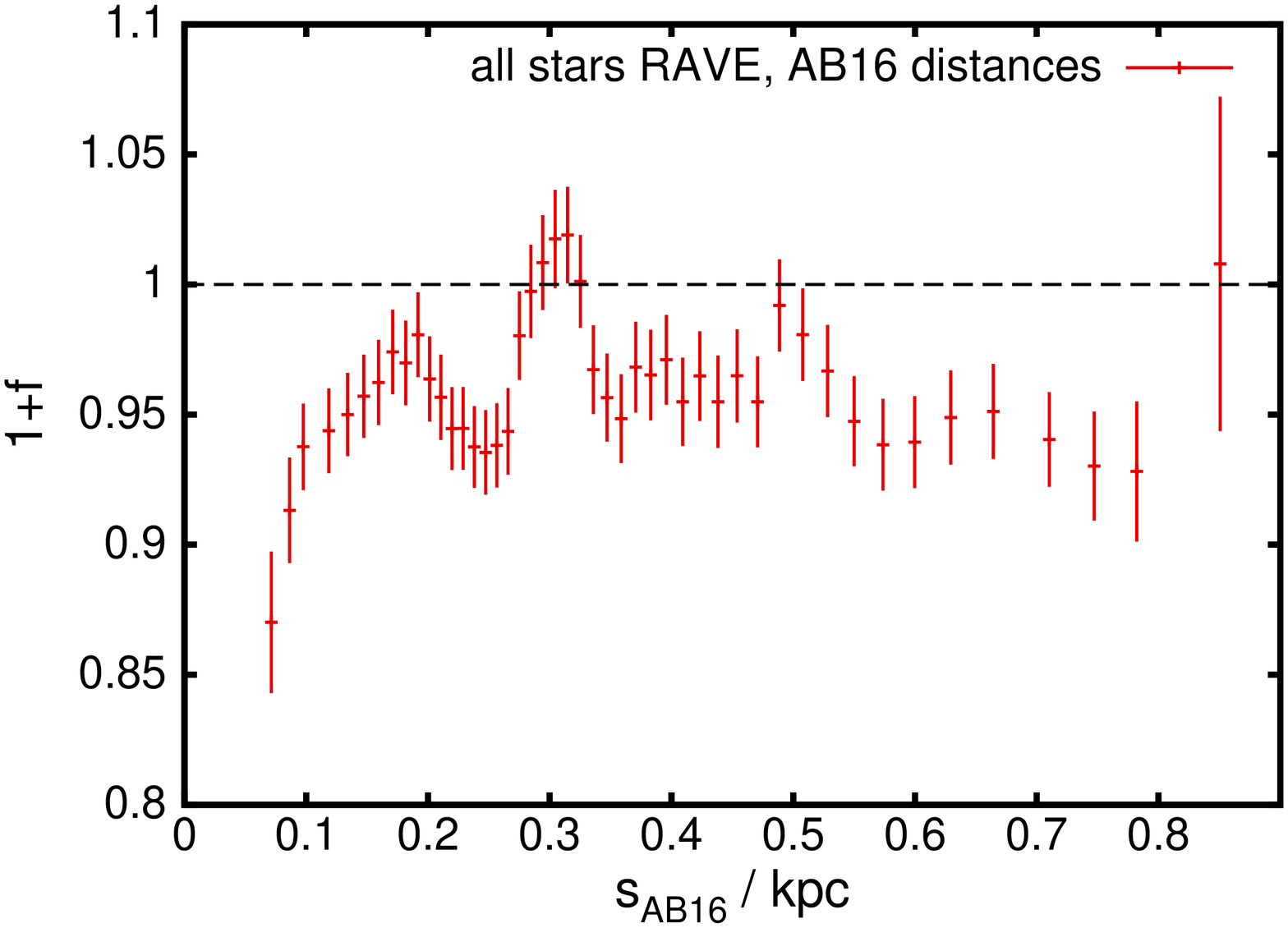,angle=-90,width=0.99\hsize}
\epsfig{file=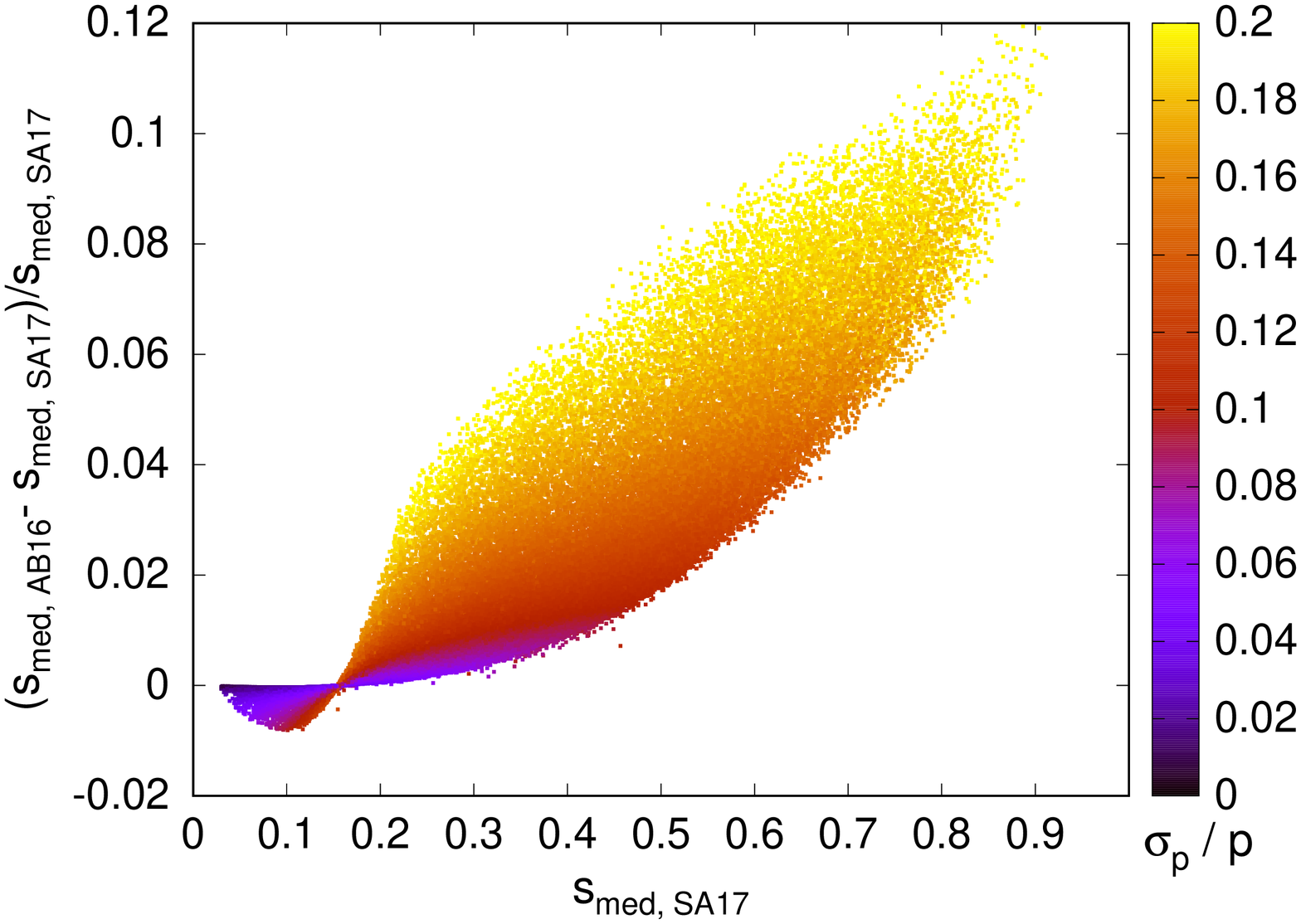,angle=-90,width=0.99\hsize}
\caption{Comparison with the distance estimates from Astraatmadja \& Bailer-Jones (2016, AB16). We could not find expectation values for their dataset, so for the statistics $f$ shown in the top panel, we follow their recommendation to use the mode of the MW-estimator. In the bottom panel, we show the fractional difference of their median distances $s_{\rm med, AB16}$ minus $s_{\rm med, SA17}$, with $s_{\rm med, SA17}$ on the ordinate, coloured in $\sigma_p / p$.}\label{fig:astraatmadja}
\end{figure}

\subsubsection{Comparison with Astraatmadja \& Bailer-Jones (2016)}

To understand how our distances compare to the estimates by \cite{Astraatmadja16} (hereafter AB16), it is useful to revisit Fig. \ref{fig:distdist}, which demonstrated two things: i) the use of median and in particular mode of the distribution will strongly underestimate the true expectation value, since the posterior probability distribution is strongly skewed, and ii) neglecting the selection function $S(s)$ will result in very strong distance overestimates.

Fig. \ref{fig:astraatmadja} examines the distance estimates by AB16 on the full RAVE sample with parallaxes better than $20 \%$. The top panel shows our distance statistics applied to the distance mode of AB16, as they suggest that their comparison to other data suggest using their mode. The bottom panel directly compares our distances with theirs. It shows the fractional difference between the two distance median determinations. Towards large distances, the difference between the respective distance estimates diverges. We ascribe this to the lack of a selection function in AB16, which leads to increasing distance overestimates by them on the far end. Note that according to Fig. \ref{fig:distdist} the difference in expectation values would be even larger, since the shown median is less vulnerable to the long tail of the distributions. For nearby stars, the bright limit of the survey dominates, and the selection function rises, which is why their median distances are shorter. We colour-coded the relative parallax error $\sigma_p/p$ in the bottom panel, which shows that this is the main explaining variable for the scatter in this plot. 

On the far end, the neglected selection function (leading to a distance over-estimate), and the use of the mode (leading to a distance under-estimate), thus nearly cancel each other, and so the distance statistics shown in the top panel of Fig. \ref{fig:astraatmadja} show only a mild distance underestimate, just in line with Fig. \ref{fig:distdist}, where for our example star also the expectation value for the correct posterior distribution and the mode of the posterior with neglected $S(s)$ coincide.
For short distances/nearby stars the selection function $S(s)$ rises, and so the two biases have the same sign, leading to comparably strong distance underestimates.

\begin{figure}
\epsfig{file=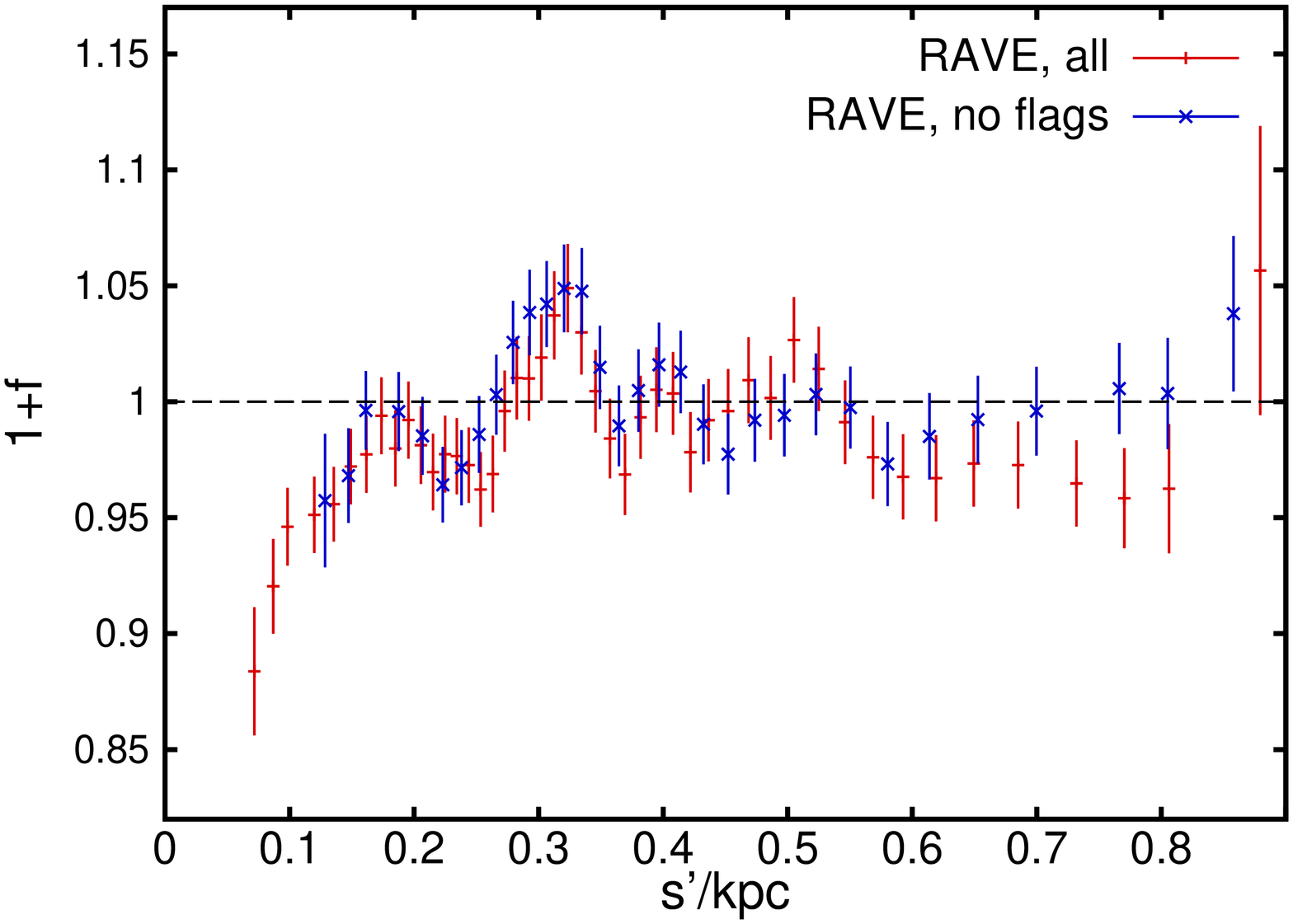,angle=-90,width=0.9\hsize}
\epsfig{file=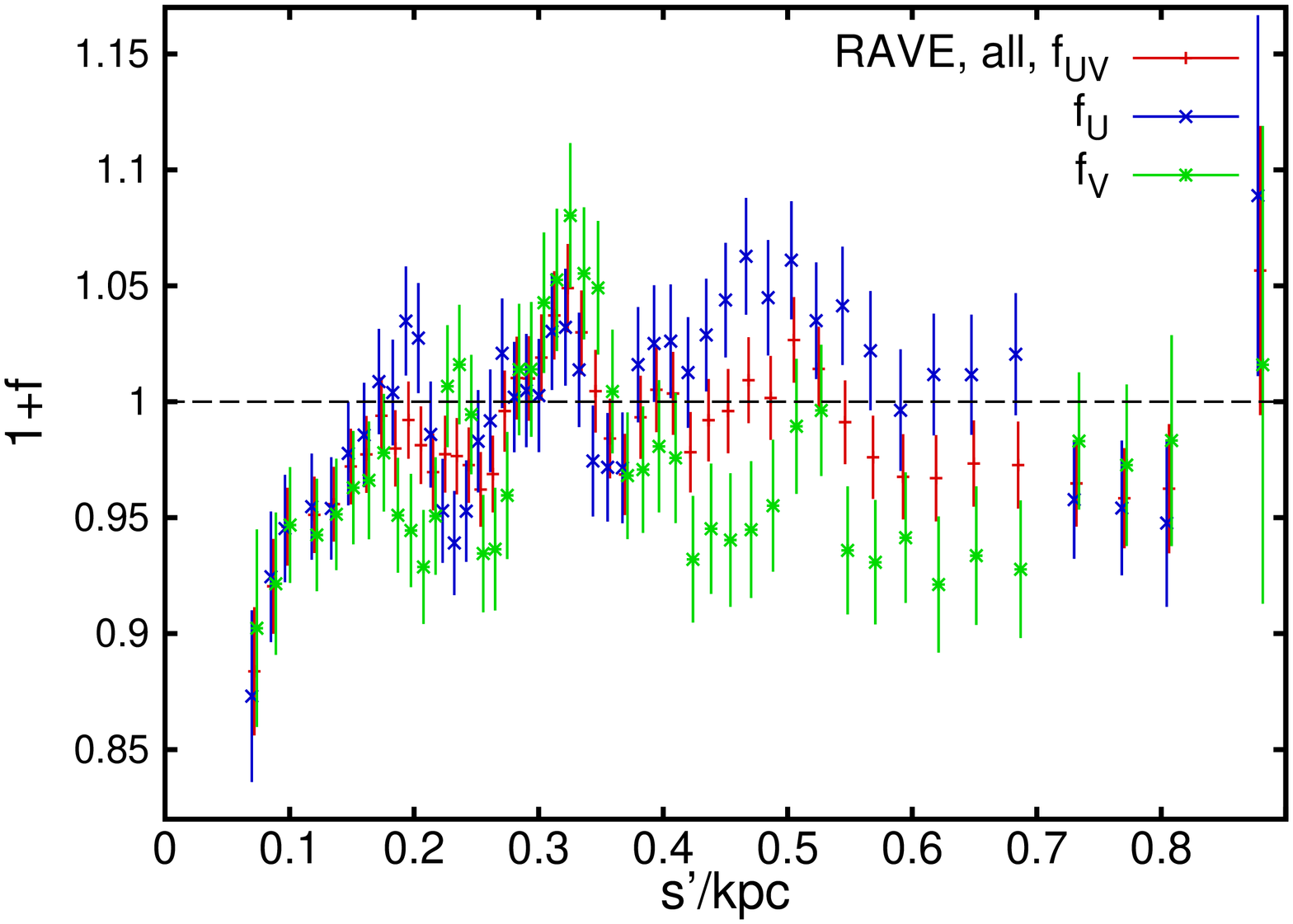,angle=-90,width=0.9\hsize}
\epsfig{file=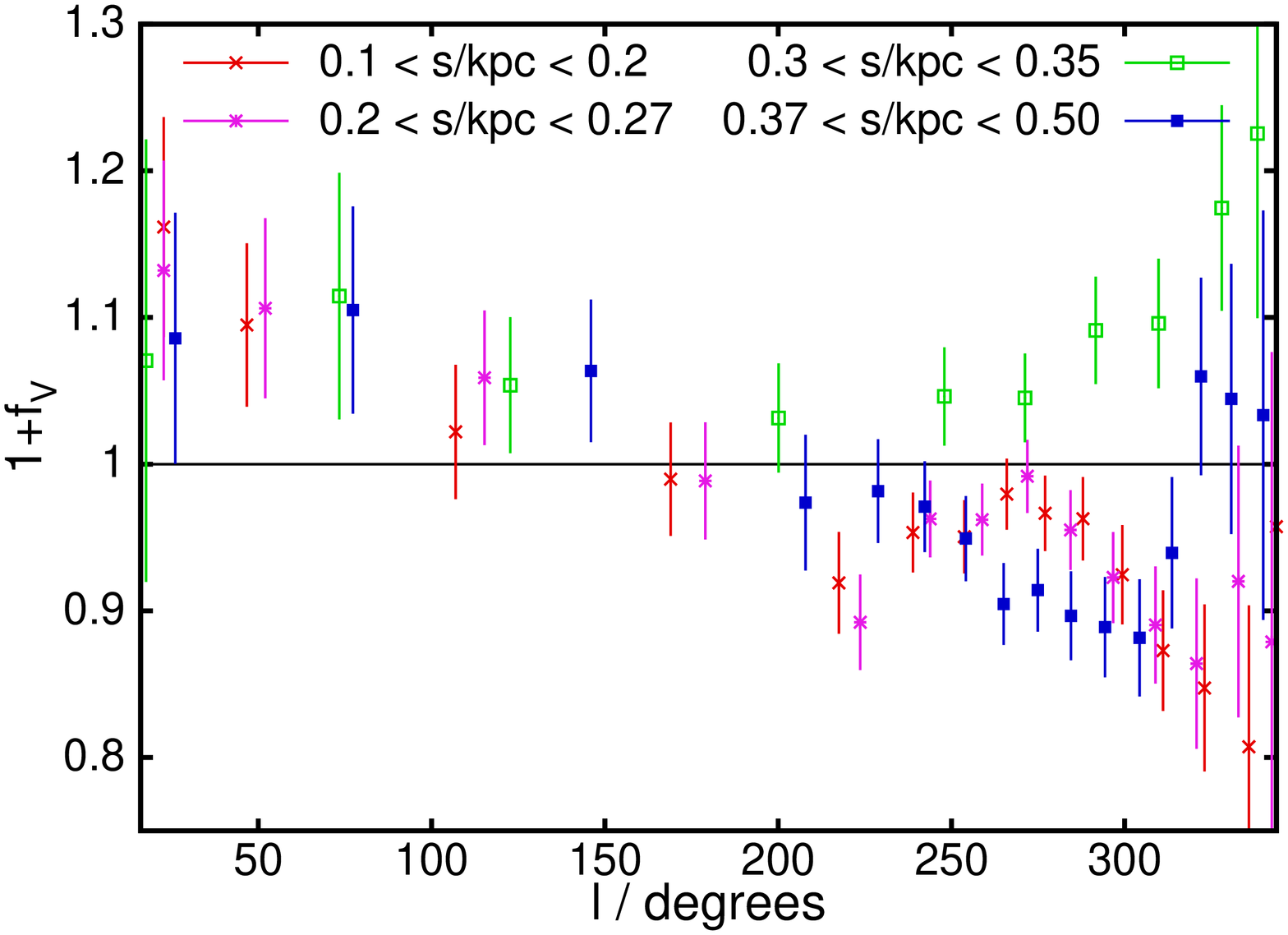,angle=-90,width=0.9\hsize}
\caption{Top panel: comparing the distance statistics for RAVE stars, when binning in distance to samples of $6000$ stars each, sliding the mask in steps of $2000$ stars. Middle panel: Comparing the full distance estimator $1+f$ (red) to the restricted estimators using only the radial $f_U$ (green), or the azimuthal $f_V$ (blue) velocity components. Bottom panel: Dissection of the distance statistics vs. Galactic longitude $\gl$ in the RAVE sample at different heliocentric distances $s$.}\label{fig:distscan}
\end{figure}

\subsection{A bump in the distance statistics}\label{sec:dist03}

The top panel of Fig. \ref{fig:distscan} shows the distance estimator $1+f$ when binning stars in the RAVE sample according to their distance $\langle s \rangle $ in bins of $6000$ stars each. Both sample selections of the RAVE-TGAS sample (red error bars: all stars, green errror bars: unflagged stars) display a marked peak in $1+f$ around $s \sim 0.3 \kpc$. The feature is very narrow, and mostly limited by the distance accuracy and our ability to sample around it. It is stronger than we would expect to be possible for a statistical fluctuation, and we have checked that changes in the adopted selection function do not alter this result. In addition, an increase of the parallax accuracy does not remove the feature, even if we use $\sigma_p/p < 10\%$. We can exclude that this is a global issue with the parallax measurements, since the LAMOST survey in the opposite direction of the sky does not show this signal. 

In the middle and bottom panels of Fig. \ref{fig:distscan} we try to dissect the sample further to find the origin of this feature. As we have discussed in Section \ref{sec:caveats}, separating the different parts of the distance estimator, and in particular separating the sample in longitude or latitude make the analysis more prone to impacts by Galactic substructure. The middle panel compares the full estimator to its restricted counterparts from equation (\ref{eq:restricted}) which exclusively use the azimuthal velocity component ($f_V$, blue) or the radial velocity component ($f_U$, green). The peak around $s = 0.3 \kpc$ is stronger in $f_V$ than in $f_U$, though both estimators show a distance overestimate. The bottom panel attempts to dissect the sample in longitude $\gl$. This is not a safe selection, and likely the general downtrend seen in $\gl$ is a consequence of the Galactic warp (the warp correlates $W$ and $V$, since the sample is mostly at $\gb < 0$ and $T_{VW}$ changes sign at $\gl = 180 \deg$, $f_V$ should change sign at this point). However, again the stars between $s = 0.3$ and $0.35 \kpc$ show a markedly different behaviour. Attempts to further narrow the origin of this feature revealed that the signal is coming from around $\gl \sim 300\deg$ and latitudes both above and below the Galactic plane. This position corresponds roughly to $\RA \sim 185 \deg$ and $\DEC = -60 \deg$. Detailed inspection of the kinematics did not reveal any evident stream at this position. Also, a stream passing through could correlate $W$ and $V$ motions, but the sign of $T_{VW}$ changes with the sign change in latitude $\gb$, and so $f$ should reverse sign as well, which it does not. Consistently, kinematic cuts to remove eventual outliers in $W$ do not remove the feature. We have checked that the bias corrections are more than an order of magnitude too small to account for this deviation, i.e. even a large mistake in our assumptions can not explain the feature. We have further drawn exclusion zones of $7 \deg$ around the positions of the Magellanic clouds and found no difference in the values of $f$. A vertical breathing mode in the disc would be the only possibility that we could currently imagine to give rise to this feature, and the reversal of $f_V$ to distances of $s > 0.4 \kpc$ could be interpreted as the location of a spiral arm or spur. However, there is no known spiral arm in this region. The next feature in this direction would be the Sagittarius/Carina arm, which is more than a $\kpc$ away \citep[see e.g.][]{Reid14}. In addition to the distance discrepancy the explanation by a spiral arm/spur would also raise the question how this feature can be so narrow/sharp in distance. Another possible Galactic substructure in the solar vicinity, which could have a localised effect on stellar kinematics, is the Gould belt \citep[][]{Torra00}; however, shape, position, and distance distribution of the Gould belt stars \citep[see e.g.][]{Guillout98} do not agree well with the observed feature.

We also note that all RAVE samples show a distance underestimate for stars with $s < 0.15 \kpc$. The parallaxes at this short distance are nominally too good to allow for such an effect by a wrong choice of the selection function $S(s)$. We think that this is either an extreme statistical outlier, or there is a problem in TGAS with large parallaxes and/or the bias against stars with large proper motions is significantly more severe than we take from the precaution in \cite{Lindegren16}. A definitive account is not possible from our side, since the bright limit of TGAS correlates with large proper motions, and so the fact that some nearby stars with large proper motions from Hipparcos are missing, can be interpreted in both ways. 

To summarize this: the more likely conclusion is that there is a problem with TGAS parallaxes near $\gl \sim 300 \deg$ and distances of $0.32 \pm 0.03 \kpc$ ($p \sim 3 \marcs$). This conclusion would be supported by \cite{Arenou17}, who report a problem possibly coinciding with the Gaia scanning regime change at a similar position on the sky (cf. their Fig. 28). If it is not a glitch in the data, the feature is unlikely to be a stream. However, it might be related to disc structure. Since our estimate correlates azimuthal and vertical velocities, this cannot be a pure in-plane feature, though the location would rather point to a not yet established interarm, or spur. This feature warrants tests of the Gaia pipeline, and it should be investigated as a possible place for Galactic substructure.

\begin{figure}
\epsfig{file=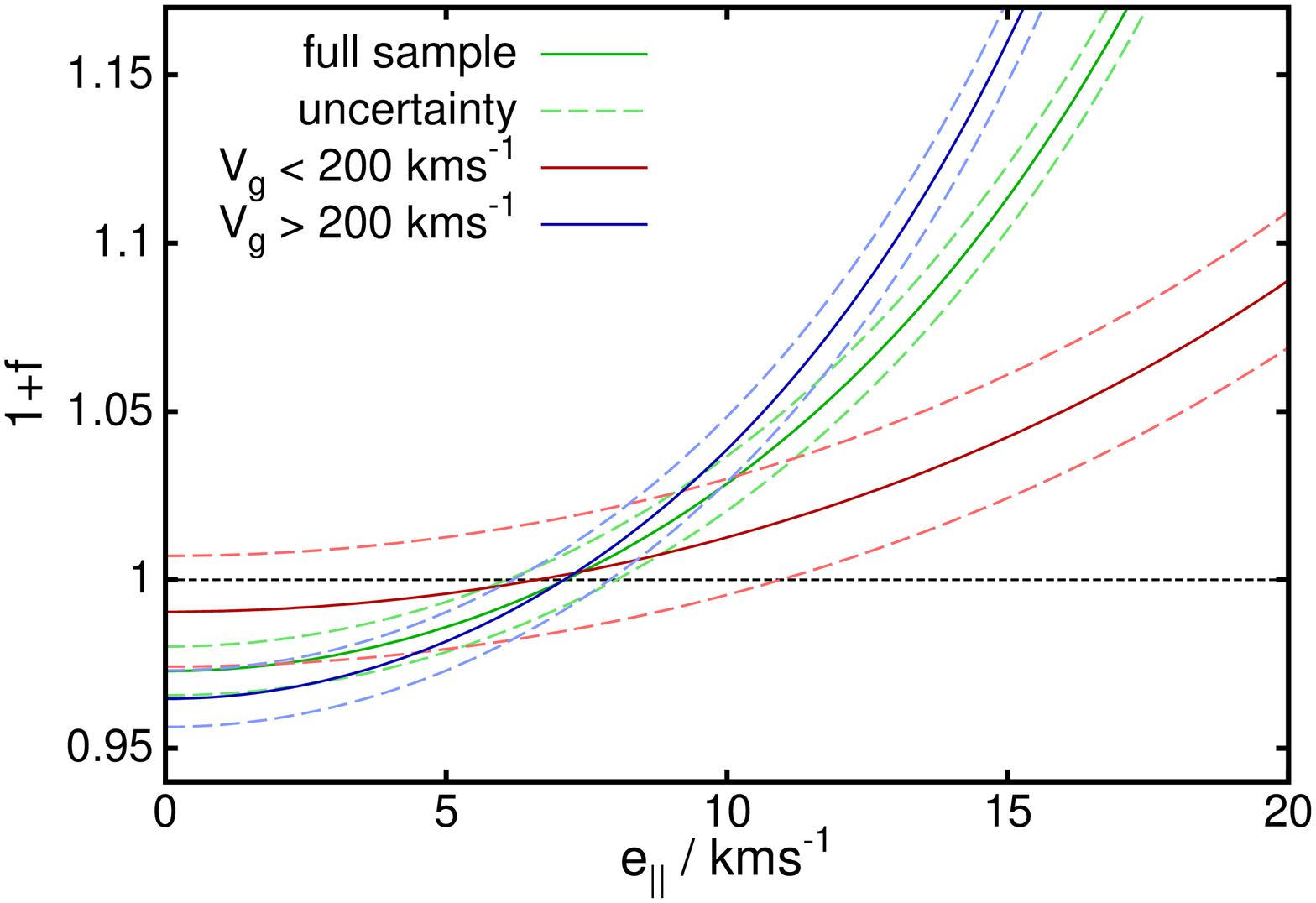,angle=-90,width=\hsize}
\epsfig{file=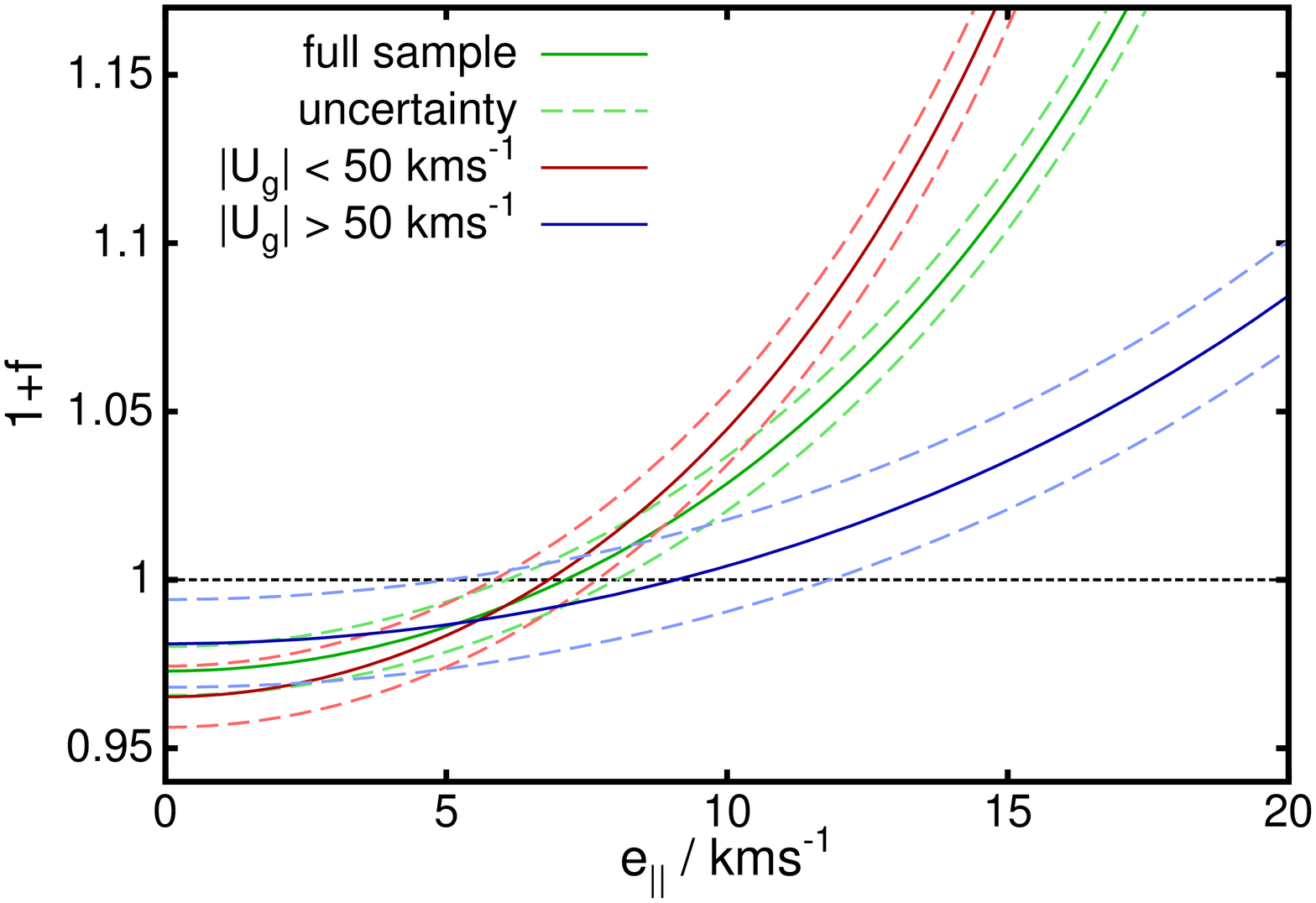,angle=-90,width=\hsize}
\caption{Both panels show the distance estimator $1+f$ against the assumed $\vlos$ source dispersion $\epar$ for the LAMOST sample. In these plots, $\epar$ can be determined by demanding an unbiased distance estimator $1+f \sim 1$. The full sample is shown with green lines, dashed lines show the $1 \sigma$ error interval. The right combination of $f$ and $\epar$ can be tested by cutting the sample into stars with a large vs. small baseline in $VT_{VW}$ or $UT_{UW}$ and demanding mutual agreement. In the top panel, we cut in galactocentric azimuthal velocity $V_g$, and in the bottom panel we cut in the radial velocity $\Ug$. These subsamples are shown with red and blue lines in each panel.}\label{fig:lamostvcorr}
\end{figure}

\subsection{Assessing the source $\vlos$ dispersion: LAMOST}\label{sec:lamostlos}

We now return back to the question of the source $\vlos$ dispersion $\epar$ and ways to differentiate it from parallax errors.
As discussed above, the bias term in equations (\ref{eq:bias}, \ref{eq:vloscorr}) can be used to measure $\epar^2$, if the distance bias $f$ in a sample is sufficiently well-constrained.

This works very well e.g. for the LAMOST dataset. In this case the parallaxes and our method to retrieve the selection function and distance priors have been validated on the RAVE sample to an accuracy better than $2\%$. The source $\vlos$ dispersion $\epar$ can now be determined by plotting the corrected distance bias estimate $1+f$, while varying the source $\vlos$ dispersion $\epar$. This is done in Fig. \ref{fig:lamostvcorr}, where the green lines show this variation for the LAMOST sample, dashed lines represent the $1 \sigma$ error intervals. We use the quality cuts derived in Section \ref{sec:vloserr}, $\SNR > 30$ and $\sigloslamost < 27 \kms$, however, do not use $\sigloslamost$ in the bias correction of $1+f$. Under the assumption that the distance estimates from TGAS parallaxes are unbiased, we obtain $\epar = 7 \pm 1.0 \kms$. This estimate contains a generous allowance for uncertainties in the velocity ellipsoid and proper motions. The systematic uncertainty from the distances is difficult to estimate, but from the previous indications it should be of order $2 \kms$. $\epar$ consists of the $\siglos$ and additional source dispersion from binary systems. However, the latter should be of order $\sigb \lesssim 3 \kms$, so according to equation (\ref{eq:epar}), their contribution to $\epar$ is small.

\subsubsection{Separating distance bias and line-of-sight velocity errors}

Can we separate the line-of-sight velocity errors from a true distance error? In a larger sample, we can. The idea behind this is tested with the red and blue lines in Fig. \ref{fig:lamostvcorr}. These lines show a separation of the LAMOST sample into subsamples with large and small $\Vg$ (top panel), or subsamples with large and small $\Ug$ (bottom panel). We recall that $f$ is estimated based on the correlation of vertical velocities $W$ with the baseline ($y$) components $T_{VW}V$ and $T_{UW}U$, measured by the covariance $\Cov(W,y)$ in equation (\ref{eq:fest}). The selection in this plot target subsamples with a larger or smaller baseline $y$, i.e. while $f$ is similar, both the covariance terms and the variance in the denominator of equation (\ref{eq:fest}) vary. In contrast, the bias on the covariance from the $\vlos$ source dispersion is approximately the same, since it only depends on the sample geometry and $\epar^2$. If the bias from $\epar$ dominates the distance statistics, we hence expect the different sample selections in Fig. \ref{fig:lamostvcorr} to have the same $1+f$ close to $1$ at the correct value of $\epar$, but the sample chosen to be closer to the Sun's velocity (blue lines in Fig. \ref{fig:lamostvcorr}) should have a steeper relationship of $1+f$ against $\epar$ with larger curvature than its counterpart with larger heliocentric velocities (red lines). As a consequence, the low velocity selection (blue) should also have a larger deviation of $1+f$ from $1$ at $\epar = 0$, exactly what is observed in Fig. \ref{fig:lamostvcorr}. In contrast, if the measured deviation is a distance error, $1+f$ of the subsamples should be in agreement at small values of $\epar$. In Fig. \ref{fig:lamostvcorr} both ways to separate the sample agree perfectly on $1+f \sim 1$ at an $\epar \sim 7 \kms$. However, we also note that the still moderate sample size limits the statistical significance: the discrimination on LAMOST without our prior knowledge from RAVE only barely excludes $\epar \sim 0$. 

The cut can be done either in $\Vg$ (top) or $\Ug$ (bottom panel), but it has to be done in galactocentric velocities, since a heliocentric selection would slice through the velocity ellipsoid at an angle and hence correlate $W$ and the horizontal components. Another strategy would be to select the sample at different values of $T_VW$ and $T_{UW}$, since the bias term has a quadratic dependence on these, or to use e.g. a metallicity cut to select kinematically hotter stars that have a larger baseline $y$ by their larger velocities. 

Each of these options has their caveats: The kinematic cuts in $\Ug$ or $\Vg$ select to some extent for the errors we are measuring. Stars with larger velocity errors (either from distance or from velocity uncertainties) are pushed towards the tails of the velocity distribution. A good position to cut in a larger sample are hence the points in the velocity distribution where it has a small or vanishing slope. If one selects in metallicity, the major caveats are a possible difference in the binary fraction, or also a correlation of $\siglos$ with the metallicity (typically $\siglos$ increases for metal-poor, hot stars). 

\begin{figure*}
\epsfig{file=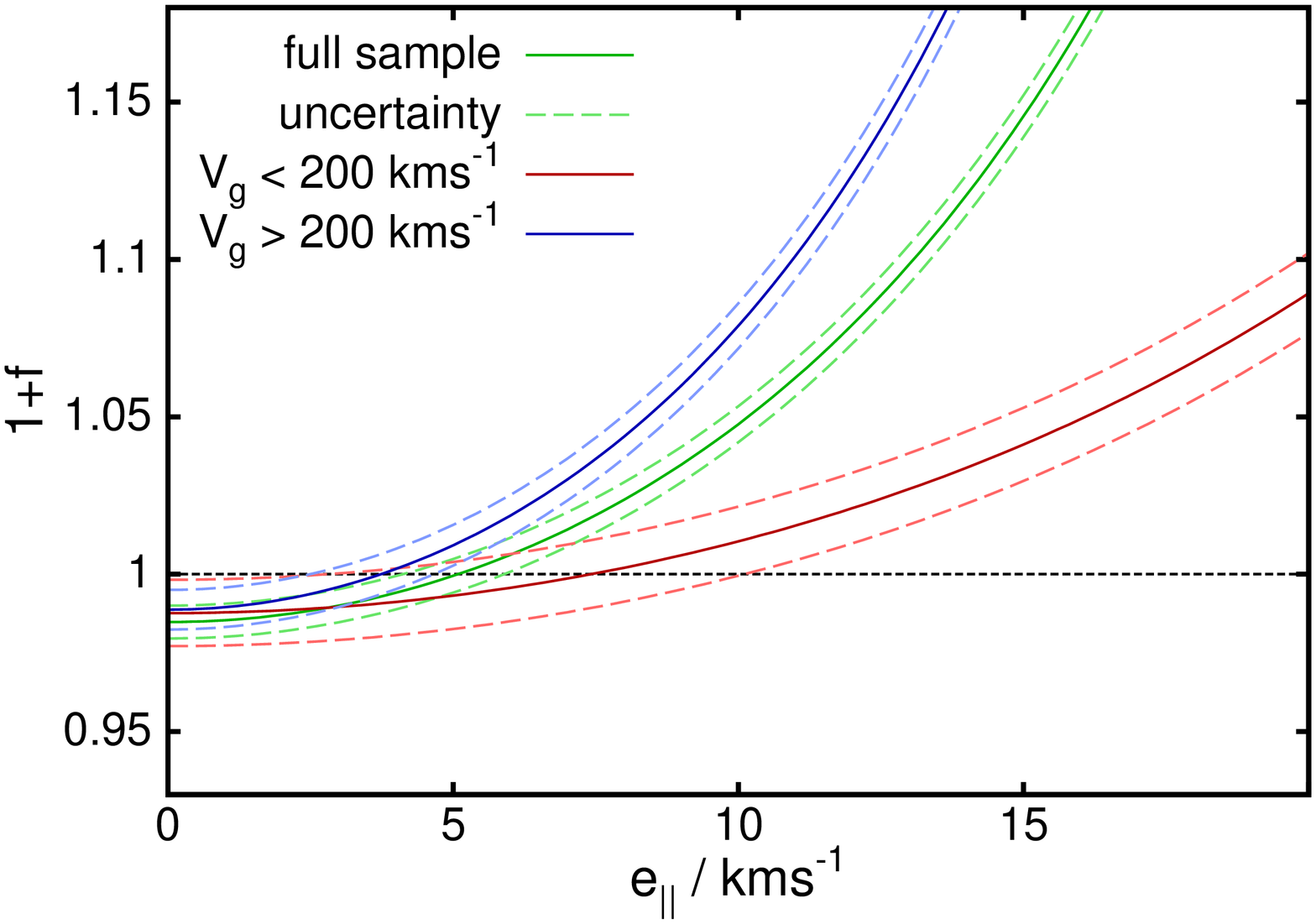,angle=-90,width=0.49\hsize}
\epsfig{file=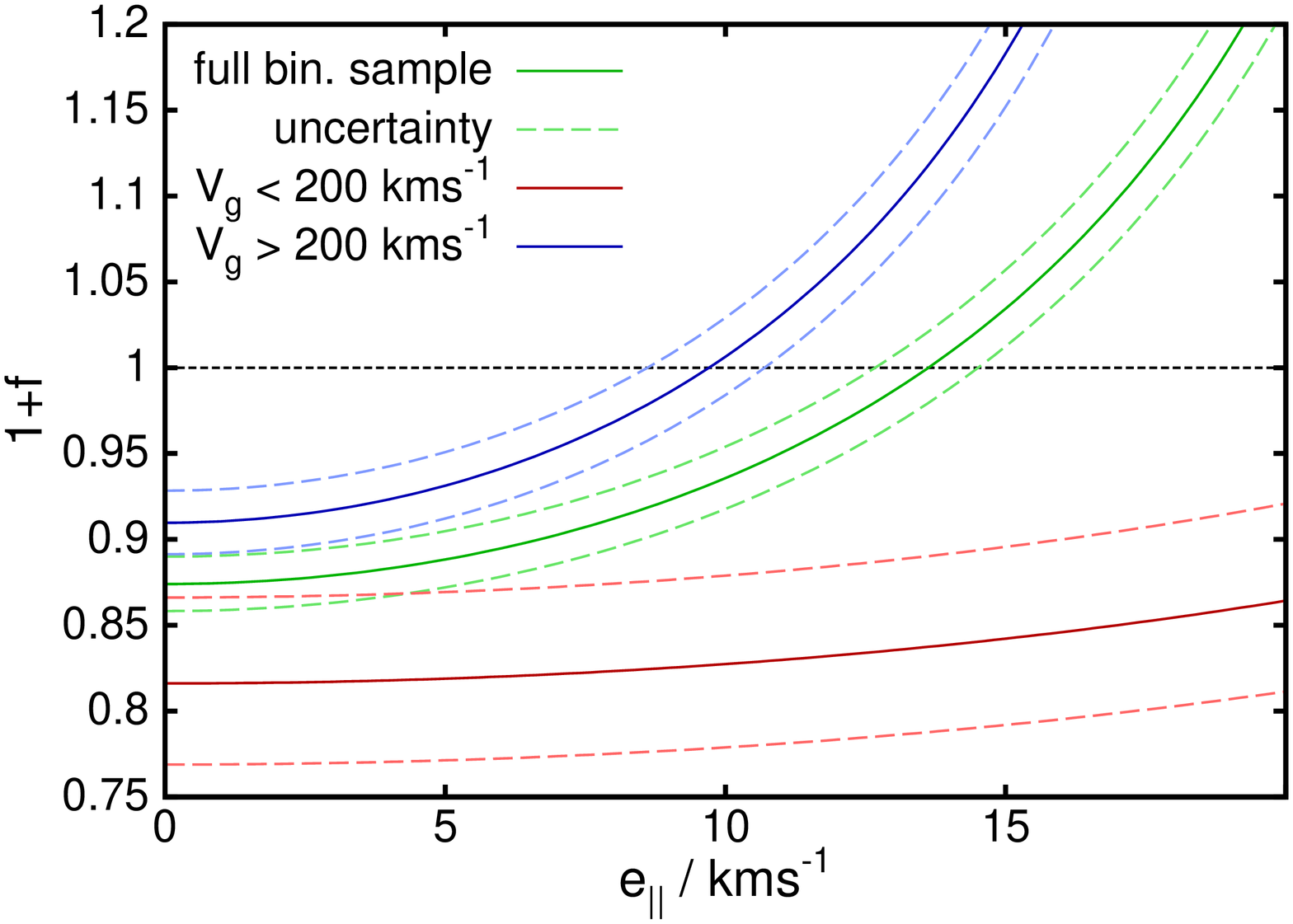,angle=-90,width=0.49\hsize}
\epsfig{file=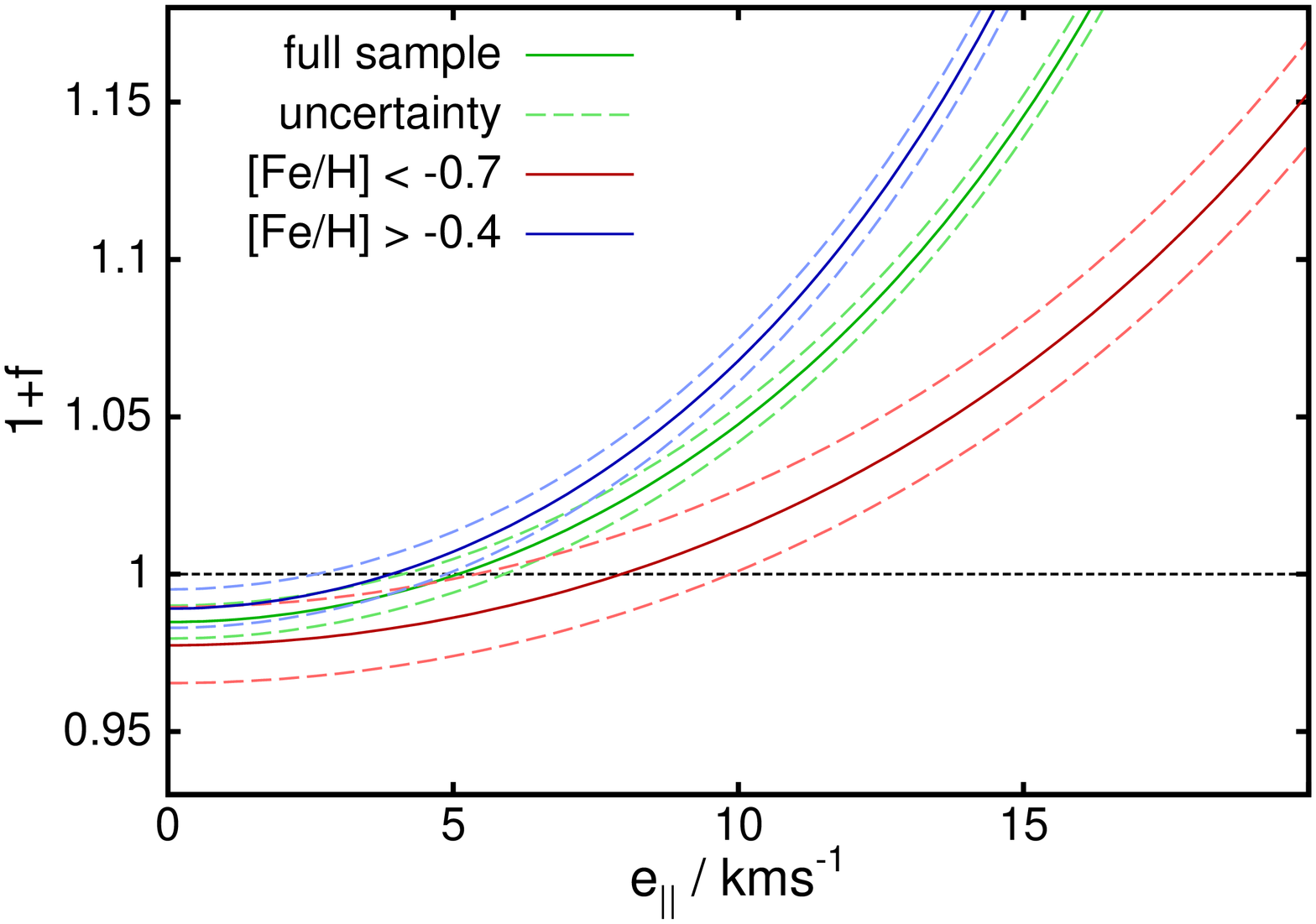,angle=-90,width=0.49\hsize}
\epsfig{file=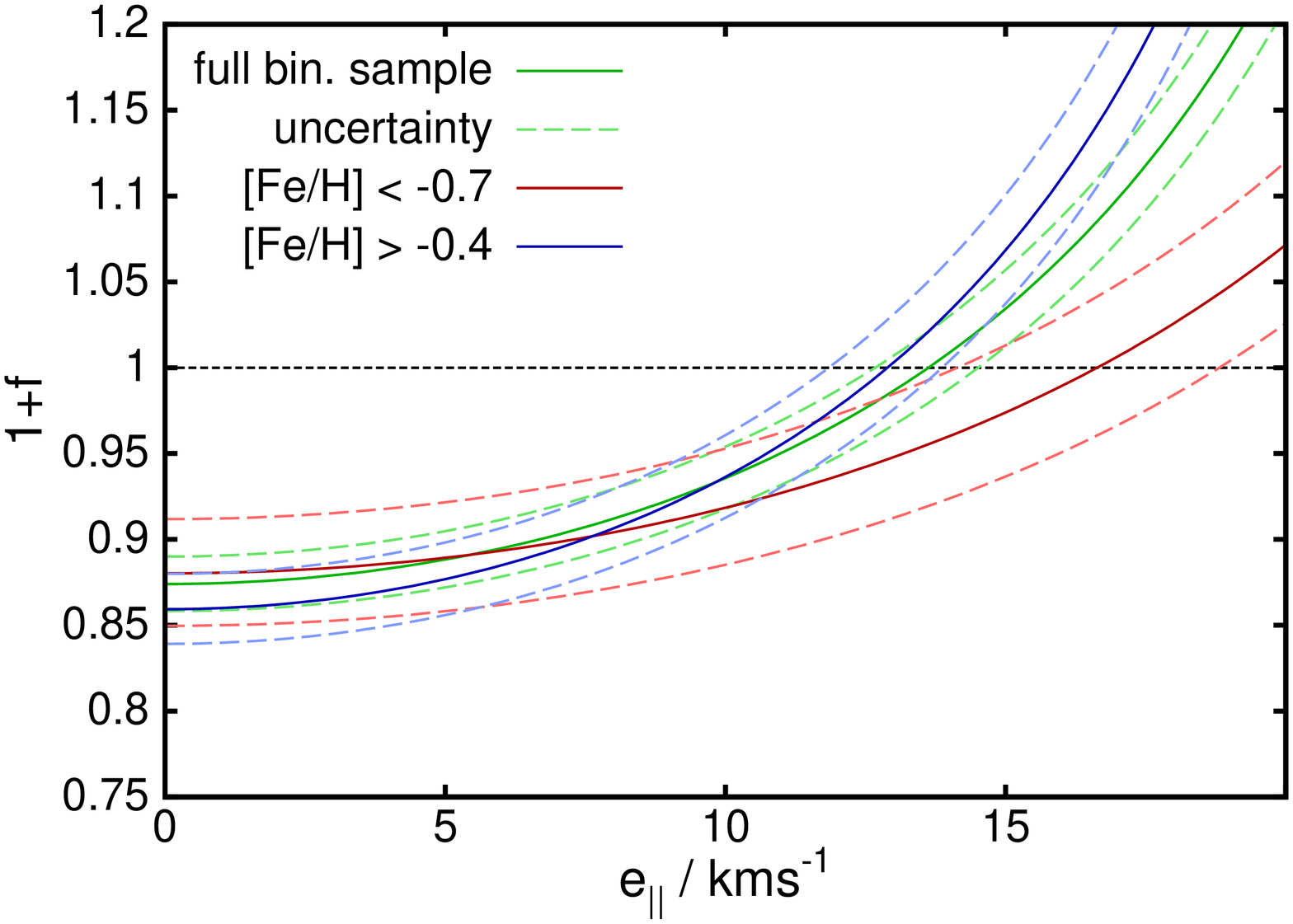,angle=-90,width=0.49\hsize}
\caption{Distance bias $1+f$ vs. $\vlos$ source dispersion $\epar$ in RAVE for the entire sample (left-hand column) and for the subsample carrying binary flags (right-hand column). The plots are organised as in Fig. \ref{fig:lamostvcorr}, the green lines show the full sample, blue lines the subsample with smaller heliocentric velocities, and red lines the subsample with larger heliocentric velocities. Here the top panels show a separation at $\Vg = 200 \kms$, the bottom row shows a separation in metallicity, selecting once stars with $\feh < -0.7\dex$ and once with $\feh > -0.4 \dex$. The top right plot should be taken with some caution, since the low-$\Vg$ binary sample only contains $\sim 400$ stars.}\label{fig:ravevcorr}
\end{figure*}

\subsection{Assessing the line-of-sight velocity uncertainty in RAVE}\label{sec:losrave}

While the main measurement for the LAMOST sample is the line-of-sight velocity errors, the situation in the RAVE sample is far less clear. RAVE line-of-sight velocities have typically errors smaller than $\siglosrave \lesssim 2 \kms$, i.e. an order of magnitude smaller than the LAMOST uncertainties. Consequently, we get a precise measurement on LAMOST line-of-sight velocity errors, but we just see that something is wrong with the RAVE sample, either stellar distances being too short, or some contamination with larger $\vlos$ errors, or binary systems contaminating the data.

Fig. \ref{fig:ravevcorr} gives an overview of the trade-off between parallax/distance bias and $\epar$, similar to Fig. \ref{fig:lamostvcorr}. From the earlier analysis we know that a major part of the problems can be ascribed to the suspected binary systems, so we show the full RAVE sample on the left-hand side and the suspected binary stars on the right-hand side. Note that the subsample of suspected binary stars with $<6000$ stars is very small, so the further separations (which have only $400$ stars in the low-$\Vg$ component) should be taken with caution due to small number statistics.

As we can see from the entire sample printed in green on the left-hand column, the full sample has a significant average distance underestimate ($1+f < 1$) if we assume $\epar \lesssim 1 \kms$ as suggested by the $\siglosrave$ estimates of the RAVE pipeline, which tested fine in the statistics of Section \ref{sec:vloserr}. If we assumed that Gaia-TGAS and our distance derivation were completely bias-free, we would conclude that $\epar \sim 5 \kms$. It would be very unlikely that this is caused by large $\vlos$ measurement error in light of our findings in Section \ref{sec:vloserr}. Just like in LAMOST, we attempt to break the degeneracy by separating the sample in $\Vg$ (top panel) and in metallicity (bottom panel). This sample, however, is biased towards younger stars with small velocity dispersions and disc populations, so we have few useful halo stars. Both separations barely scratch statistical significance in the region of interest. Demanding agreement between the sub-samples, we can only conclude that $\epar < 5 \kms$, and that the statistics favour a problem with the stellar distances. We note that the values for $\epar$ are at least at the upper end of what might be achievable by binary systems \citep[][]{Hargreaves96, Olszewski96}. We tested that a change of the parallax quality to $\sigma_p / p < 10\%$ yields no significant difference.

The two panels on the right-hand side of Fig. \ref{fig:ravevcorr} attempt the same dissection on the subsample of suspected binary stars. While assuming unbiased distances for this subsample would yield a $\vlos$ source dispersion $\epar \sim 14 \pm 2 \kms$, slicing the sample into subsets of either low and high azimuthal velocity (top panel) or low and high metallicity (bottom panel) fails to convince that a velocity dispersion $\sigb$ caused by binary stars is the sole explanation. As discussed above, the samples with large intrinsic baselines in $V T_{VW}$ and $U T_{UW}$ (i.e. the low-$\Vg$ sample, or the metal-poor subsample) should then start at a value of $1+f$ closer to $1$ at $\epar = 0$, and intersect with their small-baseline counterpart at near-zero distance bias $f$. Here we observe the opposite. The low-$\Vg$ subsamples (red lines) start already at more extreme $1+f$ than the high-$\Vg$ subsample (blue lines). There is no agreement between the subsamples anywhere near the region of $1+f \sim 1$. Both the top and the bottom panel suffer from the same discrepancy, though the bottom panel shows a milder version, while having different potential biases: the top panel possibly suffers from a potential direct kinematic bias, whereas the bottom panel could be impacted by binaries being wrongly evaluated as particularly metal-poor. This would be a reasonable suspicion in a low-resolution spectroscopic survey like RAVE. The unresolved binary broadens the lines, and while the equivalent width should not be strongly lowered, different line-shapes might have this effect. In addition the more metal-poor stars could have unaccounted $\vlos$ determination errors affecting both separations. A potential way out would be to claim that the \cite{Matijevic12} classification strongly depends on metallicity, but the most straight-forward explanation is that TGAS astrometry fails for binary systems, in addition to some contribution by $\epar$.

What we should take from this section is: Once we have a larger sample with certainly clean astrometry and a $\vlos$ determination at the quality level of RAVE and later Gaia, this method will be able to constrain even the binary velocity dispersion for different samples, and to differentiate well between $\epar$ und distance uncertainties. The sizes of the current samples are just at the boundary of offering a full solution which breaks the degeneracy. And, we can clearly see from these statistics that stars with binary flags should be excluded from any analysis.

\section{Conclusions}\label{sec:Conclusions}

In this work we have derived distance expectations and kinematics for stars in the RAVE-TGAS and LAMOST-TGAS samples.
\footnote{Please find the datasets with distances and kinematics and our source code at \small{http://www-thphys.physics.ox.ac.uk/people/RalphSchoenrich/data/ tgasdist/data.tar.gz}, 
or request them directly from the authors.}
 We have applied the statistical methods developed by SBA to estimate the distance bias $f$ in the RAVE-TGAS and the LAMOST-TGAS samples, and provide a set of validated stellar distances and kinematics. While there are several papers testing the accuracy and precision of parallaxes from Gaia-TGAS, even perfectly unbiased parallaxes do not guarantee unbiased distances, since there is a large uncertainty in estimating the right priors and selection functions, which enter a Bayesian distance estimate. Our method uses correlations of derived velocities with position in the sky, and hence directly tests the derived distances. We have validated this method using a realistic Galaxy simulation, for which we know the exact distances, as shown in the Appendix.

We find that the entire RAVE-TGAS sample shows statistically a global distance underestimate by of order $2-3 \%$. However, more than half of this signal disappears when we restrict the sample to a subset of stars that are flagged as fully normal objects according to the classifications of \cite{Matijevic12}, which detect binaries or peculiar stars, e.g. with emission lines. We have demonstrated that the objects with at least one binary flag show distance statistics equivalent to distance underestimates in excess of $15\%$.

While it would be tempting to put all the blame of the distance bias on TGAS parallaxes, we have shown that distance statistics $f$ are biased proportional to the source $\vlos$ variance $\epar^2$. This comprises $\vlos$ measurement errors and contamination with binary stars, which act like a distance underestimate on $f$. On the one hand this requires that we find a way to differentiate $\epar$ from a true distance bias. On the other hand it is a useful way to measure $\epar$ and hence determine $\siglos$ on samples where parallaxes are good and the $\epar$ dominates the $f$ estimates. We can use this dependence of $f$ on $\epar$ in two ways: i) we can vary $\epar$ with the condition $f \sim 0$, and thus measure $\epar$; ii) we can probe the accuracy of $\siglos$ estimates in a pipeline by separating each sample in $\siglos$ and looking for trends in $f$. 

We have also examined the distances provided by \cite{Astraatmadja16}. The comparison underlines the importance of applying the right selection function. Neglecting this selection function leads to a relative distance overestimate in AB16. However, their choice to use the mode of their distributions (over-)compensates this effect, which leads to mildly short distance estimates. We advise to use expectation values, which are well-defined statistical quantities.

The LAMOST sample demonstrates this capability. Our method has revealed that the uncertainties given by the LAMOST pipeline have no detectable information content for uncertainties below $\sigloslamost < 27 \kms$, while stars with larger $\sigloslamost$ show a sudden increase in $\epar$. More importantly, when comparing the full distance estimator $f$ with its counterpart $f_V$ that uses only the azimuthal $V$ velocity vs. the vertical motion $W$, these two estimators yield highly inconsistent results. We also find that the LAMOST sample shows a near-perfect relationship between $W$ and $\sin(\gb)$, indicating a $5 \kms$ offset in the $\vlos$ measurements \citep[a similar offset has been suggested in][]{Tian15}, while RAVE has no significant correlation. When we correct LAMOST for this $\delta \vlos = 5 \kms$, we measure $\epar = 7.1 \pm 1 \kms$ with a systematic uncertainty of about $2 \kms$, i.e. the LAMOST $\vlos$ measurements are far more precise than indicated by their nominal errors.  

In RAVE, the full sample starts showing signs of increased $\epar$ above $\siglosrave \ge 1.5 \kms$, and these reach an unacceptable level above $5 \kms$. This behaviour is not detectable when we restrict the sample to unflagged/normal stars, i.e. the flags indicate the most problematic $\vlos$ determinations.

We have demonstrated how to measure the selection function $S(s)$ and distance prior of each sample from the data directly in Section \ref{sec:prior}. We start from a flat $S_0(s)$, measure a new $S_1(s)$ by comparing the stellar distance distribution to the expectation in the prior, and then iteratively insert this selection function into the distance determinations. This procedure converges within $5$ iterations. By separating the sample in distance, we can show that neglect of this selection function (i.e. a flat $S(s)$) results in catastrophic distance bias towards larger distances, while the derived selection function provides unbiased distances within the measurement accuracy. 

It is generally inadvisable to use the naive parallax estimator $s'=1/p$ for stellar distances, because it neglects the proper transformation of the parallax error distribution into $s$, which gives a large distance underestimate against the true expectation value $\langle s \rangle $. However, to our surprise, we find that on both RAVE and LAMOST-TGAS this estimator results only in a mild distance underestimate. The reason for this is that the selection function/distance prior falls steeply towards larger $s$, compensating partly for the missing transformation.

We note that the distance estimator $f$ for RAVE stars with very short distances ($\langle s \rangle \lesssim 0.15 \kpc$) consistently indicate distance underestimates. This can either be an extreme statistical fluctuation, or it might indicate either a problem with the largest TGAS parallaxes, or a stronger selection against stars with large proper motions than indicated in \cite{Lindegren16}. 

We have detected an anomaly either in the stellar kinematics or in the parallaxes for the RAVE-TGAS sample around a galactic longitude $\gl \sim 300 \deg$, and at a distance of $0.3 < s/\kpc < 0.35$. The feature is found on both sides of the Galactic plane, and the correlation between vertical and azimuthal velocities is neither found in its foreground nor in its background. The same sign on both sides of the plane argues strongly against a (halo) stellar stream as source of this. The feature is also robust against outliers, and cannot be explained by eventual anomalies in the selection function. A likely explanation is substructure at this position in the Galaxy, e.g. a breathing mode near a spiral arm or spur, though we are not aware of such a feature at this position. Further analysis will be required to understand if this is a possible localised failure of TGAS astrometry, or in the other case, to unravel the exact nature of this structure.

We have further shown that the degeneracy between $\vlos$ source dispersion $\epar$ and a true distance bias can be broken by selecting subsamples with different lengths of baseline in $VT_{VW}$ or $UT_{UW}$. This is achieved by splitting the sample either with a metallicity selection or directly in stellar kinematics. The larger upcoming Gaia data releases will be sufficient to make this method work. In the current samples, this ability is just borderline significant. It shows that while we have a consistent picture for LAMOST being dominated by line-of-sight velocity uncertainties, the statistics suggest that the deviations in the RAVE survey, in particular on the binary subsample, derive from a combination of astrometric problems and intrinsic $\vlos$ dispersion.

\section*{Acknowledgements}
It is a pleasure to thank the referee, L. Eyer, J. Binney, W. Dehnen, and P. McMillan for helpful discussions and comments. This work used the DiRAC Data Centric system at Durham University, operated by the Institute for Computational Cosmology on behalf of the STFC DiRAC HPC Facility (www.dirac.ac.uk. This equipment was funded by a BIS National E-infrastructure capital grant ST/K00042X/1, STFC capital grant ST/K00087X/1, DiRAC Operations grant ST/K003267/1 and Durham University. DiRAC is part of the National E-Infrastructure. This work was supported by the European Research Council under the European Union's Seventh Framework Programme (FP7/2007-2013)/ERC grant agreement no.~321067. 
{This  work  has made use of data from the European Space Agency (ESA)  mission Gaia (http://www.cosmos.esa.int/gaia), processed by the Gaia Data Processing  and  Analysis  Consortium  (DPAC, http://www.cosmos.esa.int/web/gaia/dpac/consortium). Funding  for  the  DPAC  has  been  provided  by  national institutions, in particular the institutions participating in the Gaia Multilateral Agreement.}

\appendix
\section{Validation on galaxy simulations}\label{sec:appendix}

To perform a sanity check on our method and to confirm its robustness against galactic substructure, we employ galaxy simulations that resemble the Milky Way. The simulations are described in detail in \citet{abs16a} and \citet{ab17}. They feature a reasonable bar, have appropriate thin and thick disc components and an active spiral pattern. 

We focus on model $\Vmsechs$. To build this model, we continuously add stellar particles on disc orbits to a model galaxy within a live dark matter halo over the course of $12$ Gyr. The birth velocity dispersions of the stellar populations decline continuously in this model, from $>40\kms$ during the first $1.5$ Gyr to essentially near-circular orbits after $4$ Gyr. This creates a disc galaxy with reasonable thin and thick disc components. 

Moreover, the model grows inside-out with a declining star formation history. $\Vmsechs$ has a final baryonic mass $M_{\rm baryons}=6\times10^{10}\Msun$, and a live dark matter halo with mass $M_{\rm dm}=1\times10^{12}\Msun$, which is set up with an initial concentration parameter of $c=6.5$.

At our evaluation time 12 Gyr, the vertical profile is double-exponential. At $R=8\kpc$, it can be fitted with exponential scaleheights $h_{z, {\rm thin}}\approx285\pc$ and $h_{z, {\rm thick}}\approx 990\pc$. The thick disc contributes $\sim 20 \%$ to the local surface density. The final model has a bar of length $\sim 4\kpc$. Its $R=8\kpc$ circular speed, dark matter density and baryonic surface density compare reasonably with observational constraints for the Solar neighbourhood. The level of radial migration in the model agrees with constraints from MW chemical evolution and its Solar neighbourhood velocity distributions show good agreement with RAVE+TGAS data if one corrects for selection effects. Overall, the model is well suited to create mock Solar neighbourhood samples.

\renewcommand{\tabcolsep}{4pt}
\begin{table}
\caption{Distance error estimates $f$ in simulations with perfect distances. Each simulation has $25$ samples around 
$R= 8 \kpc$ at equidistantly spaced positions in azimuth. The first three columns give statistics for the full distance estimator $f$. $\langle f \rangle$ is the mean value of $f$ from averaging over all $25$ samples, $\sigma_f$ is the residual dispersion of $f$, and $\sigma_{f,{\rm st}}$ is the average of the statistical errors derived with $f$. If the estimates are true, then $\sigma_f$ should be close to $\sigma_{f,{\rm st}}$. Systematic biases in $f$ depending on the sample position would show as $\sigma_f$ significantly larger than $\sigma_{f,{\rm st}}$. The last three columns give the same statistics for $f_V$.}
\begin{tabular}{l||r|r|r|r|r|r|}
model&$\langle f \rangle$&$\sigma_f$&$\sigma_{f,{\rm st}}$&$f_V$&$\sigma_{f,V}$&$\sigma_{f,V,{\rm st}}$ \\ \hline
$\Vmsechs$&$-0.0004$&$0.014$&$0.014$&$0.0015$&$0.033$ &$0.022$ \\
V$\alpha$5$\lambda$&$-0.0073$&$0.018$&$0.013$&$-0.0064$&$0.024$&$0.019$ \\
M$\alpha$1$\zeta$*&$0.0026$&$0.019$&$0.017$&$-0.0013$&$0.030$&$0.027$ \\
P2&$-0.0064$&$0.020$&$0.017$&$-0.0066$&$0.035$&$0.028$ \\
EHR2&$-0.0072$&$0.010$&$0.008$&$0.0009$&$0.014$&$0.013$\\
\end{tabular}\label{tab:diststatsim}
\end{table}

\begin{figure}
\epsfig{file=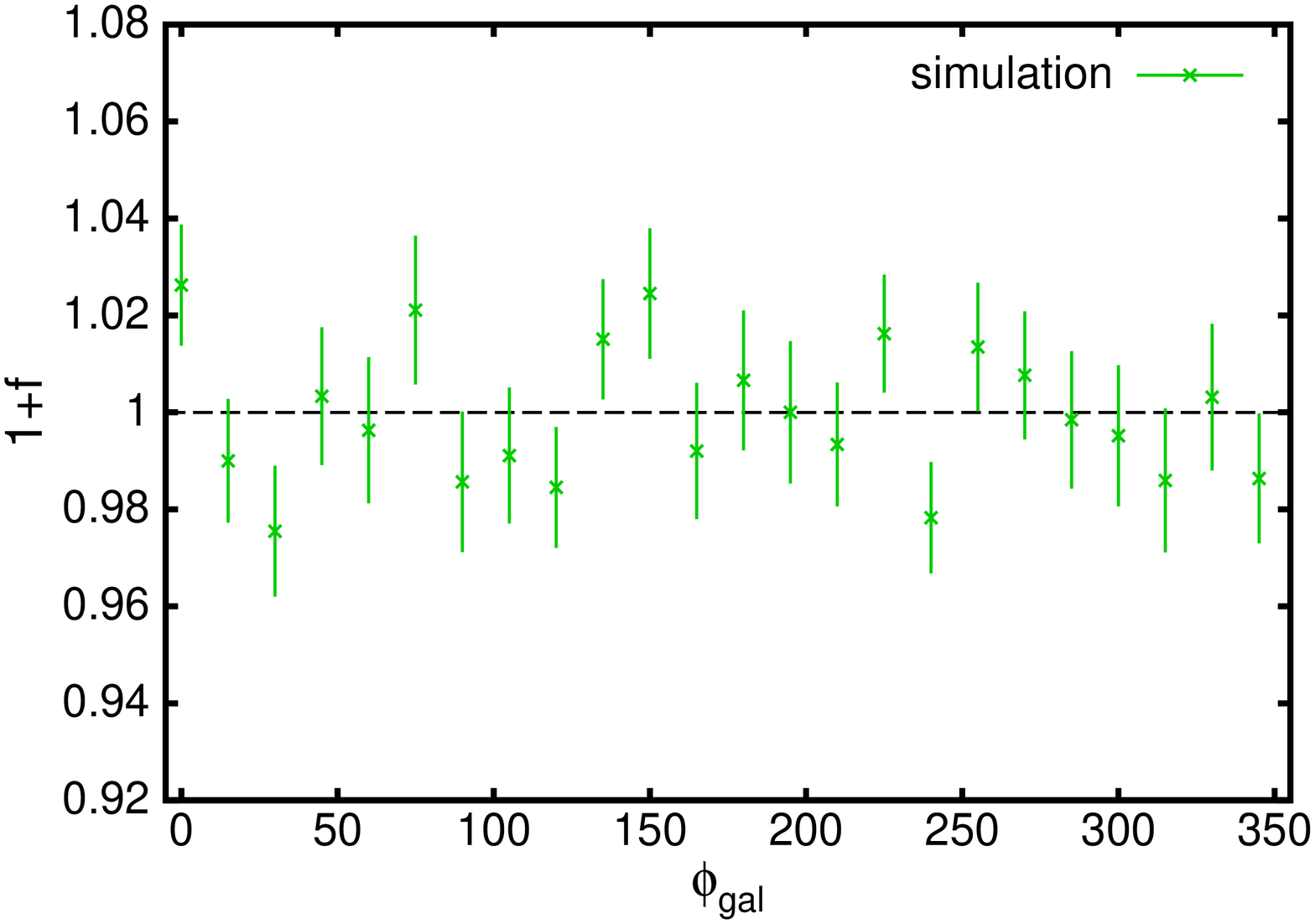,angle=-90,width=\hsize}
\epsfig{file=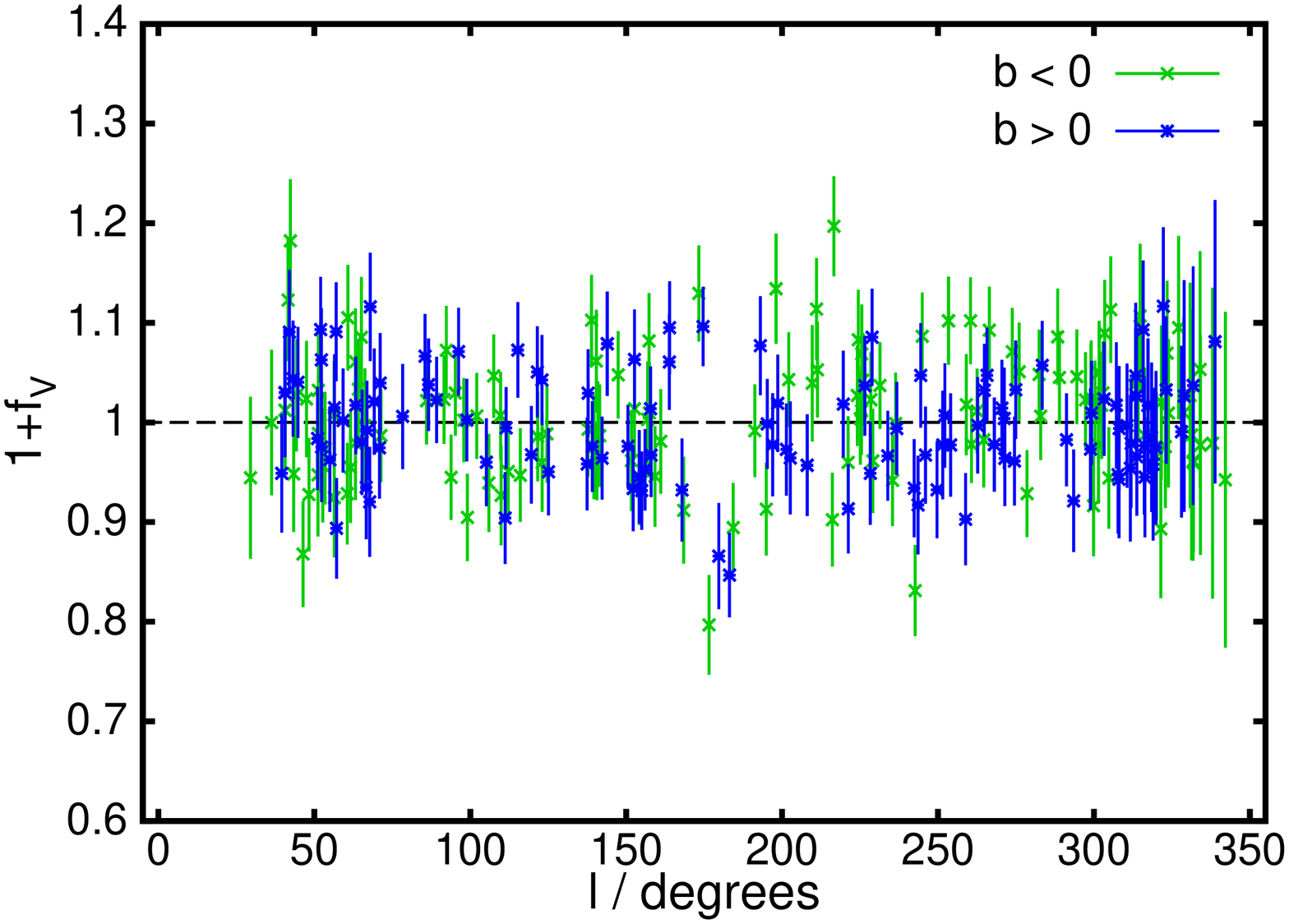,angle=-90,width=\hsize}
\caption{Testing the distance estimator on N-body galaxy model $\Vmsechs$. The top panel shows the distance statistics for $1 \kpc$ sized samples taken around simulated solar positions at $(R,z) = (8.3, 0) \kpc$, equally spaced in azimuth $\Phi_{\rm gal}$ (x-axis). The bottom panel shows the restricted distance estimator $f_V$ when we bin each sample in Galactic longitude $\gl$ (x-axis) and separate the sample into positive (blue) and negative (green) galactic latitude $\gb$.}\label{fig:nbodytest}
\end{figure}

To create these samples, we place the observer at $(R,z) = (8.3, 0)\kpc$ choosing $25$ different positions, which are equally distributed in galactocentric azimuth $\Phi_{\rm gal}$. We then select all stars with distances $s<1\kpc$, obtaining samples of about $15-20000$ stars each. In these samples, we calculate the stellar positions, proper motions and line-of-sight velocities using a solar motion of $(\Usun,\vsun,\Wsun) = (11,250,7)\kms$ (we note that the only effect of the set solar motion is to alter the length of the baseline component $T_{VW}V$ and thus the precision of $f$). Fig. \ref{fig:nbodytest} shows the tests of our distance statistics on these quite realistic samples. The top panel shows the full distance estimator $1+f$ for each of the $25$ observer positions in $\Phi_{\rm gal}$. The test statistics are consistent with a perfectly unbiased estimator. There are $9$ datapoints beyond $1 \sigma$, which is just the expected number of outliers. 

The mean distance estimator $<f>$ in each sample, the dispersion of the distance estimates $\sigma_f$, and the expected dispersion from shot noise and uncertainty in the correction of the velocity ellipsoid tilt $\sigma_{f,{\rm st}}$ for model $\Vmsechs$ are summarized in the first line of Table \ref{tab:diststatsim}. We also show the numbers for four additional Milky Way like models. Models V$\alpha$5$\lambda$ and M$\alpha$1$\zeta$* are of the same type as $\Vmsechs$, but differ in model parameters and consequently in structural details at final time. Models P2 and EHR2 also contain both thin and thick disc components, but the thick disc components are created as initial conditions and the thin disc is populated with stars born on near-circular orbits over 10 Gyr.

The values for all five models included in Table \ref{tab:diststatsim} demonstrate that, on average, the distance statistics show no bias at the levels needed for this work (generally less than $1 \%$). Moreover, the statistically expected scatter almost matches the observed scatter in $1+f$, i.e. any impact by unaccounted systematics in the samples is significantly below $1 \%$. We note that the dispersion values are lowest in EHR2 as this model has a higher resolution and thus larger sample sizes than the other four. $f_V$ is in principle more vulnerable to disc structure, but we see no clear indication for a significant additional scatter in its statistics provided in the three right-hand columns. 

As detailed in Section \ref{sec:caveats} the fits become more risky when we separate samples in longitude and/or latitude, because effects caused by galactic structure do not cancel out any more to first order. This is particularly the case for simulation EHR2, which has a longer bar than the Milky Way. As the bar thus has a larger impact at $R=8\kpc$, in that simulation, cutting the test samples at $\gl = 180 \deg$ results in a scatter of $\sim 5 \%$ in $f_V$. As expected, most of this gets compensated in the full estimator $f$, which has just $2 \%$ dispersion (compared to about $1.2 \%$ statistically expected dispersion). This simulation shows a discernible $m = 2$ pattern when the $f_V$ statistics are plotted against $\Phi_{\rm gal}$. However, as we see in Table \ref{tab:diststatsim} even this simulation does not yield a discernible error if we use a full sky coverage. 

The bottom panel in Fig. \ref{fig:nbodytest} looks for trends in the restricted distance estimator $f_V$ when separating simulation $\Vmsechs$ from the top panel in galactic longitude $\gl$, selecting subsamples of $3000$ stars each. Before ordering the sample in $\gl$ we have additionally cut it into northern and southern hemisphere ($\gb > 0$ and $\gb < 0$). There is no detectable systematic deviation, apart from a minor issue with the error determination near $\gl = 180 \deg$. At that position $T_{VW} \sim 0$, i.e. there is virtually no signal for $f_{V}$, which also affects the error determinations in a small region. 

Similarly, the increase in uncertainties of $f_{V}$ is due to the vanishing of $T_{VW}$. Overall, the statistics of this sample look fine - due to the cold stellar kinematics, and the small sample sizes, the statistically expected errors are $\sim 12\%$, close to the observed dispersion.

\label{lastpage}

\end{document}